\documentclass[pof,reprint,amsmath,amssymb]{revtex4-2}

\usepackage{graphicx}
\usepackage{dcolumn}
\usepackage{bm}

\usepackage[utf8]{inputenc}
\usepackage[T1]{fontenc}
\usepackage{mathptmx}
\usepackage{etoolbox}

\makeatletter
\def\@email#1#2{%
	\endgroup
	\patchcmd{\titleblock@produce}
	{\frontmatter@RRAPformat}
	{\frontmatter@RRAPformat{\produce@RRAP{*#1\href{mailto:#2}{#2}}}\frontmatter@RRAPformat}
	{}{}
}%
\makeatother

\usepackage{capt-of}
\usepackage{float}

\draft 
\begin{document}
	\title{An open source MATLAB\textsuperscript{\textregistered} package to perform basic and advanced statistical analysis of turbulence data and other complex systems.}
	
	
	\author{André Fuchs}
	\email{andre.fuchs@uni-oldenburg.de}
	\affiliation{Institute of Physics and ForWind, University of Oldenburg, Küpkersweg 70, 26129 Oldenburg, Germany}
	\author{Swapnil Kharche}
	\affiliation{IRIG-DSBT, CEA Grenoble, 17 rue des Martyrs, 38054 Grenoble, France}
	\author{Matthias W\"achter}
	\affiliation{Institute of Physics and ForWind, University of Oldenburg, Küpkersweg 70, 26129 Oldenburg, Germany}
	\author{Joachim Peinke}
	\affiliation{Institute of Physics and ForWind, University of Oldenburg, Küpkersweg 70, 26129 Oldenburg, Germany}


	\date{\today}
	
	\begin{abstract}
		We present a user-friendly open-source MATLAB\textsuperscript{\textregistered} package developed by the research group Turbulence, Wind energy and Stochastics (TWiSt) at the Carl von Ossietzky University of Oldenburg. 
		Firstly, this package helps the user to perform a very basic statistical analysis of a given turbulent data set which we believe to be useful to the entire turbulence community.
		It can be used to estimate the statistical quantities of turbulence such as the spectrum density, turbulent intensity, integral length scale, Taylor microscale, Kolmogorov scale and dissipation rate.
		Different well-known methods available in the literature were selected so that they can be compared.
		Secondly, this package also performs an advanced 
		analysis which includes the scale-dependent statistical description of turbulent cascade using the Fokker-Planck equation which consequently leads to the assessment of integral fluctuation theorem.
		This is utilized to estimate velocity increments, structure functions and their scaling exponents, drift and diffusion coefficients of the Fokker-Planck equation and consequently the total entropy production of the turbulent cascade.
		As a precondition for the stochastic process approach, Markovian properties of the turbulent cascade 
		in scale are tested.
		The knowledge of a Fokker-Planck equation allows to determine
		for each  independent cascade trajectories a total entropy production. 
		The estimation of total entropy production allows to verify a rigorous law of non-equilibrium stochastic thermodynamics, namely the integral fluctuation theorem, which must be valid if Markov properties hold and the Fokker-Planck equation is correct.
		This approach to the turbulent cascade process has the potential for a new way to link the statistical description of turbulence, 
		non-equilibrium stochastic thermodynamics and local turbulent flow structures. 
		At last we want to emphasize that the presented package can be used also for the analysis of other data with turbulent like complexity.
	\end{abstract}
	
	
	
	\maketitle 

	\section{Introduction} \label{sec:intro}

	The phenomenon of turbulence has been known to mankind for many centuries.	
	One of the best-known models for describing turbulent flows is the phenomenologically inspired energy cascade model proposed by Richardson \cite{Richardson22}.
	In this model, turbulence is understood as the evolution of turbulent structures on different spatial or temporal scales. 
	The central assumption of this cascade model is that the kinetic energy is transferred through all scales of the inertial range by the repeated random breakup of eddies, to increasingly smaller eddies in a cascade-like process. Dissipation of kinetic energy into heat will not take place in this inertial range of the cascade but on smaller scales in the so-called dissipation range.
	Based on this assumption, in the famous dimensional analysis initiated by the work of Kolmogorov \cite{kolmogorov1941dissipation,Kolmogorov1941,kolmogorov1941dit} and Obhukov \cite{Obukhov1941,Obukhov1941a} in 1941 (known as the K41 theory), the cascade model by Richardson is extended with a theory for fully developed, isotropic and homogeneous turbulence (HIT).
	This kind of turbulence, assumed to be stationary, represents an idealized case that should have universal features. 
	The intermittency phenomenon is considered as one of the key signatures of turbulence, which is still not fully understood \cite{Sreenivasan_1997}.
	Kolmogorov himself \cite{kolmogorov1962refinement} and Obukhov 	\cite{Oboukhov_1962} refined the K41 theory by allowing fluctuation of the transferred energy leading to a log-normal distribution for the local energy rate and consequently to intermittent, i.e. non-Gaussian velocity fluctuations
	This theory 
	is known as K62 theory.
	An overview of various intermittency correction models along with fractal and multifractal models can be found in \cite{Frisch_1995,Sreenivasan_1997}. 
	To characterize various types of turbulent flows such as atmospheric flows, jet flows, boundary layer flows, grid turbulent flows, wake flows and von Kármán flows, in a statistical way it is beneficial to have a common data post-processing tool in turbulence. 
	For this purpose, this open-source package is created to perform a basic statistical analysis of turbulence data (here we focus on one-dimensional velocity time series).
	We are not aware of such a comprehensive compilation of standard analyses commonly used in turbulence research in a single user-friendly package.
	Along with the basic statistical analysis of turbulence data the application to Fokker-Planck equations and Integral fluctuation theorem, which provides a comprehensive statistical description in terms of the complexity of turbulent velocity time series, is implemented into this package.
	This analysis is very complex and time-consuming by means of developing a post-processing algorithm.
	With this package, for the first time, the application of this new method in practice is possible for everyone easily and quickly.

	All in all, this 
	open source package 
	enhances the practicability and availability of both the standard analyses already established in turbulence research and the extended analysis via Fokker-Planck equation and integral fluctuation theorem.  
	We believe that this package should be shared with the community as it is of great interest, especially to young scientists and those who start with this topic. It possesses a high reuse potential. 
	
	Some efforts are put into a collection of different alternative methods for the estimation of quantities. This allows to see if consistent results are obtained, or to see which method works better. It would be of benefit if other researcher add further methods to our collection.\\
	
%
%
%
	The open source package  can also be used by researchers outside of the field of turbulence.
	Using this package may contribute to new insights into the complex characteristics of 
	scale-dependent processes (in space or time), as it is known for many fractal systems, ranging from physics to meteorology, biology systems, finance, economy, surface science and medicine.
	The package can be applied to these types of data (one-dimensional time series), but it is important to keep in mind that a variety of turbulence-specific assumptions like the theory of HIT or the Taylor hypothesis of frozen turbulence are assumed as the analyses proceed. The meaning of turbulent specific aspects for other systems has to be evaluated critically. 
	A central feature is the Markov property (see details in Section \ref{markov_est}). 
	It has turned out that the method of analysis based on the Markov processes can also be successfully applied
	to the characterization of rough surfaces \cite{Jafari_2003,Fazeli_2008,Waechter_2003},
	Rogue waves \cite{Hadjihosseini_2014,Hadjihosseini_2016,hadjihoseini2018rogue}, and financial time series \cite{RENNER_2000,Renner_2001,Nawroth_2010} to name just a few.
	Such spatially disordered structures can be analyzed as scale-dependent stochastic processes (see also \cite{Friedrich_2011,Peinke2018}).
	It has to be verified whether the investigation of entropy also contributes to the characterization of extreme events in the data the user investigates.\\

\begin{figure}[h]
		\centering
		\includegraphics[width=0.5\textwidth]{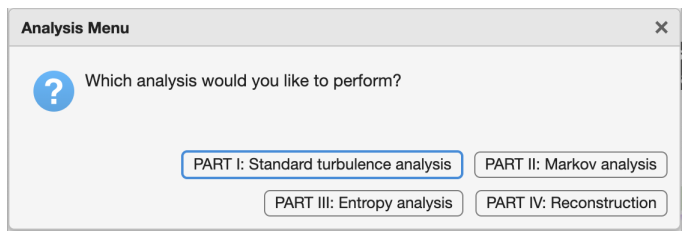}
		\caption{Question dialog box that allows the user to select the analyses to be performed.}
		\label{fig:question} 
	\end{figure}

	The central part of this paper is a detailed discussion of the procedure of how to use this package.
	This paper covers only the discussion of turbulent flows satisfying the assumption of homogeneous, isotropic turbulence.
	In principle the tools of this package can be applied to turbulence data in other configurations and turbulent flows with non ideal HIT conditions as shown for non fully developed turbulence \cite{luck1999experimental} or
	turbulence with shear \cite{Reisner_1999,Ali_2019}. 
	Here again, a careful interpretation of the results is advised.
	A general overview of all functions to carry out various analyses is given as part of this discussion in Section	\ref{part1}-\ref{part4}. 
	All subroutines that can be accessed in this package and the setting the relevant parameters are called in a logical order using the \texttt{main} function.
	Overall, the analysis consists of four parts: PART I: Standard turbulence analysis, PART II: Markov analysis, PART III: Entropy analysis and Part IV: Consistency check. 
	A question dialog box is displayed where the user can select the analyses to be performed (see Fig.~\ref{fig:question}).
	
	\newpage
	To use this package, it is beneficial if the user has an adequate knowledge about turbulent flows. Nevertheless, for the fundamental understanding of turbulent flows we recommend standard books \cite{pope2001turbulent,Frisch_1995}.
	For a quick and simple overview of the theoretical background associated with the turbulent cascade process, Fokker-Planck Equation (FPE), Integral fluctuation theorem (IFT) and its interpretation, we recommend the following review article by Peinke, Tabar \& Wächter \cite{Peinke2018}.\\
	
	Note, all subroutines are also accessible from the command line and can be included by the user in other applications. Additionally all abbreviations used in this document are listed at the end in the list of nomenclature/abbreviations. Also, it should be kept in mind that many of the equations discussed in this paper refer to the theory of fully developed, isotropic and homogeneous turbulence. Thus, the use is to be done with caution, as the assumption in the experimental data may not be fulfilled and can lead to incorrect/misleading results.
In general, velocity is a vectorial quantity (with respect to its coordinate system), and it is dependent on the physical location in space. 
In this paper, the streamwise component of velocity is considered and features are discussed for  this single velocity component. 
Other components may be characterized similarly yet their behavior may largely vary.
Note that the analysis discussed in \mbox{PART  II} can be extended to more component data sets like it is done in \cite{Siefert_2004,SIEFERT_2006,Siefert}.

	\subsection{Quality control}
	\label{sec:reuse}
	This package has been used continuously within our lab since 2018. 
	It also has been successfully used by a large number of students (practical exercises which were part of the fluid dynamics lecture at the University of Oldenburg) to ensure stability across different machines and operating systems. 
	%
	%
	Furthermore, this package has proved its value in a number of research publications \cite{Reinke_2018,Ali_2019,Peinke2018,cascades_II_2019,Fuchs_2020,fuchs2021entropy,fuchs2021instantons,iti2021progress}.
	%
	The package itself or parts of it, as well as results obtained by using the package for analyzing turbulent data, have also been presented at several international conferences.

	\subsection{Implementation and architecture}
	This software is implemented in MATLAB\textsuperscript{\textregistered} (2022a), which is a high-level, matrix-based programming language
	designed specifically for engineers and scientists to analyze data, develop algorithms, and create models.
	The package is available as free software, under the GNU General Public License (GPL) version 3. 
	The package (source code and standalone applications (64-bit) for Windows, macOS and Linux) and a typical dataset can be downloaded from the repository on GitHub or Matlab File Exchange Server to replicate all the results presented in this article. 
	Support is available at  \mbox{\url{github.com/andre-fuchs-uni-oldenburg/OPEN_FPE_IFT}}, where questions can be posted.
	%
	%
	
	Before using this script the following toolboxes should be included in your MATLAB license.
	\begin{itemize}
		\itemsep0em 
		\item Curve Fitting Toolbox
		\item Optimization Toolbox
		\item Parallel Computing Toolbox
		\item Signal Processing Toolbox
		\item Statistics and Machine Learning Toolbox  
	\end{itemize}
	As these MATLAB toolboxes are essential for the application of the package, the compatibility to Octave cannot be provided.  
	But to enhance the accessibility, standalone applications (64-bit) for Windows, macOS and Linux are also created to run the MATLAB code on target machines that do not have a MATLAB license.

	\subsection{Typical dataset and system requirements}
	To demonstrate the application of this program an typical dataset obtained in a turbulent air jet experiment by \mbox{by Renner \textit{et al.} \cite{renner2001}} is used within this document. 
	The local velocity was sampled by a time-resolved single hot-wire measurement \mbox{(Dantec 55P01)} with a spatial resolution of 1.25 mm and a maximal time resolution of 30 kHz.
	This dataset is composed of 10 independent measured data sets and the data acquisition comprises in total \mbox{$1.25 \times 10^7$ samples} with a sampling frequency of 8\,kHz.
	The package additionally includes this typical dataset. 
	We also provide the generated plots/results allowing the user to verify if the software is operating correctly.
	Table~\ref{tab:overview} lists all parameters that the user must enter during the analysis to reconstruct the results shown in this paper.
	\begin{table}[h]
		\centering
		\begin{tabular}{c|c|c|c|c}
			$Fs$  & $L$ &  $\lambda$ &  $inc\_bin$ & $\Delta_{EM}$\\ 
			\hline 
			8000\,Hz & 0.067\,m	& 0.0066\,m  & 93  & 22 samples\\ 
		\end{tabular} 
		\caption{\label{tab:overview} Information that the user must enter during the analysis to reconstruct the results shown in this paper. With the sampling frequency $Fs$, integral length scale $L$, Taylor length scale $\lambda$, number of bins to be used to divide the velocity increment series $inc\_bin$ and Einstein-Markov length $\Delta_{EM}$.}
	\end{table}
	
	The system requirements (memory, disk space, processor, total run time) demanded by the script depend very much on the size of the data set to be examined, the resolution in relation to the number of bins as well as the available number of CPUs.
	The required memory will be allocated to each MATLAB computational engines during parallel computations using multicore CPUs and computer clusters.
	A typical processing time for this data set using 4 physical cores and 16 GB ram is about 60 minutes.

	\newpage
	\section{Part I: Standard turbulence analysis}
	\label{part1}
		In this section, various subroutines to perform basic statistical analysis of the turbulence data are presented. 
		This mainly includes the verification of stationarity of data set, filtering of the signal, estimation of turbulence length scales and dissipation rate along with the assessment of structure functions from velocity increments.
	
	\subsection{Loading data and variables}
		In the very first step, the user is asked to load the data set and to choose the folder path to save the results. Furthermore, the user has to input the sampling frequency of the data and the kinematic viscosity of the fluid.\\
	
	\texttt{uiimport} is the first command in the program which asks the user to select interactively the data file which will be used for the analysis. 
	At this point, it is possible to specify the percentage of the total data that will be used to perform the analysis (for example, the first 20 \% of the data). 
	This feature is sometimes of help if one wants to make a first fast check of data and program.
	Note that this parameter has a significant effect on the overall performance of the script.
	\begin{figure}[h]
		\centering
		\includegraphics[width=0.5\textwidth]{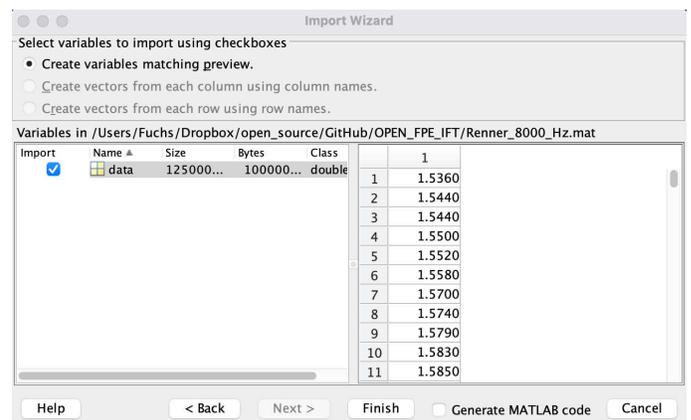}
				\caption{The import tool lets the user interactively select the data file which will be used for the analysis.}
		\label{fig:import} 
	\end{figure}	
	
	\texttt{save\_path} opens a dialog box to navigate through the local directories in order to select a folder for saving figures and files.\\
	
	\texttt{save\_name} generates a pop-up dialog box to enter the name for the analyzed data and figures to be saved.\\
	
	\texttt{Fs} generates a pop-up dialog box to enter the sampling frequency of the data in Hz.\\

	\texttt{kin\_vis } generates a pop-up dialog box to enter the value of kinematic viscosity $\nu$ in $m^2/s$ of the fluid for which the experimental data has been acquired.\\
	
	\texttt{increment\_bin} generates a pop-up dialog box to specify the number of bins to be used to divide the velocity increment series. 
	Note that this parameter has a significant effect on the overall performance of the script.
	A first estimation of the number of bins is made using
	\begin{eqnarray}
		inc\_bin=10\frac{max(data)-min(data)}{u'},
	\end{eqnarray}
	with $u'$ is the standard deviation of the data.  

	\subsection{Test of stationarity and filtering of the time series}
		For many statistical quantities to characterize a data set it is erequired that the data are stationary. 
		Thus before performing the basic turbulence analysis of the data set the stationarity of the data using different functions and parameters as mentioned below is examined.\\
	
\begin{figure*}
		\centering
		\includegraphics[width=0.8\textwidth]{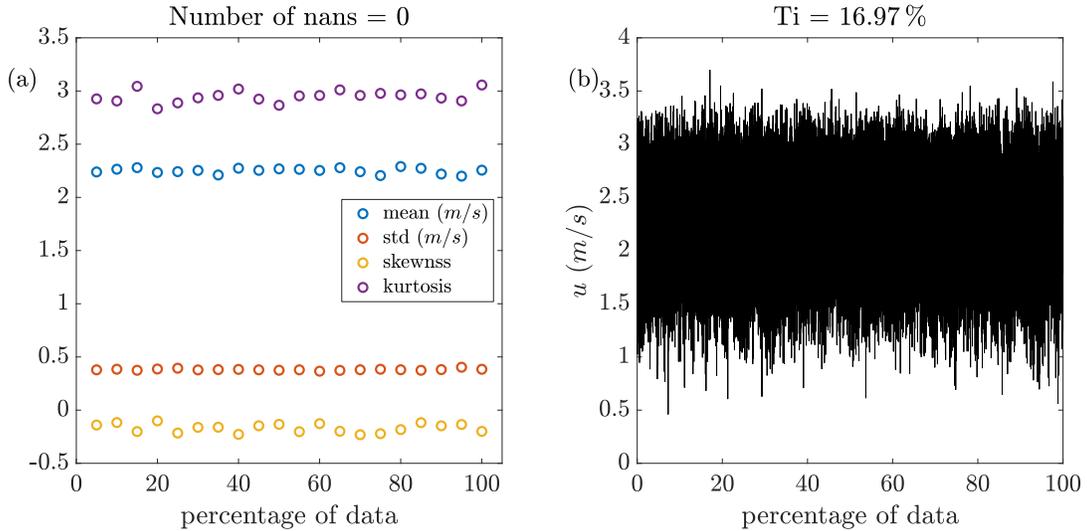}
		\caption {(a) For fixed subdivision in 20 sections (each section corresponds to a length of 5\% of the data) the mean, standard deviation, skewness, and kurtosis are plotted. (b) plot of the complete data itself. In the title the number of nans and the turbulence intensity is printed.}
		\label{fig:statio} 
	\end{figure*}

	\texttt{plot\_stationarity} With this function the stationarity of the data is verified. 
	For this purpose, the following analysis checks whether the statistical properties 
	do not change over time (stationary process). 
	Therefore the data is subdivided into 20 sections 
	and the mean, standard deviation, skewness, and kurtosis are plotted respectively in Fig.~\ref{fig:statio}(a).
	In the title of the Fig.~\ref{fig:statio}(a), the number of NaNs (not a number) is printed.
	The origin of these NaNs may be, for example, due to the measurement apparatus. 
	Regardless of the origin of these, no preprocessing is performed in the package up to this point, which can lead to a NaN. 
	If the data to be analyzed contains NaNs a pop-up dialog box is displayed, which allows the user to choose 5 different methods to fill these entries using \texttt{fillmissing} function of MATLAB. 
	The user must choose one of the following options to fill missing entries: ``nearest non-missing value'', ``linear interpolation of neighboring, non-missing values'', ``piecewise cubic spline interpolation'', or ``moving average'' and ``moving median'' using a moving window of a length, which can also be set by the user.
	%
	
	In Fig.~\ref{fig:statio}(b) the complete data itself is plotted. 
	This figure is used for the qualitative validation of stationarity, as it is very easy to detect by eye already, for example, outliers or drift in the data. 
	In the title of this figure, the turbulence intensity
	\begin{eqnarray}
		Ti=100\frac{u'}{\left<u\right>}
	\end{eqnarray}
	is printed, with $\left<u\right>$ is the mean value of a velocity time series.
	
\begin{figure}
		\centering
		\includegraphics[width=0.4\textwidth]{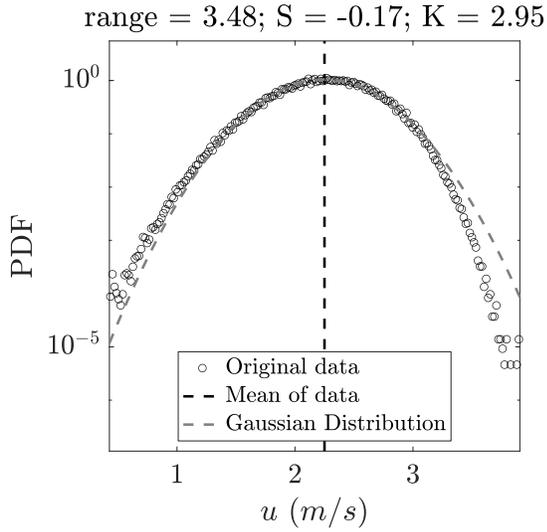}
		\caption {Probability density function (PDF) of the data. The grey dashed line corresponds to a Gaussian distribution with the same standard deviation and mean value (vertical black dashed line) as of the data.}
		\label{fig:pdf} 
	\end{figure}

	\texttt{plot\_pdf} This function plots the probability density function (PDF) of the data with the number of bins specified in the function \texttt{increment\_bin} in Fig.~\ref{fig:pdf}. 
	It also plots the Gaussian distribution with the same standard deviation and mean value as of the data. 
	In the title the range of the data (the difference between the maximum and minimum values of sample data), the skewness and flatness of the data are printed.\\

    \newpage
	\texttt{spectrum} This function calculates the energy spectral density (ESD) of the time series using the \texttt{fft} function implemented in  MATLAB, which computes the discrete Fourier transform of the time series using a fast Fourier transform algorithm.
	The energy spectral density is normalized so that	
	\begin{eqnarray}
		u'^2&=&\int_{0}^{\infty} E(f) df,
	\end{eqnarray}
	with $u'^2$	is the variance of the timeseries.  
	In this context, Parseval's theorem can be mentioned.
	A commonly used interpretation of this theorem is that the total energy of a signal can be calculated by integrating the spectral density across frequency.
	In terms of turbulence, $u'^2$ is often understood as being one velocity component of the turbulent kinetic energy
	\begin{eqnarray}
		E_{turb}=\frac{1}{2}\left(u'^2+v'^2+w'^2\right).
	\end{eqnarray}
	
\begin{figure}
		\centering
		\includegraphics[width=0.39\textwidth]{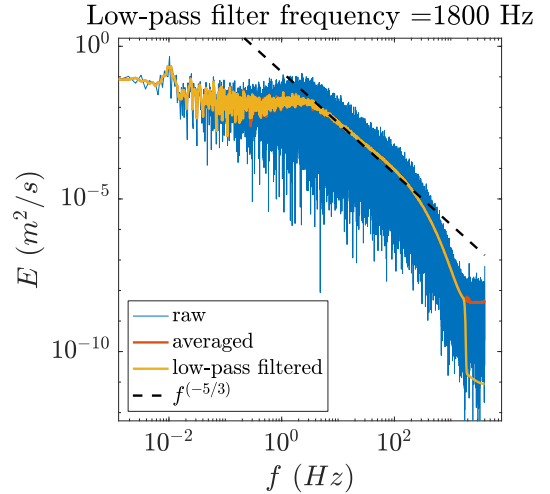}
		\caption {Energy spectral density (ESD) in the frequency domain. The yellow solid line corresponds to the averaged and filtered energy spectral density in the frequency domain using the low-pass filter. 
		The black dashed line corresponds to the Kolmogorov's $f^{-5/3}$ prediction \cite{kolmogorov1941dissipation}.}
		\label{fig:spec_filter} 
	\end{figure}

	In Fig.~\ref{fig:spec_filter}, the ESD with and without averaging (moving average with equally spaced frequency interval in log-space) as a function of frequency is plotted.
	In addition, the user can choose the range (called inertial range) of the spectrum to be used to fit Kolmogorov's $f^{-5/3}$ prediction \cite{kolmogorov1941dissipation}  (represented by a black dashed line).\\
	
		In the literature, the term power spectral density (PSD) is often used for stationary turbulence signals.
		We distinguish between PSD and ESD based on the following Matlab code fragment: \mbox{PSD: $abs(fft(data))^2/L_{data}$}, \mbox{ESD: $abs(fft(data))^2/(Fs \cdot L_{data})$}, where $abs()$ returns the absolute value, $L_{data}$ is the number of samples of the data and $Fs$ is the sampling frequency.\\
	
	

	\texttt{low\_freq} generates a pop-up dialog box to select whether the data should be filtered (low-pass filter) or not. 
	If a filter is to be applied then in the next step the frequency in Hz at which the data will be filtered by a low-pass filter have to be specified (for example 1800 Hz). 
	If the pop-up dialog box is denied, it is set to the value \texttt{low\_freq=Fs/2}.\\
	
	\texttt{frequency\_filter} This function returns the filtered data and the filtered energy spectral density in the frequency domain (see yellow curve Fig.~\ref{fig:spec_filter}).
	This function uses the \texttt{butter} function, which returns the transfer function coefficients of an nth-order lowpass digital Butterworth filter, and the \texttt{filtfilt} function of \mbox{MATLAB} to low-pass filter the data at the previously set \texttt{low\_freq} frequency.  
	For a fixed cut-off frequency, there is a threshold of nth-order of Butterworth filter at which the filter can be applied successfully to the data.
	We have implemented a routine that tries to design the highest order for a fixed cut-off frequency.
	The filtered data is named "data\_filter" (a variable in MATLAB) and will be used for all the further data post-processing. The variable $u'$ is from this point on the standard deviation of the filtered data.  
	If the filtering was not performed in the previous step, the variable  "data\_filter" and "data" (which is the unfiltered data) are equal.
	
\begin{figure*}
		\centering
		\includegraphics[width=0.9\textwidth]{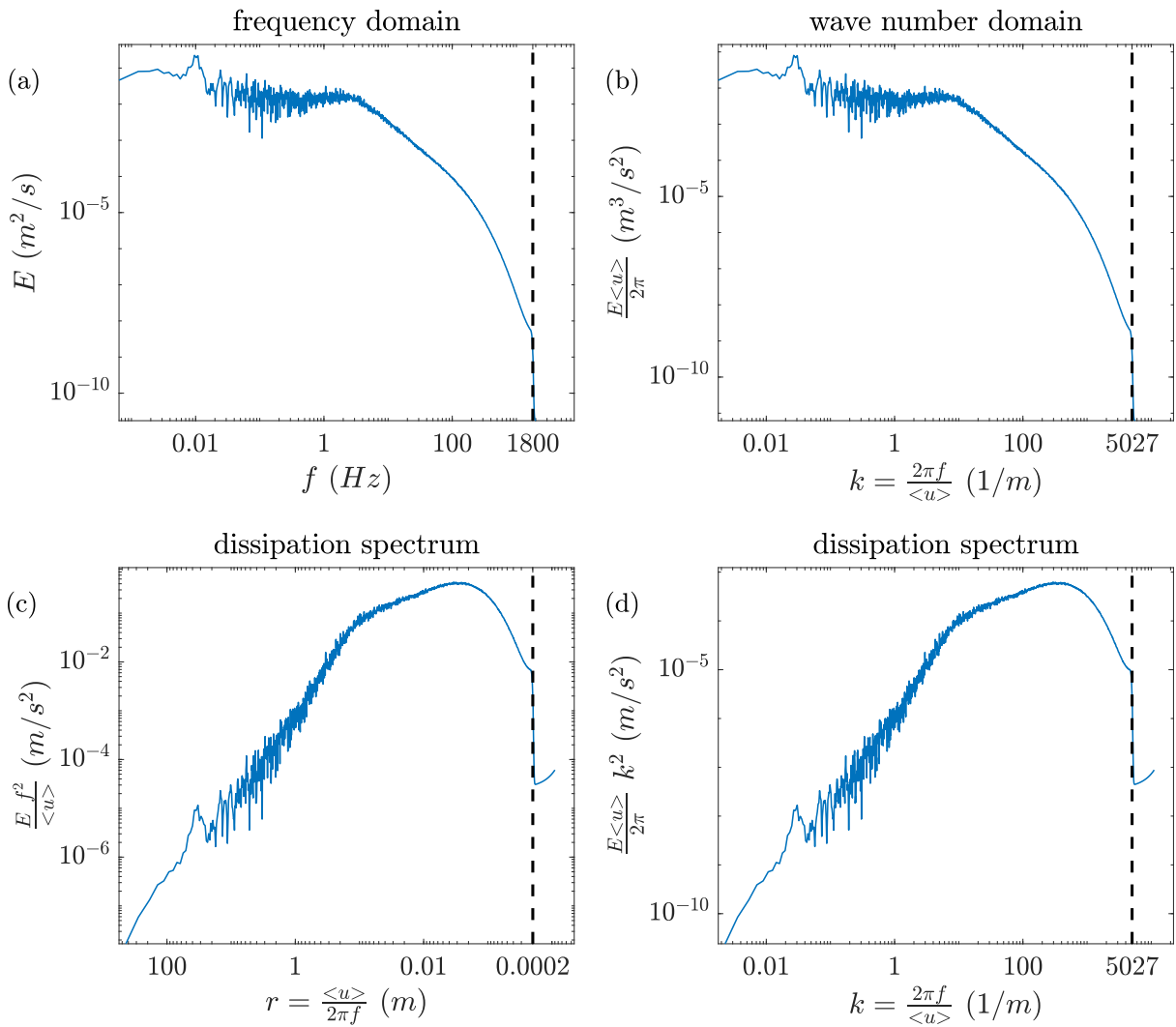}%
		\caption {Different representation/normalization of the energy spectral density with respect to (a) frequency $f$ and (b) wave number $k$ and dissipation spectral density with respect to (c) scale $r$ and (d) wave number $k$. The vertical dashed lines indicate the low-pass filter frequency in the different representation.}
		\label{fig:spec__diss_filter} 
	\end{figure*}

	\newpage
	By means of Taylor’s hypothesis of frozen turbulence (discussed in detail in the next section), the spectral energy density as a function of frequency can be transformed into the energy spectrum in the wave number domain
	\begin{eqnarray}
		k &=& \frac{2\pi f }{\left<u\right>},\\
		E(k) &=& \frac{E(f) \left<u\right>}{2\pi}.
	\end{eqnarray}
	In addition, different representations/normalization of the energy respectively dissipation spectrum density with respect to frequency $f$, scale $r$, wave number $k$ will be plotted in Fig.~\ref{fig:spec__diss_filter}. 

	\subsection{Estimation of fundamental turbulence length scales}
	\label{sec:length_scale}
	Length scales are important for any analysis of turbulent data. 
	For scales larger than the "large length scale", called integral length scale, it is expected that only uncorrelated noise is present in the flow. 
	The "small length scale" determine the end of the cascade where dissipation smooths out all fluctuations. 
	Here the Taylor length scale is about there where dissipation starts to play a role, the dissipation or Kolmogorov length is where the flow field becomes totally dissipative. 
	Thus interesting features of turbulence are expected in the interval between large and small length scale, which is called inertial range.	
	As these scales are important for the data analysis and interpretation, we present in the following different methods to estimate each of these scales. 
	We select those that are well-know to us. 
	Note there are definitely more methods suggested in the literature. 
	It would be of benefit if other researcher add further methods to our collection.\\

	Before scales are discussed in detail, the equivalence of scales in time and space by the use of Taylor's hypothesis of frozen turbulence \cite{taylor1938spectrum} is mentioned.  
	A spatial distance $r$ is related  to temporal separation $\tau$ by the mean velocity 
	\begin{eqnarray}
		r=-\tau \cdot \langle u\rangle.
		\label{eq:taylor}
	\end{eqnarray}
	Thus, the requirement for the data is that it is non-normalized local velocity data. 
	For other data for which no mean value can be determined, the mean value can be set to 1\,m/s.
	The consequences for the changed dimension have to be worked out by the user.
	The minus sign in Eq.~\eqref{eq:taylor} is a consequence of the typical time measurement with a hot wire at one location of the flow. 
	In a time step $\tau$ the flow from a location upstream ($-r$, direction $\hat{x}$ is defined down stream) is transported to the sensor. 
	With the knowledge of the sampling frequency, all scales can also be expressed in units of samples, or, respectively sample steps. 
	Thus all scales can be represented either as the number of samples, seconds or meters.
	For example consider a hot wire signal with a sampling frequency of 50\,kHz which is characterized by a mean velocity of 10\,m/s and an integral length scale of 0.1\,m. 
	According to the basic equation of velocity = distance/time, the integral length scale of 0.1\,m 
	corresponds to a given duration 0.01\,s.
	Considering the sampling frequency, this time is equal to 500 samples. 
	In order to successfully apply the Taylor hypothesis, a turbulence intensity of less than 20\,\% is often referred to in the literature. 
	\newpage

	From the data series $u(t)$ or $u(x)$ also of the longitudinal velocities (the component of  $u$ in direction of the mean flow) we construct their longitudinal increments 
	\begin{eqnarray}
		u_{\tau}(t)=u(t)-u(t+\tau)
		\label{eq:increment}
	\end{eqnarray}
	labeled by the time-separation $\tau$.
	Accordingly a spatial velocity increments	is \mbox{$u_r = u(x+r) - u(x) = - u_{\tau}$.} From the velocity increments n-th order structure functions 
	\begin{eqnarray}
			S^{n}(r)=\left\langle u_{r}^n \right\rangle,
		\end{eqnarray}	
	are determined. $n$ is an integer.\\
	%

	\texttt{length\_scales} In this function, the integral length scale $L$, Taylor length scale $\lambda$, Kolmogorov length scale $\eta$, mean energy dissipation rate $\langle \epsilon \rangle$, normalized energy dissipation rate $C_{\epsilon}$ and the local Taylor-Reynolds number $Re_{\lambda}$ are estimated using different methods of calculation. 
	These parameters are used for the further processing of data (solving the FPE and extracting cascade trajectories). 
	If the fundamental length scales are already known from previous analysis, this calculation can be skipped and the values of the integral, Taylor, Kolmogorov length scale in $m$ and energy dissipation rate in $m^2/s^3$ can be entered in the pop-up dialog boxes.
	This also enables to study of the effect of different chosen scales on further results of the analysis.
	The entered length scales will be rounded towards the nearest integer in sample units.
	Here the Taylor hypothesis is used, so the link between mean velocity, sampling frequency and a time interval is used to convert this interval into a length scale (see details in Section \ref{sec:length_scale}). 
	The proposed value in the pop-up dialog box is the median length scale for all methods.\\
	

		Note, that the following collection of the methods for the estimation of the fundamental length scales and the dissipation rate does not claim to be complete. 
		When creating the package, the methods known to us have exclusively been taken into account. 
		However, this list can be extended.
		The different methods for estimating these parameters are numbered sequentially. 
		While each method may share the same assumptions and hypotheses, the different methods are not necessarily consistent/coherent with one another. 
		For this reason, for example, the number of methods for determining the Taylor length and dissipation rate differs.
		At the end of the estimation of the respective length scale, the results are presented in a single figure to compare them with each other. 
		The comparison of the results of different methods allows the user to evaluate the consistency of the methods and adjust the proposed value in the pop-up dialog boxes (median length scale for all methods) accordingly.
		In an ideal case, all methods for one scale should give the same result, but in reality, different aspects of experimental data may change results. In such cases, it is the work of the user to find out the most reliable value.

	\newpage
	\noindent
	\textbf{The integral length scale $L$ is estimated by using:}
	\begin{enumerate}
		\item the energy spectrum density which requires the range of frequency that will be used to linearly extrapolate the value of ESD at a frequency of 0\,Hz \cite{Hinze1975,roach1987generation} (see Fig.~\ref{fig:int_spec}). Therefore the user will be asked to enter the $f\_start$ and $f\_end$ in Hz ($f\_start<f\_end$). 
		For example: $f\_start = 0.03$ Hz and $f\_end = 2.7$ Hz.
		\begin{eqnarray}
			L=\lim\limits_{f \rightarrow 0}\left[\frac{E(f)\left<u\right>}{4u'^2}\right]
		\end{eqnarray}
	\begin{figure}[h]
		\centering
		\includegraphics[width=0.43\textwidth]{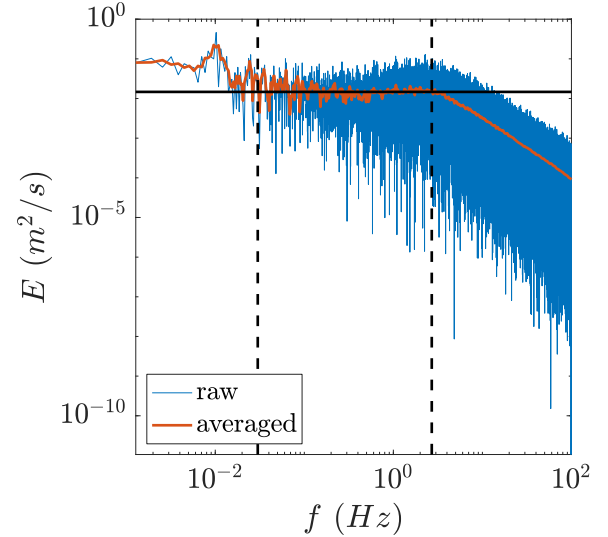}
		\caption{Representation of the linear extrapolation of ESD at a frequency of 0\,Hz for estimating the integral length scale with the method according to Hinze \cite{Hinze1975} \& Roach \cite{roach1987generation}. The two vertical dashed lines correspond to the range of frequency that will be used to linearly extrapolate (solid black line).}
		\label{fig:int_spec} 
	\end{figure}

		\item the autocorrelation coefficient $R_{\widetilde{u}\widetilde{u}}$ with respect to scales \cite{Frisch_1995,pope2001turbulent,Bourgoin_2017} plotted in Fig.~\ref{fig:autocorr_expfit}
		\begin{eqnarray}
			L&=&\int_{0}^{\infty}R_{\widetilde{u}\widetilde{u}}(r)dr,\\
			&=&\int_{0}^{\infty}\frac{\langle \widetilde{u}(x)\widetilde{u}(x+r)  \rangle}{\langle (\widetilde{u}(r))^{2} \rangle} dr,
			\label{eq.auto_int}
		\end{eqnarray}
		where $r$ is the scale in meter and $\widetilde{u}= u - \langle u \rangle$ is the fluctuating component of streamwise velocity.
		The cumulative integral gives the asymptotic value at a specific scale $r$ which is characterized by the integral length scale $L$. 
	\end{enumerate}

	In the case of non-ideal data or experimental data, Eq.~\eqref{eq.auto_int} may not converge. 
	As a result, the aforementioned method can lead to large errors in the estimation of the integral length. 
	In the case of a non-monotonic decrease of $R_{\widetilde{u}\widetilde{u}}(r)$, the autocorrelation function is
	\begin{enumerate}
	\addtocounter{enumi}{2}
		\item integrated up to the first zero-crossing of the autocorrelation function \cite{o2004autocorrelation}.
		\item integrated up to the first $1/e$ crossing of the autocorrelation function \cite{tritton2012physical}.
	\end{enumerate}
	In particular, for experimental data, there may be additional measurement noise present in the data that may cause no zero crossing of the autocorrelation function. In this case $R_{\widetilde{u}\widetilde{u}}(r)$ is 
	\begin{enumerate}
	\addtocounter{enumi}{4}
		\item fitted by an one-term exponential function \cite{Hinze1975,tritton2012physical} 
		\begin{eqnarray}
			R_{\widetilde{u}\widetilde{u}}(r) = a \; e^{-b \; r}.
		\end{eqnarray}
		The fit region correspond to the range of scales r=$\left[0 : r_e\right]$, with $r_e$ is the scale of the first $1/e$ crossing of the autocorrelation function. 
		This fit range is  indicated by the two black, vertical dashed lines in Fig.~\ref{fig:autocorr_expfit}.		
		The integral is calculated by using the coefficients $a$ and $b$ of the exponential fit
		\begin{eqnarray}
			L &=& \int_{0}^{\infty} a \; e^{-b \; r}dr= \frac{a}{b}.
		\end{eqnarray}
		\begin{figure}[h]
			\centering
			\includegraphics[width=0.38\textwidth]{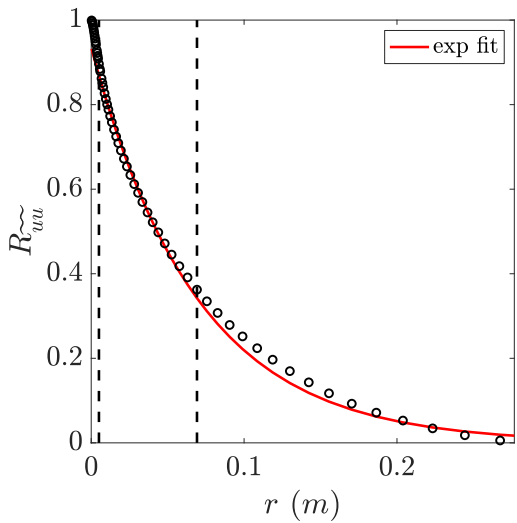}
			\caption{Representation of the autocorrelation coefficient $R_{\widetilde{u}\widetilde{u}}$ as a function of scale in meter. The two vertical dashed lines correspond to the range of scale that will be used for the exponential fit (red solid line) for estimating the integral length scale with the method according to Hinze \cite{Hinze1975} \& Tritton \cite{tritton2012physical}.}
			\label{fig:autocorr_expfit} 
		\end{figure}
\end{enumerate}
\noindent
Furthermore, the integral length scale $L$ is estimated via:
\begin{enumerate}
	\addtocounter{enumi}{5}
		\item the second order structure function 
		\begin{eqnarray}
			S^{2}(r)=\left\langle u_{r}^2\right\rangle,
		\end{eqnarray}
		which holds the link to the autocorrelation coefficient $R_{\widetilde{u}\widetilde{u}}$ such as
		\begin{eqnarray}
			R_{\widetilde{u}\widetilde{u}}(r) = 1-\left[\frac{S^{2}(r)}{2u'^2}\right],
		\end{eqnarray}
		where $u'$ can be calculated directly from the second order structure function. 
		At a sufficiently large length scale compared to the energy injection scale of the experiment the second order structure function truncates to the asymptotic value of $2u'^{2}$ \cite{Mordant_2001_Ruu_with_S2}.
		\item the zero crossings proposed by \mbox{Mora \& Obligado \cite{mora2020estimating}}.
		To verify that zero crossing are well resolved, the signal has to be filtered with a low-pass filter with a characteristic length $\eta_c$. 
		This method consists of estimating the Voronoi tessellation of the 1D zero-crossings positions data set. 
		It is compared with a Random Poisson Process, which has no correlations between the scales. 
		The method proposes that the integral length scale is equal to the characteristic length $\eta_c$ for which
		\begin{eqnarray}
			\frac{\sigma_{voro}}{\sigma_{RPP}}=1.
		\end{eqnarray}
		$\sigma_{voro}$ is the standard deviation of the Voronoi cells normalized by their mean value and $\sigma_{RPP}$ is the equivalent value for an Random Poisson Process, that is equal to $\sqrt{(1/2)}$.
		In Fig.~\ref{fig:zero_crossing_int_L} the standard deviation (normalized) of the Voronoi cells as a function of the characteristic length $\eta_c$ is plotted.
		Finally, the integral length scale is defined as the value of $\eta_c$ that correspond to $\sigma_{voro}/\sigma_{RPP}=1$. 
		If we observe that $\sigma_{voro}/\sigma_{RPP}>1$ for all values of  $\eta_c$, the method cannot provide the value of the integral length scale and longer signals are needed (nevertheless the extrapolation of the value remains possible).
		\begin{figure}[h]
			\centering
			\includegraphics[width=0.41\textwidth]{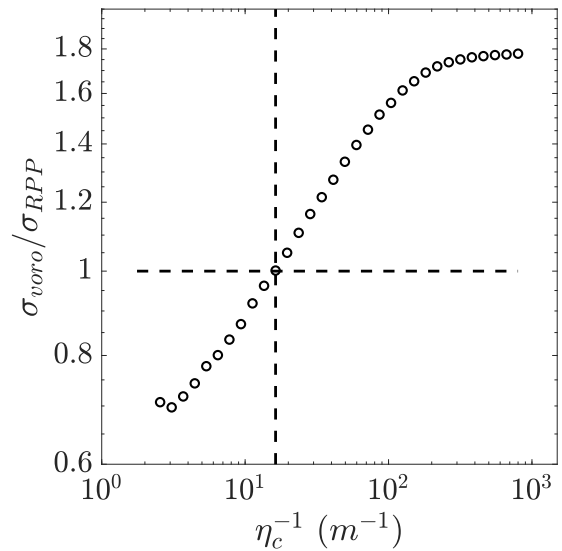}
			\caption{Standard deviation (normalized) of the Voronoi cells as a function of the characteristic length $\eta_c$ (not to be confused with the Kolmogorov length scale $\eta$) of a low-pass filter for estimating the integral length scale with the method according to \mbox{Mora \& Obligado \cite{mora2020estimating}}.}
			\label{fig:zero_crossing_int_L} 
		\end{figure}
	
		Note, the characteristic length $\eta_c$ should not be confused with the Kolmogorov length scale $\eta$. To provide a direct link to \cite{mora2020estimating}, exactly the same letters were used for the variables.
	\end{enumerate}

\newpage
\noindent
\textbf{The Taylor length scale $\lambda$ is estimated by using:}
	\begin{enumerate}
		\item a parabolic fit 
		\begin{eqnarray}
			R_{\widetilde{u}\widetilde{u}}(r)=1-\frac{r^2}{\lambda^2}
		\end{eqnarray}
	 	to the autocorrelation function, to estimate $\lambda$ at the origin $r=0$ (see Fig.~\ref{fig:autocorr_parabolic}). 
		The range of the positive curvature is therefore used for the estimation (close to $r=0$ the auto-correlation function has an inflection point). 
		Since this method needs a well-resolved auto-correlation function it strongly depends on the resolution of the sensor. 
		It is also affected by low pass filtering. 
		\begin{figure}[h]
			\centering
			\includegraphics[width=0.405\textwidth]{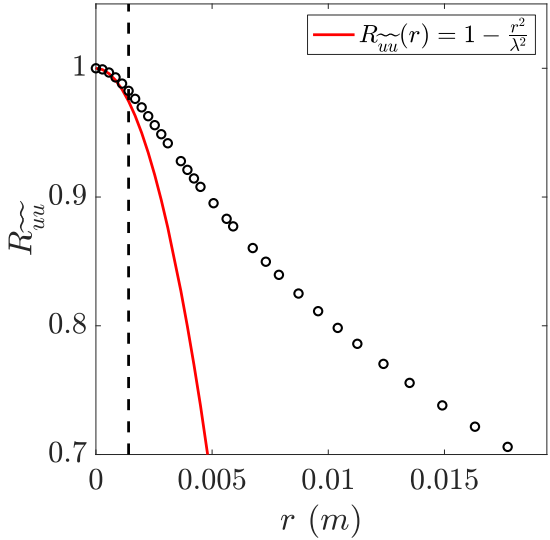}
			\caption {Autocorrelation function with parabolic fit as a function of the scale $r$ for estimating the Taylor length scale. The vertical dashed line correspond to the range of scale that will be used for the extrapolation (red solid line) by a parabolic fit.}
			\label{fig:autocorr_parabolic} 
		\end{figure}
	\end{enumerate}
	
\noindent
	Assuming isotropy at small scales and Taylor's frozen turbulence hypothesis the Taylor length scale $\lambda$ is estimated using the relation 
	\begin{eqnarray}
		\lambda^2 = \frac{u'^2}{\left<\left(\partial{\widetilde{u}}/\partial x\right)^2\right>}.
	\end{eqnarray}
	The numerical differentiation in the denominator is approximated by:
	\begin{enumerate}
		\addtocounter{enumi}{1}
		\item the simple difference quotient. 
		Due to the finite spatial resolution of the measuring sensor used in this typical experimental dataset and measurement noise,
 		this method will yield an incorrect result. 
		In order to correctly compute the derivatives, the spatial resolution must be of the order of the Kolmogorov microscale $\eta$ (see \cite{hussein1990influence}). 
		$\eta$ estimated via Eq.~\eqref{eq:kol}.
		\newpage
		\item the procedure proposed by \mbox{Aronson \& Löfdahl \cite{aronson1993plane}}. 
		Here the derivative of the velocity is approximated via the second order structure function $S^2$ in the inertial range. 
		In Fig.~\ref{fig:Aronson_Löfdahl} the development of 
		\begin{eqnarray}
			\lambda = \lim\limits_{r \rightarrow 0} \sqrt{\frac{u'^2 r^2}{S^{2}(r)}}.
			\label{eq:Aronson_Löfdahl} 
		\end{eqnarray}
		as a function scale $r$ is plotted.
		For the extrapolation we use a linear fit. The fit-region is used for spatial lags that are larger than 4 times the scale that corresponds to the low-pass filter frequency \texttt{low\_freq}. The larger limit of the fit-region must be set by the user. 
		\begin{figure}[h]
			\centering
			\includegraphics[width=0.43\textwidth]{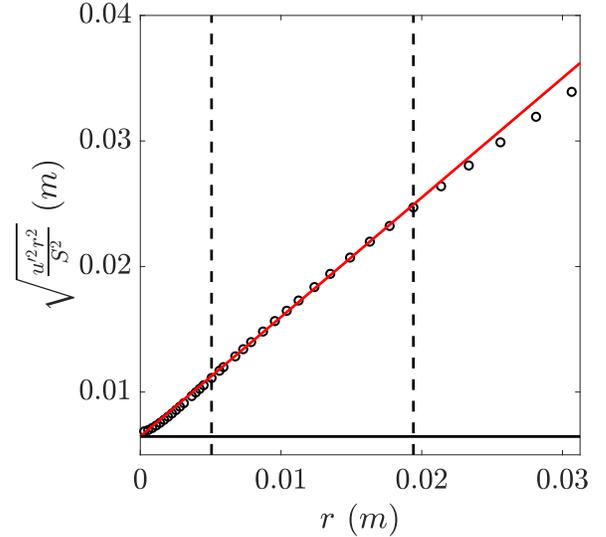}
			\caption {Development of Eq.~\eqref{eq:Aronson_Löfdahl} with linear extrapolation (red solid line) for estimating the Taylor length scale with the method according to \mbox{Aronson \& Löfdahl \cite{aronson1993plane}}. The two vertical dashed lines correspond to the range of scale that will be used for the extrapolation.}
			\label{fig:Aronson_Löfdahl} 
		\end{figure}
		\item using the dissipation spectrum 
		\begin{eqnarray}
			\left<\left(\frac{\partial{\widetilde{u}}}{\partial x}\right)^2\right> &=& \int_{0}^{\infty} k^2 E(k) dk.
			\label{eq:diss_spec}
		\end{eqnarray}
		This procedure has been proposed by \mbox{Hinze \cite{Hinze1975}}.
		The upper limit of the integration is set to the low-pass filter frequency \texttt{low\_freq}. 
		\item using the dissipation spectrum but here the upper limit of the integration of Eq.~\eqref{eq:diss_spec} is set to infinity (or the largest available wave number $k$) so that the full dissipation spectrum is used.		
		\end{enumerate}
	
		\noindent
		Furthermore, the Taylor length scale $\lambda$ is estimated via:
		\begin{enumerate}
		\addtocounter{enumi}{5}
		\item the zero crossings of the fluctuating velocity,
		\begin{eqnarray}
			\label{eq:zero_crossing_lambda} 
			\lambda=\frac{l}{C \pi},
		\end{eqnarray}
		with $l$ is the average distance between zero-crossings. 
		$C$ is a constant in the order of unity that quantifies the level of non-Gaussianity of the derivative $\partial{\widetilde{u}}/\partial x$. 
		This method is discussed in \cite{mazellier2010turbulence,Sreenivasan_1983,Mora_2019}. 
		In Fig.~\ref{fig:zero_crossing_lambda} the density of zero-crossings times the average distance between zero crossings as a function of the characteristic length $\eta_c$ is plotted.
		\begin{figure}[h]
			\centering
			\includegraphics[width=0.4\textwidth]{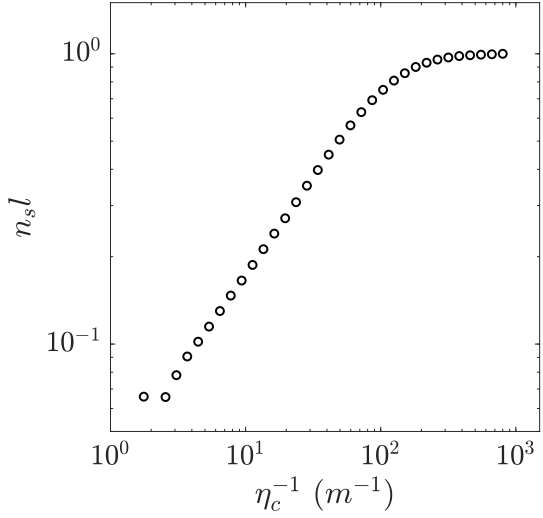}
			\caption {Density of zero-crossings times the average distance between zero crossings as a function of the characteristic length $\eta_c$ of a low-pass filter for estimating the Taylor length scale using the method discussed in \cite{mazellier2010turbulence,Sreenivasan_1983,Mora_2019}.}
			\label{fig:zero_crossing_lambda} 
		\end{figure}
	
		For values of $\eta_c$ within the inertial range, a power law 2/3 is expected \cite{mazellier2010turbulence}, and eventually for smaller filter sizes (or large $1/\eta_c$) a plateau is reached. 
		The presence of this plateau, related to the dissipative range, implies that the density of zero-crossings $n_s$ are well resolved, and therefore $l$ can be deduced using the trivial relation $n_s \cdot l =1$. 
		If the plateau is not reached, small scales are not resolved and the method cannot estimate the Taylor length scale. 
		On the other hand, if after the plateau the value of $n_s \cdot l$ increases again, it means that the cut-off frequency is too high and the analysis is affected by small scale noise.
		Initially, the constant $C=1$ gives a good approximation of $\lambda$ using Eq.~\eqref{eq:zero_crossing_lambda}.
		\item the zero crossings of the fluctuating velocity, but here $C$ is defined as
		\begin{eqnarray}
			C &=& \sqrt{\frac{2}{\pi}} \frac{\sigma_{\partial_x\widetilde{u}}}{ \left<|\partial_x\widetilde{u}| \right>}
		\end{eqnarray}
		with $\left<|\partial_x\widetilde{u}| \right>$ is the mean and $\sigma_{\partial_x\widetilde{u}}$ is the standard deviation of $\partial{\widetilde{u}}/\partial x$, where $\widetilde{u}$ is filtered with the largest frequency within the plateau of $n_s \cdot l$.
		With the use of this method a better estimation of $\lambda$ can be obtained if $\partial{\widetilde{u}}/\partial x$ is resolved.\\
	\end{enumerate}

\newpage
\noindent
\textbf{The mean energy dissipation rate $\langle \epsilon \rangle$ is estimated by its one-dimensional, isotropic surrogate using:}
	\begin{enumerate}
		\item \& 2. method: either $2^{nd}$ or $3^{rd}$ order structure function. 
		The estimation of dissipation using this method relies on the transfer of energy within the inertial range. 
		This method is particularly useful when the higher frequency content present in the flow is not fully resolved by the measurement device. 
		This is generally the case where for example the length of the hot wire is larger than the Kolmogorov length scale $\eta$ (see Eq.~\eqref{eq:kol}) of the flow under consideration. 
		This function generates a pop-up dialog box to enter the value of Kolmogorov constant $C_{2}$ (typically within 2.0 - 2.4) used in the relation between second order structure function $S^{2}(r)$ and energy transfer rate $\epsilon(r)$ (transfer from one scale to another one) based on the assumption of homogeneous isotropic turbulence (HIT) \cite{pope2001turbulent,taylor1938spectrum}. 
		The mean energy dissipation rate $\langle \epsilon \rangle$ is calculated by finding the mean amongst 5 points closest to the peak (i.e. small plateau) value of $\epsilon(r)$. 
		We expect that with increase in Reynolds number this plateau becomes more clear and evident.
		In Fig.~\ref{fig10} the development of
		\begin{eqnarray}
			\label{eq.epsi_2}
			\epsilon(r) &=& \frac{1}{r}\left[\frac{S^{2}(r)}{C_{2}}\right]^{3/2},\\
			\label{eq.epsi_3}
			\epsilon(r) &=& -\frac{5}{4}\left[\frac{S^{3}(r)}{r}\right],
		\end{eqnarray}
		using either $2^{nd}$ or $3^{rd}$ order structure function, \mbox{$S^{2}(r)=\left\langle u_{r}^2\right\rangle$} respectively $S^{3}(r)=\left\langle u_{r}^3\right\rangle$, is plotted.
		\begin{figure}[h]
			\centering
			\includegraphics[width=0.4\textwidth]{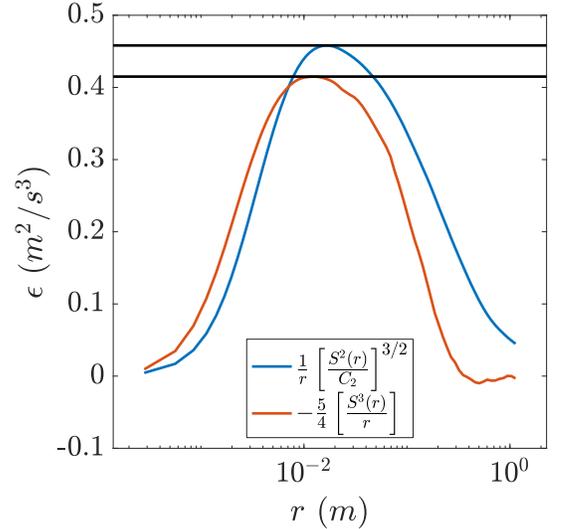}
			\caption {Development of $\epsilon(r)$ using either $2^{nd}$ or $3^{rd}$ order structure function (see Eq.~\eqref{eq.epsi_2} and Eq.~\eqref{eq.epsi_3}). The black solid line marks the peak value of $\epsilon(r)$.}
			\label{fig10} 
		\end{figure}
	 
		Basically, the energy transfer mechanism in the turbulent cascade happens in three different ways.
		Firstly, the energy is injected at large scales in the turbulent flow.
		Secondly, the injected energy up to the integral length scale transfers across within the inertial range down to the scale at which viscous effects start being significant.
		Thirdly, the transferred energy across the inertial scales starts to dissipate into heat because of the dominant effects of fluid viscosity over the scales of the dissipation range.
		From the conservation of energy principle, it follows that in steady state the rate of injection of energy at large scales, the energy transfer rate across inertial scales and the energy dissipation rate at small scales in the dissipation range must be equal.
		We note that Eq.~\eqref{eq.epsi_3} is an exact result in the inertial range and it requires stationarity, homogeneity and isotropy of the turbulent flow. 
		The lack of isotropy in turbulent flows invokes the modification of the prefactor $-5/4$. 
		For further details the corresponding literature should be studies, see for example \cite{ZHOU_2000}.
		In the present context we assume the underlying flow to be isotropic. 
 		Eq.~\eqref{eq.epsi_2} and Eq.~\eqref{eq.epsi_3} leads to the same estimation of the dissipation rate by choosing the right value of the constant $C_{2}$.  
		
		\addtocounter{enumi}{1}
		\item by using the chosen Taylor length scale and the following relation
		\begin{eqnarray}
			\label{eq.HIT_Taylor}
			\langle \epsilon \rangle = 15 \nu \frac{u'^2}{\lambda^2},
		\end{eqnarray}
		which is valid for isotropic turbulence \cite{Taylor_1935}.

		\item As a consequence of the K41 phenomenology, the second order structure function implies an energy spectrum density of the form
		\begin{eqnarray}
			\label{eq.Kolfit_a}
			E(k)=C_k\langle \epsilon \rangle^{2/3}k^{-5/3}.
		\end{eqnarray}
		$C_k$ is the so-called Kolmogorov constant that remains undetermined in Kolmogorov's theory (typically \mbox{$C_k\approx0.53\pm0.01$} \cite{Sreenivasan_1995,Oboukhov_1962}). 
		Following Kolmogorov's $k^{-5/3}$ prediction a fit in the inertial range ($\lambda<r<L$) according to Eq.~\eqref{eq.Kolfit_a} is used to estimate the mean energy dissipation rate $\langle \epsilon \rangle$ (see Fig.~\ref{fig:spec_fit}).
		\item a fit in the inertial range ($\lambda<r<L$) according to 
		\begin{eqnarray}
			\label{eq.Kolfit_b}
			E(k)&=&C_k\langle \epsilon \rangle^{2/3}(k-k_0)^{-5/3},
		\end{eqnarray}
		with $k_0$ is a shift for the argument of the function, i.e. a shift along the wave number axis, which is now used to estimate the mean energy dissipation rate $\langle \epsilon \rangle$ (see Fig.~\ref{fig:spec_fit}).
		\begin{figure}[h]
		\centering
		\includegraphics[width=0.4\textwidth]{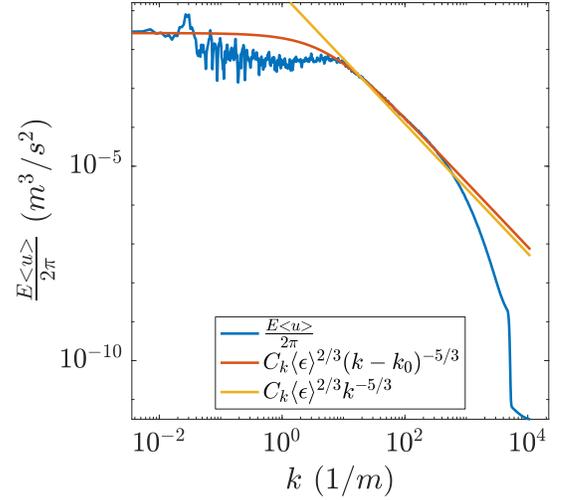}
		\caption {Energy spectral density (ESD) in the wave number domain. The red and yellow solid line corresponds to the fit according to Eq.~\eqref{eq.Kolfit_a} and Eq.~\eqref{eq.Kolfit_b} for estimating the mean energy dissipation rate $\langle \epsilon \rangle$.}
		\label{fig:spec_fit} 
	\end{figure}
		\item via the dissipation spectrum
		\begin{eqnarray}
			\label{eq.epsi_diss}
			\langle \epsilon \rangle =  \int_{0}^{\infty} 15 \nu k^2 E(k) dk.
		\end{eqnarray}
		Eq.~\eqref{eq.epsi_diss} is valid for isotropic turbulence. 
		As proposed by \mbox{Mora \textit{et al.} \cite{Mora_2019}} the dissipation spectrum is modeled for large wave number $k$ using a power law (see Fig.~\ref{fig:model_dissipation})
		\begin{eqnarray}
			15 \nu k^2 E(k) = a \; k^b,
		\end{eqnarray} 
		with the coefficients $a$ and $b$ of the power law fit.\\
%
%
		\begin{figure}
			\centering
			\includegraphics[width=0.4\textwidth]{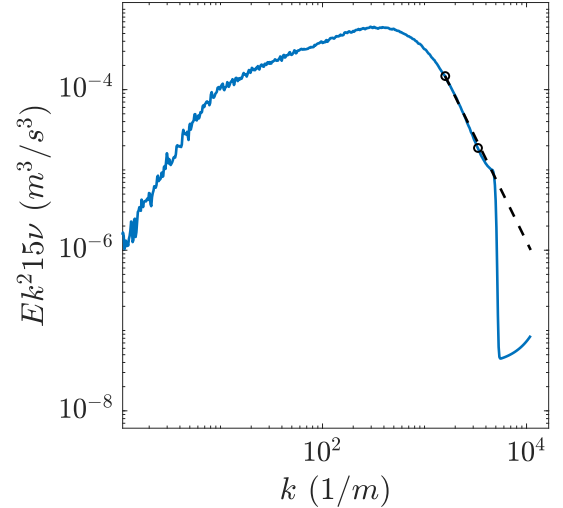}
			\caption {Dissipation spectral density in the wave number domain. The black dashed line corresponds to the power law fit to model the dissipation spectrum for large wave number $k$.}
			\label{fig:model_dissipation}		
		\end{figure}
	\end{enumerate}

	\noindent
	The \textbf{Kolmogorov length scale $\eta$} (the smallest size of the eddy in a given turbulent flow) is estimated by using the classical relation \cite{Frisch_1995} given by 
	\begin{eqnarray}
		\label{eq:kol}
		\eta = \left(\frac{\nu^{3}}{\langle \epsilon \rangle }\right)^{1/4}.\\
	\end{eqnarray} 
	
	\newpage
	\noindent
	In addition, the \textbf{normalized energy dissipation rate $C_{\epsilon}$} will be returned \cite{batchelor1948decay,Tennekes_1972}
	\begin{eqnarray}
		C_{\epsilon_1} &=& \frac{	\langle \epsilon \rangle  L}{u'^3},\\
		C_{\epsilon_2} &=& 15 \frac{L}{\lambda} \frac{1}{Re_{\lambda}},
	\end{eqnarray}
	with the \textbf{local Taylor-Reynolds number}
	\begin{eqnarray}
		Re_{\lambda} = \frac{u'\lambda}{\nu}.
	\end{eqnarray}
	$Re_{\lambda}$ allows for reasonable comparisons of experiments with different boundary conditions or energy injection mechanisms ($Re_{\lambda}$ is independent of the integral length scale $L$).\\

\begin{figure}[h]
		\centering
		\includegraphics[width=0.48\textwidth]{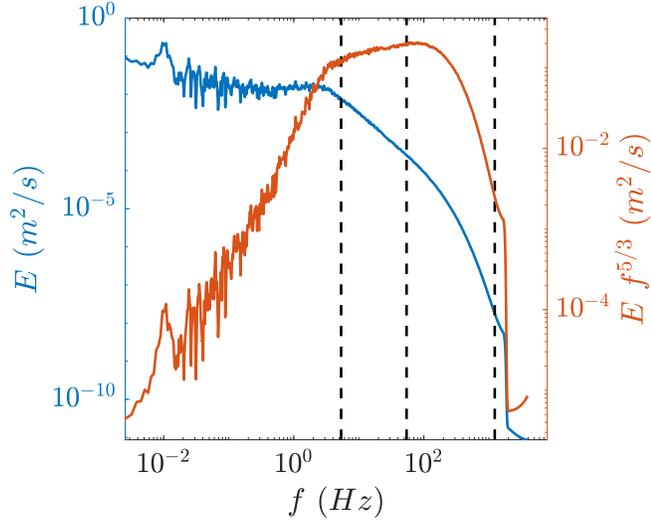}
		\caption {ESD and compensated ESD in the frequency domain. Black dashed vertical lines correspond to the respective frequency of the fundamental length scales.}
		\label{fig:spec_length_scale} 
	\end{figure}

	At the end of this function, in Fig.~\ref{fig:spec_length_scale}, a vertical dashed line at the integral $L$, Taylor $\lambda$ and Kolmogorov length scale $\eta$ will be added to energy spectral density and the compensated energy spectral density.

	%
	

	\subsection{Forward cascade and normalization of the data}
		In the following, a test corresponding to the forward cascade using the third-order structure function is performed.
	Furthermore, on the account of the comparability of various turbulent data sets, a common data normalization strategy is proposed, as mentioned below. Such normalization can be proposed, as we have now introduced the necessary scales.\\
	
	\texttt{struc\_flip\_test} This function tests whether the data have to be flipped or not. 
	Flipping the data here is referring to reversing the order of the elements.
	The decision of flipping of data depends on a simple relation of $3^{rd}$ order structure function $S^{3}(r)$ with the dissipation based on the assumption of homogeneous isotropic turbulence (HIT). 
	In the present analysis, we assume the forward turbulent cascade. 
	Thus, in the forward turbulent cascade, the $3^{rd}$ order structure function becomes negative as a consequence of irreversibility (i.e. no inverse cascade) within the turbulent cascade. 
	Based on this we propose a thumb rule that the quantity $S^{3}(r)$ must be negative in the inertial range, taking into account that the increment time series is obtained according to Eq.~\eqref{eq:increment}.
	Vortex stretching is the physical phenomenon that causes the third structure function to be negative.
	To verify this,
	$S^{3}(r)$ as a function of the scale $r$ is plotted, from which it is possible to decide whether it is essential to flip the data or not.\\

	\texttt{normalization} This function is mainly to perform the normalization of the filtered data. 
	Before doing so, it generates the pop-up dialog box which asks the user whether to flip the data or not (based on the previous investigation). 
	After that, the entire data is normalized by the quantity
	\begin{eqnarray}
	 	\sigma_{\infty} = \sqrt{2}u'.
	 \end{eqnarray}
 	This normalization is proposed \mbox{by Renner \textit{et al.} \cite{renner2001}}. 
	This function also returns the variable in MATLAB named "m\_data", which is the mean of the filtered data before normalization and the variable  "siginf", which is equal to $\sigma_{\infty}$.
	%
	In addition the user is asked whether the scale $r$ should be given in units of Taylor length scale $\lambda$. 
	The normalization 
	\begin{eqnarray}
		u_r&=&\frac{u_{r_i}}{\sigma_\infty},\\
		r&=&\frac{r_i}{\lambda},
	\end{eqnarray}
	is used in the next part of this paper to obtain non-dimentionalized Kramers-Moyal coefficients (KMCs) as a function of velocity increment and scale in order to compare the results of different data sets.
	The indices $i$ describes the velocity increment and scale without normalization. As $u_{r_i}$ and $\sigma_\infty$ are of the same unit (m/s) and respectively $r_i$ and $\lambda$ have the same unit (m) , $u_r$ and $r$ become dimensionless.
	As the normalization affects the KMCs, the indices $i$ describe the initial parameter of the Kramers-Moyal coefficients without normalization:
	\begin{eqnarray}
		D^{(1)} &=& \frac{D^{(1)}_i \lambda}{\sigma_{\infty}},\\
		D^{(2)} &=& \frac{D^{(2)}_i \lambda}{\sigma^2_{\infty}},\\
		d_{11} &=& d_{11_i}\lambda,\\
		d_{20} &=& \frac{d_{20_i}\lambda}{\sigma^2_{\infty}},\\
		d_{21} &=& \frac{d_{21_i}\lambda}{\sigma_{\infty}},\\
		d_{22} &=& d_{22_i}\lambda.
	\end{eqnarray}

\newpage
		Note, the estimation of the Kramers-Moyal coefficients (KMCs) will be introduced in section \ref{KMC}. The two functions $D^{(1,2)}$ defining the Fokker-Planck equation are called drift and diffusion coefficients, respectively, and can be estimated directly from measured data by an optimization procedure proposed in \cite{kleinhans2005iterative,Nawroth2007,Reinke_2018}.

	\subsection{Statistics of velocity increments $u_r$}
		In this section, the properties of the turbulent structures estimated from the velocity increments using the probability density function and scaling exponents of the higher-order structure functions are examined.\\
	
\begin{figure}[h]
		\centering
		\includegraphics[width=0.45\textwidth]{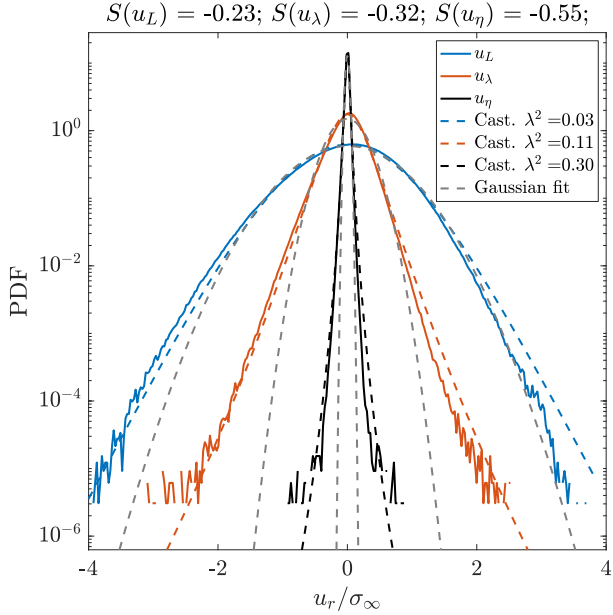}
		\caption {PDF of the velocity increments $u_r$ at the scale $r={L,\lambda,\eta}$. The colored dashed line correspond to Castaing fits (form factor $\lambda^2$ \cite{Castaing_1990}) and grey dashed line to Gaussian fits. In the title, the skewness for each of the 3 scales is displayed.}
		\label{incr_pdf}
	\end{figure}

	\texttt{plot\_increment\_pdf} This function plots in Fig.~\ref{incr_pdf} the probability density function of the velocity increments at the scale $r=L$, $r=\lambda$ and  $r=\eta$.
	
\begin{figure*}
		\centering
		\includegraphics[width=0.87\textwidth]{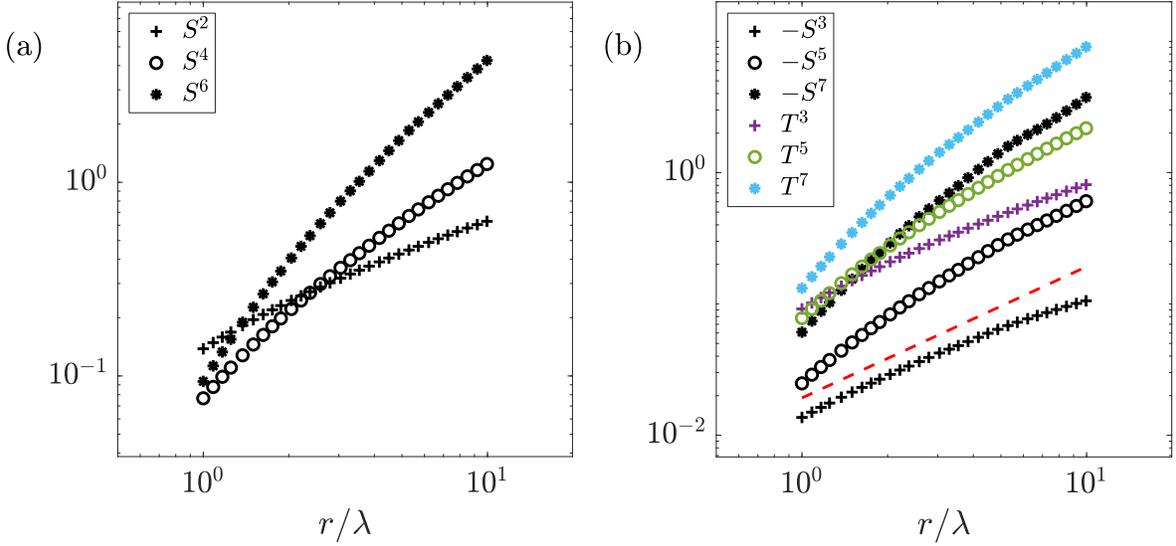}
		\caption {Course of the $k$-th order structure function $S^{k}$ and $T^{k}$ for (a) $k={2,4,6}$ and (b) $k={3,5,7}$ as a function of scale. The dashed red line in (b) represents $-4/5$ law.}
		\label{fig:struc_flip} 
	\end{figure*}

	\texttt{plot\_struc\_function} This function plots in Fig.~\ref{fig:struc_flip} the $k$-th order structure function of the velocity increments and of their absolute values
	\begin{eqnarray}
		S^{k}(r)&=&\left\langle u_{r}^k\right\rangle,\\
		T^{k}(r)&=&\left\langle |u_{r}|^k\right\rangle,
	\end{eqnarray}
	with $k={2-7}$ for scales $\lambda \leq r \leq L$.

\begin{figure}[h]
		\centering
		\includegraphics[width=0.4\textwidth]{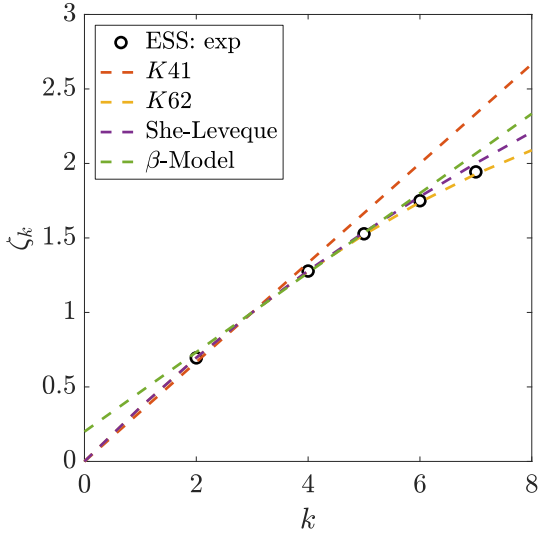}
		\caption {Course of scaling exponent $\zeta_k$ as a function of order $k$ of structure function. Dashed lines represent the scaling of a selected set of known intermittency models.}
		\label{fig:struc_ESS} 
	\end{figure}

	\newpage
	In addition this function plots in Fig.~\ref{fig:struc_ESS} the scaling exponent $\zeta_k$, with 
	\begin{eqnarray}
		S^{k}(r) \propto r^{\zeta_k},
	\end{eqnarray}
	estimated using the extended-self similarity (ESS) method introduced \mbox{by Benzi \textit{et al.} \cite{benzi1993extended}}.
	This method investigate the relative scaling laws between structure functions of different orders with $k \neq p$. 
	The most comfortable way is to apply this method for $p = 3$ and thus $\zeta_3 = 1$, leading to the investigation of $S^k(S^3(r))\propto (S^3(r))^{\zeta_k}$.
	This method allows for determining the scaling exponent $\zeta_k$ for flows with small or even not-existing inertial ranges.
	
	The scaling of a selected set of known intermittency models is also included in Fig.~\ref{fig:struc_ESS} for comparison.
	For this, the user is asked to specify the intermittency coefficient $\mu$ (experiments suggest a value of $\mu\approx0.227$ \cite{Frisch_1995} and \mbox{$\mu\approx0.26$ \cite{Arneodo_1996})} and the coefficient $D=2.8$ \cite{Anselmet_1984,novikov1964intermittency,Frisch_1995,Frisch_1978} of the $\beta$-Model proposed by \mbox{Novikov \& Stiuart \cite{novikov1964intermittency}} and introduced \mbox{by Frisch \textit{et al.} \cite{Frisch_1995,Frisch_1978}}.


	\section{PART II: Markov analysis}

		In the phenomenological model by R. Friedrich and J. Peinke  \cite{friedrich1997description} the turbulent energy cascade is interpreted as a Markov process of the velocity increments $u_r$ evolving
		in \mbox{scale $r$}.
		The chaotic property of a turbulent flow implies that the evolution of turbulent structures in the cascade exhibits a randomness. 
		For this reason, the turbulent cascade is taken as a stochastic process described by a Fokker-Planck equation and its Kramers-Moyal coefficients. 
		The distinctive feature of this approach is that the stochastic processes are not considered in time but as a process evolving in scale $r$ \cite{Friedrich_1997,friedrich1997description,renner2002universality,stresing2010towards,VanKampen2007}. 
		A realization of this stochastic process is a sequence of velocity increments,  $u_{r_0},u_{r_1},...,u_{r_n}$ with $r_0>r_1>...>r_n$ at one location $x$ or, respectively $t$, see definition Eq.~\eqref{eq:increment}.
		This hierarchical ordering of scales/increments may be considered as a so-called cascade trajectory (see Fig.~\ref{fig:traje} and the associated detailed discussion).\\
		
		To use a Fokker-Planck equation for the cascade, the process has to show the Markov property in scale. 
		The Markov property reduces a general description of cascade trajectories by the joint multiscale pdf $p(u_{r_0},u_{r_1},...,u_{r_n})$ to the knowledge of the one step or single conditioned pdfs $p(u_{r}|u_{r_0})$, with $r < r_0$ which can be described by a general Kramers-Moyal forward expansion 
		\begin{eqnarray}
			\label{eq.KM_expansion}
			-\partial _r  p\left(u_{r}| u_{r_0} \right) =  \sum_{k=1,n} \left(-\partial _{u_r}\right)^k \left[D^{(k)} (u_{r},r)p\left(u_{r}| u_{r_0} \right) \right].
		\end{eqnarray}
		Furthermore, if Gaussian noise is dominated, the general Kramers-Moyal forward expansion of the Markov process reduces to the first two terms $D^{(1,2)}$. 
		This can be tested by the Pawula Theorem and the proof of a vanishing fourth order Kramers-Moyal coefficient $D^{(4)}  $(see \cite{Risken}) as worked out for turbulence in \cite{friedrich1997description,renner2001,renner2002universality,Tutkun_2004}.\\
		
		\newpage
		A stochastic description of the energy cascade process by a Fokker-Planck equation in scale can be expressed with $r<r_0$
		\begin{eqnarray}
			\label{eq.fpe+1_theo}
			-\partial _r  p\left(u_{r}| u_{r_0} \right) = &-&\partial _{u_r} \left[D^{(1)} (u_{r},r)p\left(u_{r}| u_{r_0} \right) \right]\\
			&+&\partial^2 _{u_r}  \left[D^{(2)} (u_{r},r) p\left(u_{r}| u_{r_0} \right) \right].\nonumber
		\end{eqnarray}
		This equation is also referred to as backwards or second Kolmogorov equation \cite{renner2001}. 
		The two functions $D^{(1,2)}$ defining the Fokker-Planck equation, that describes the diffusion processes, are called drift and diffusion coefficients, respectively. 
		Both coefficients can be estimated directly from data, as shown below.
		The drift coefficient determines where the process will go to on average (purely deterministic). 
		The diffusion coefficient determines the strength of the stochastic influence (uncertainty in the evolution of the process). 
		In other words, the drift coefficient characterizes the deterministic evolution 
		from large scale to small scales, whereas the diffusion coefficient expresses the interaction of additive and multiplicative noise within the turbulence cascade.
		
		This approach hereafter referred to as Markov Analysis, achieves a comprehensive and effective characterization of the complexity of turbulence including/preserving the phenomenon of intermittency by stochastic equations.
		The correct estimation of the coefficients $D^{(1,2)}$ is crucial to a good description of the underlying scale dependent process by a Fokker-Planck equation.
		The validity of these coefficients is subsequently tested via the reconstruction of structure functions and probability density functions of velocity increments (see Section \ref{part4}) and by validating the integral fluctuation theorem (see Section \ref{sec:entropy}).

	\subsection{Examination of the Markov Property/Determination of the Markov-Einstein Length}
	\label{markov_est}
	A central assumption of the Markov Analysis is that the turbulent cascade process (more precisely the statistics of the scale-dependent velocity increments) possesses a Markov process evolving in scale. 
	Experimental evidences show that the Markov property can be assumed to hold for the cascade coarse-grained by the Einstein-Markov length $\Delta_{EM} \approx 0.9 \lambda$ \cite{Lueck2006markov,renner2001}, which suggests that molecular friction causes the break-down of the Markov assumption. 
	This finite step length can be seen in close analogy to the free mean path length of a Brownian diffusion process, which has to be so large, that two successive steps of the process can be considered as independent events \cite{Einstein_1905}.
	
	\newpage
	For Markovian processes the following relation
	\begin{eqnarray}
		\label{eq:markov}
		p\left(u_{r_n}| u_{r_{n-1}}, ...,u_{r_0} \right) = p\left(u_{r_n}| u_{r_{n-1}} \right)
	\end{eqnarray}
	holds, which means that conditional PDF of velocity increments depends only on the increment at the next larger scale $r_{n-1}$. 
	From Eq.~\eqref{eq:markov} one sees that the Markov property corresponds to a two increment closure for the joint multi-scale statistics. 
	The two increments $u_{r_n} = u(x+r_n) - u(x)$ and $u_{r_{n-1}} = u(x+r_{n-1}) - u(x)$ are given for three velocity values at three locations $x, x+r_n, x+r_{n-1}$ and thus we see that the Markov property is not only a two scale but also a three point closure. A direct relation to n-point statistics can be achieved if a further condition on $u(x)$, i.e $p\left(u_{r_n}| u_{r_{n-1}}, ...,u_{r_0}, u(x) \right)$ are included, for more details see \cite{Peinke2018}.
	%
	Motivated by the cascade picture from largest to smallest scale, this can be interpreted as that a large eddy determines the next smaller eddy directly and the following eddies only indirectly. 
	Or expressed in a different way, the future state of the process does not depend on the entire past, but only on the present state of the process (the scale evolution of the increments is memoryless).\\
	
	There are several methods to verify the Markov property using experimental data \cite{Marcq_1998,Friedrich_1998,Callaham_2021}.
	In the following, the Wilcoxon test, which is one of the most reliable procedure, and a visual test that is linked to a distance measure for the difference between two probability distributions is discussed. 
	The advantage of these two methods is that they do not depend on the estimation of the Kramers-Moyal coefficients instead both use exclusively the underlying data for the verification.\\

\begin{figure*}
		\centering
		\includegraphics[width=0.88\textwidth]{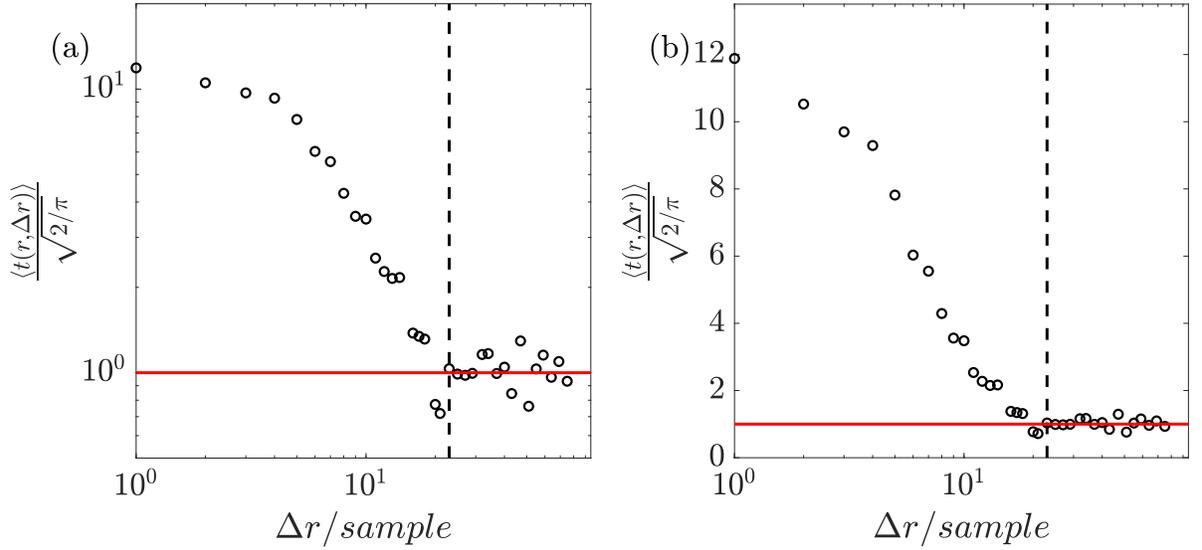}
		\caption {Development of the normalized expectation value $\left<t(r,\Delta r)\right>/\sqrt{2/\pi}$ (see \cite{Lueck2006markov} for details) as as function $\Delta r$ in terms of samples (a) on log-log scale and (b) lin-log scale. The vertical dashed line corresponds to the Taylor length scale $\lambda$ in terms of samples.}
		\label{wilcoxon} 
	\end{figure*}

	\texttt{wilcoxon\_test} This function determines the Einstein-Markov length $\Delta_{EM}$ \cite{renner2001}. 
	Above this length scale, the Markov properties hold and below this length scale, the Markov properties cease to hold. 
	The Wilcoxon test is a parameter-free procedure to compare two empirically determined probability distributions (two multi-scale statistics of velocity increments respectively two data sets of conditioned velocity increments). 
	As described above, Eq.~\eqref{eq:markov} is valid for a Markovian processes.
	For finite datasets,
	\begin{eqnarray}
		p\left(u_{r_2}|u_{r_1}\right) = p\left(u_{r_2}|u_{r_1}, u_{r_0}\right) ,
		\label{eq:markov_2}
	\end{eqnarray}
	is commonly assumed to be a sufficient condition. 
	Therefore Eq.~\eqref{eq:markov} is validated for the three different scales $r_0>r_1>r_2$, each separated by $\Delta r= \Delta_{EM}$.
	For this chosen set of scales the normalized expectation value of the number of inversions of the conditional velocity increment as a function of $\Delta r$ is calculated (see Fig.~\ref{wilcoxon}).
	If the Markov properties holds the expectation value is equal to 1. 
	A sufficient resolution in measurement below Taylor's length scale is expected to perform this test.
	A detailed description of the test is given in \cite{renner2001,Lueck2006markov}.
	
	\texttt{markov} generates a pop-up dialog box to enter the number of samples which corresponds to Einstein-Markov length (proposed value in the pop-up dialog box is
	$\Delta_{EM} \approx 0.9 \lambda$).
	Note, if the resolution of the used sensor ceases at $\Delta_{EM}$, it is possible to enter the number of samples which correspond to a larger scale than $\Delta_{EM}$ at which the data might be resolved in scale (for example samples corresponding to $\lambda$ or $5\lambda$). 
	In addition, a red vertical dashed line at the Einstein-Markov length will be added to the spectrum in the frequency domain.\\

	\texttt{min\_events} generates a pop-up dialog box to enter the minimum number of events/counts to occur in a single bin, which will make that specific bin valid for further processing. 
	If the minimum number of events is equal to 400 all the information in those specific bins in which the number of events/counts is less than 400 will be excluded for further post-processing of data. 
	The provision of the minimum number of events/counts is for avoiding the appearance of noise and hence for better fitting of parameters. 
	Based on the experience, we have fixed the minimum value of the $min\_events$ to 400. 
	Based on the length of the data and the statistics, it is possible to increase/decrease this number.\\

\begin{figure*}
		\centering
		\includegraphics[width=0.4\textwidth]{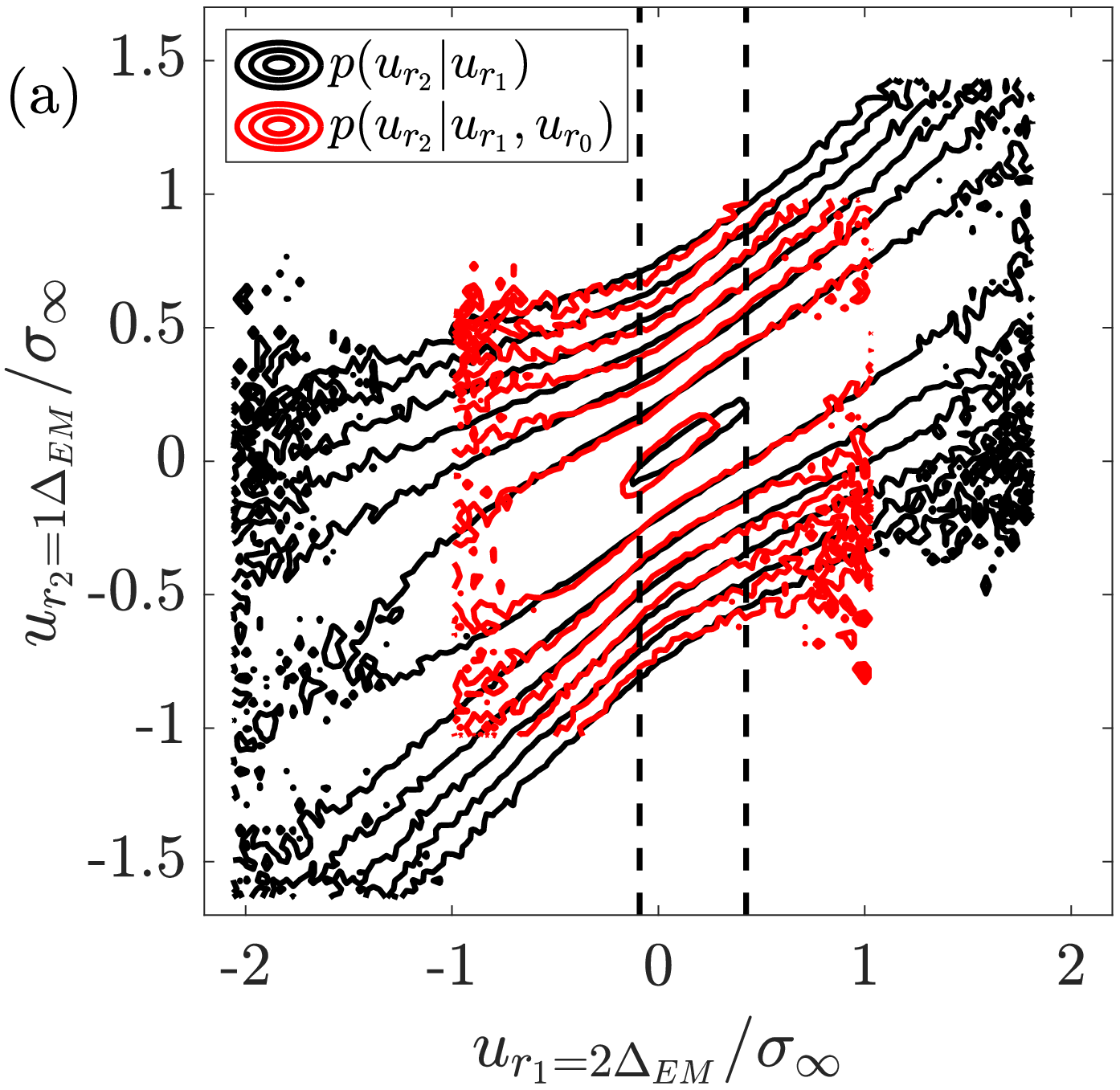} \hspace{1cm}
		\includegraphics[width=0.35\textwidth]{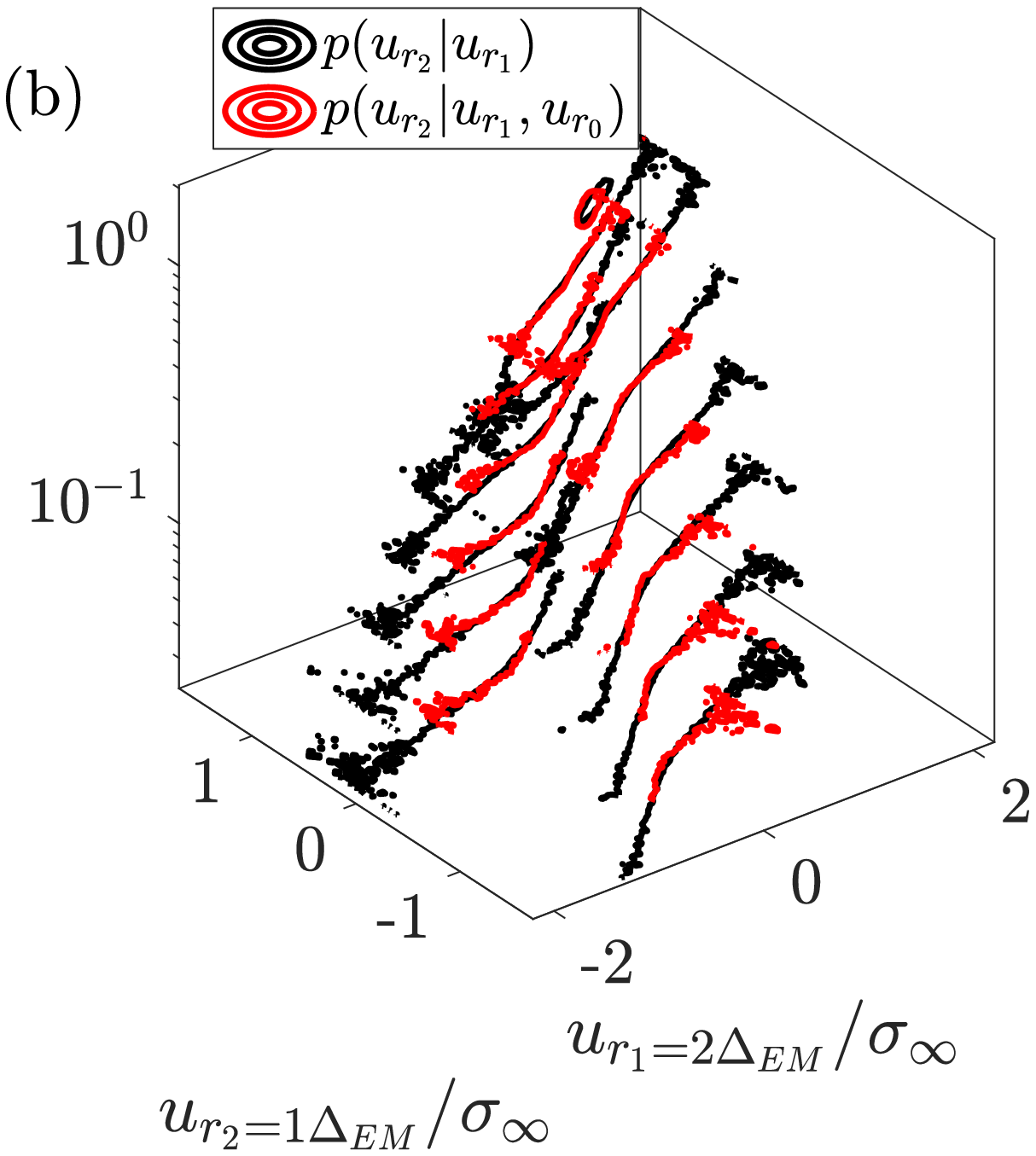}\\
		\includegraphics[width=0.4\textwidth]{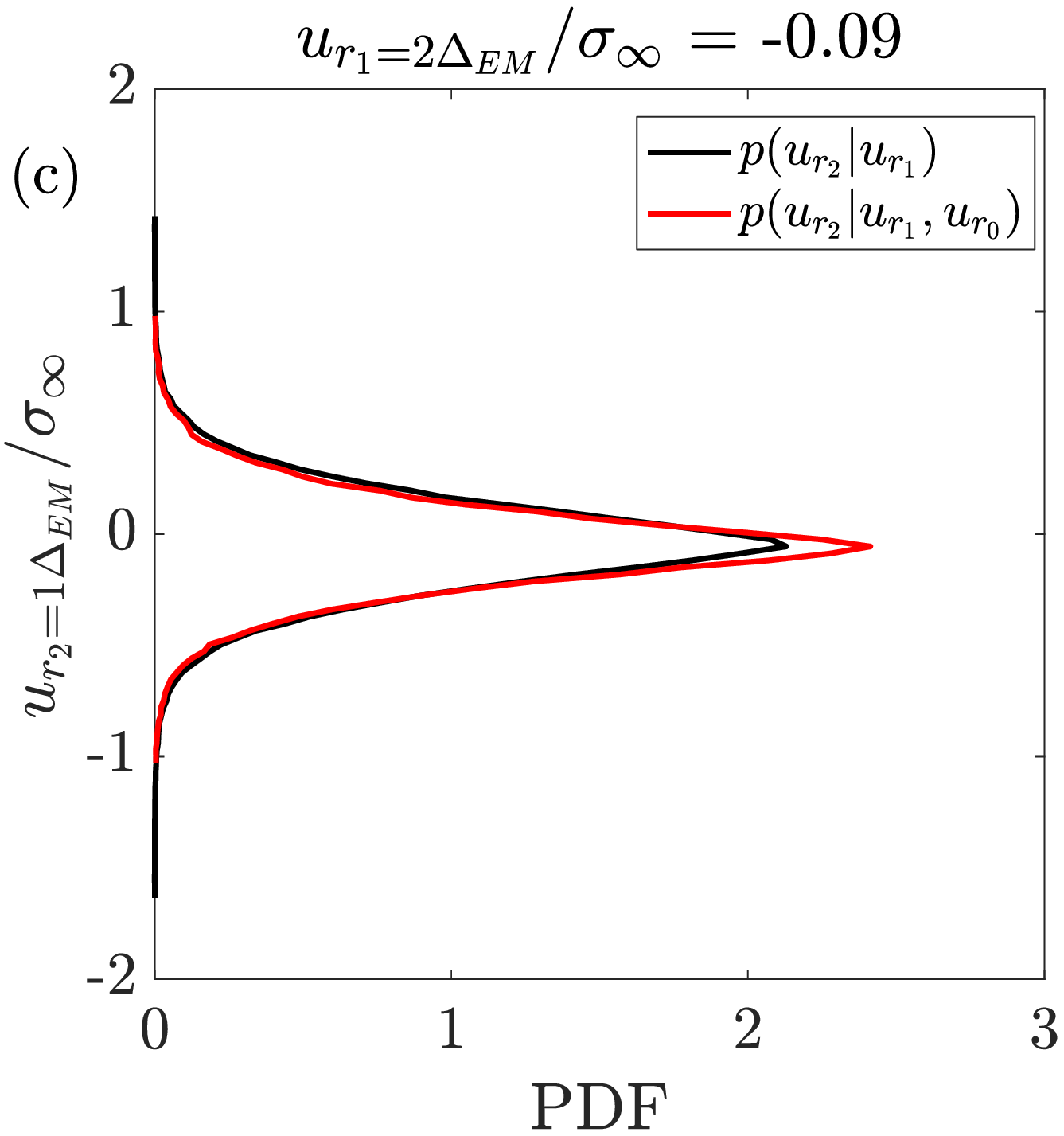}
		\includegraphics[width=0.4\textwidth]{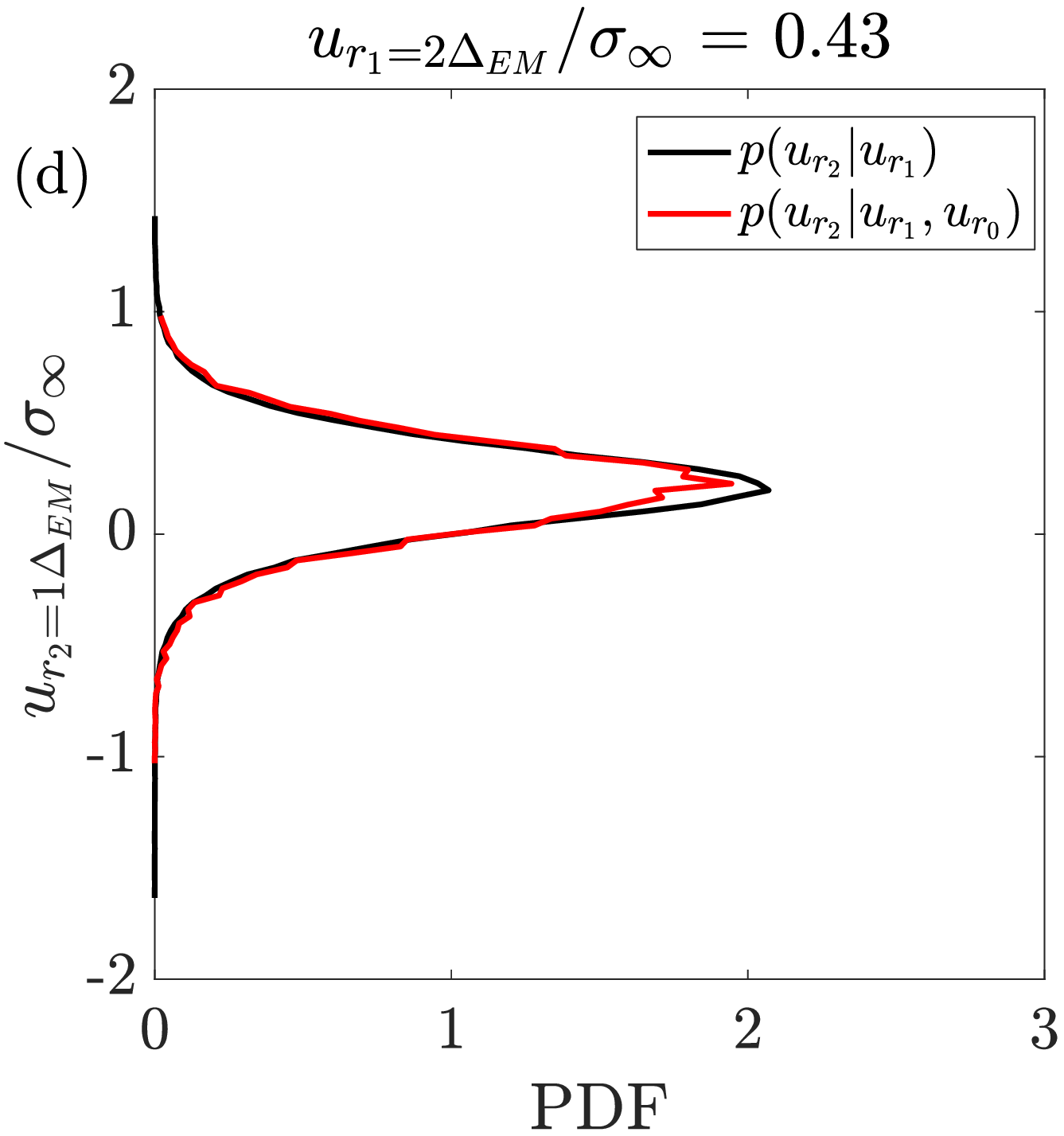}\\
		\includegraphics[width=0.4\textwidth]{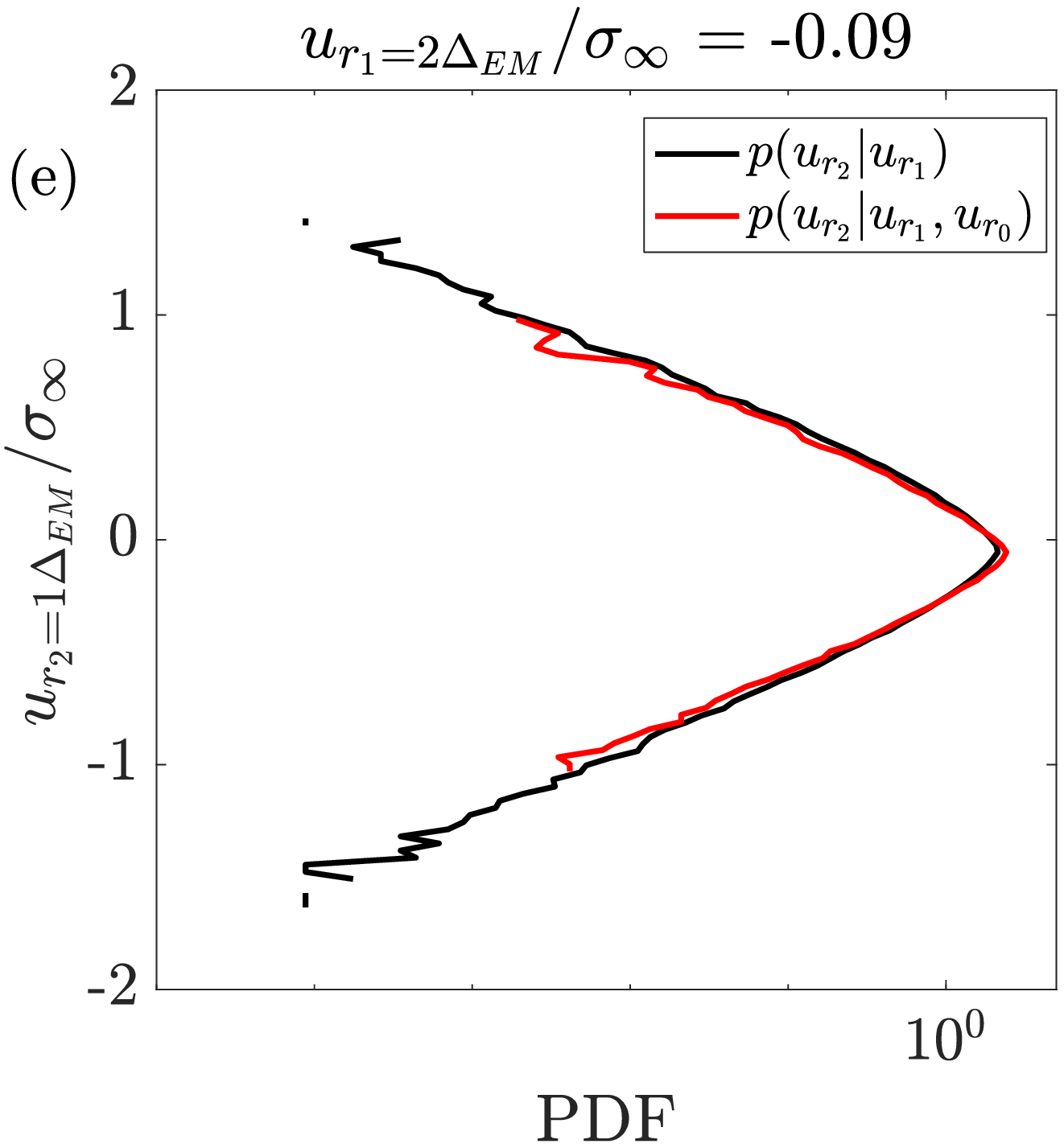}
		\includegraphics[width=0.4\textwidth]{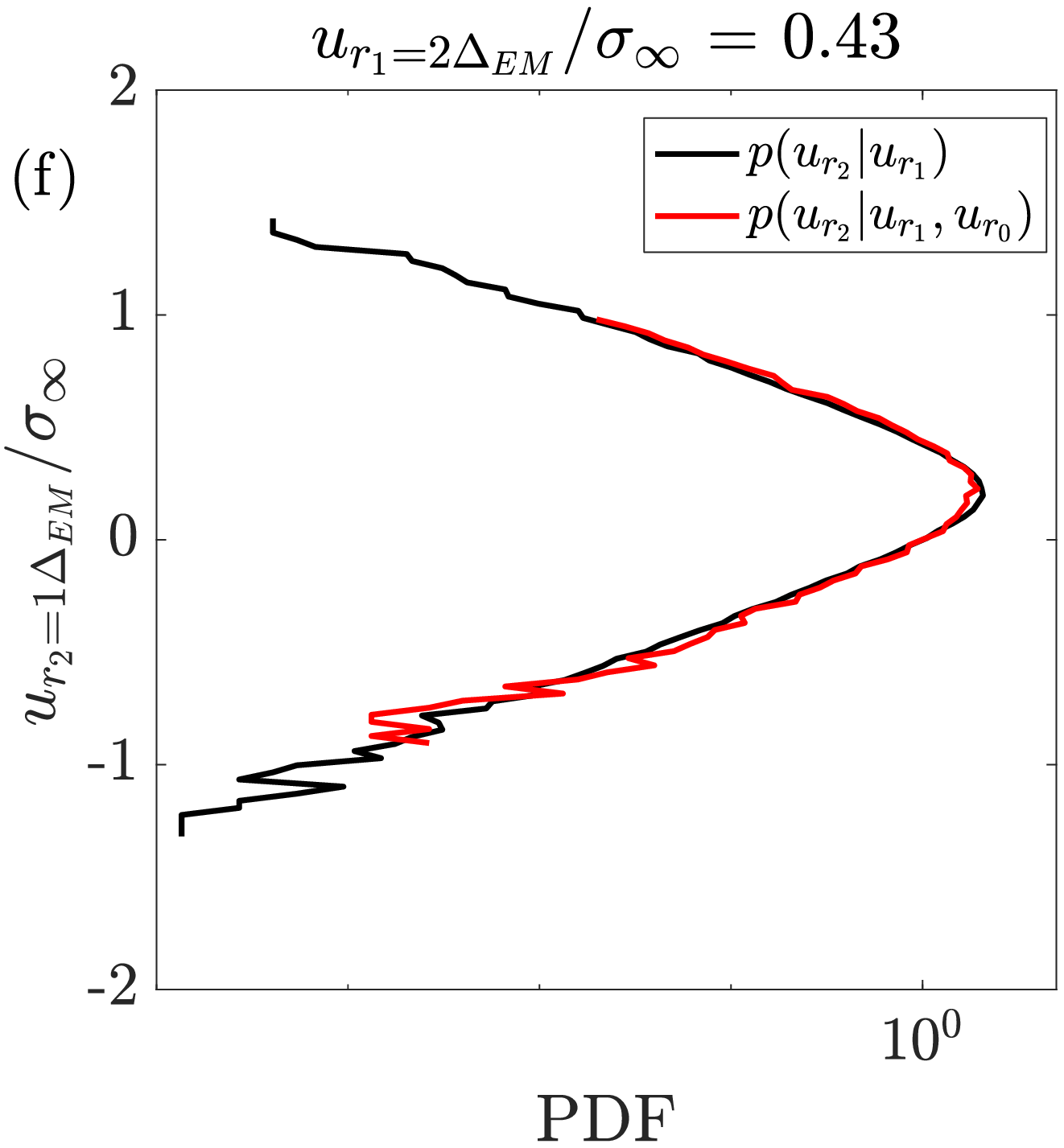}
		\caption {Visualization of Markov properties. Top: Contour plots showing single (black solid lines) and double conditioned PDFs (red solid lines, $u_{r_{0}}=0$) of velocity increments 
			for three different scales $r_0>r_1>r_2$ each of which is separated by $\Delta_{EM}$. The right figure is a three-dimensional view. 
			The dashed black lines in (a) correspond to cut through the single and double conditioned PDFs at the marked points of $u_{r_2}$. 
			(c-f): Cut through the single (black) and double (red) conditioned PDFs at the marked points of $u_{r_2}$ (c, d: double linear plot, e,f: semi logarithmic plot).}
		\label{fig:con_pdf_a} 
	\end{figure*}

	\texttt{conditional\_PDF\_markov} This function performs a qualitative/visual check for the validation of the Markov property based on the alignment or misalignment of the single conditioned $p\left(u_{r_2}|u_{r_1}\right)$ and double conditioned $p\left(u_{r_2}|u_{r_1}, u_{r_0}\right)$ probability density functions (PDFs) of datasets of velocity increments for a chosen set of three different scales $r_0>r_1>r_2$
	each of which is separated by $\Delta_{EM}$. 
	To do this, a pop-up dialog box is generated to enter the conditioned value for large scale increment $u_{r_{0}}$, for example $u_{r_{0}}=\pm 1$. Note, the condition $u_{r_{0}}=0$ corresponds to the maximum number of statistics.
	This function also plots various representations of the single and double conditioned PDFs (shown in Fig.~\ref{fig:con_pdf_a} are only two).
	
	If there is not a good agreement between the single conditioned and double conditioned PDF of velocity increments, it is possible to modify the Einstein-Markov length and/or the minimum number of events and repeat this qualitative/visual check for the validation of Markov property.
	
	In order to support this qualitative/visual check additionally in the title of Fig.~\ref{fig:con_pdf_a} a weighted mean square error function in logarithmic space \cite{feller1968} (analogous to Kullback–Leibler entropy of conditional probabilities)
	\begin{eqnarray}
		\xi =\frac{\sum_{-\infty}^{\infty}\sum_{-\infty}^{\infty}\left(A + B\right)\left(
			ln\left(A	\right) - ln\left(B\right)\right)^2}
		{\sum_{-\infty}^{\infty}\sum_{-\infty}^{\infty}\left(A + B\right)\left(
			ln^2\left(A	\right) + ln^2\left(B\right)\right)}.
	\end{eqnarray}
	is given. 
	This error function is a logarithmic measure of the difference between the single conditioned $A=p\left(u_{r_2}|u_{r_1}\right)$ and double conditioned $B=p\left(u_{r_2}|u_{r_1}, u_{r_0}\right)$ conditional probability density function. 
	The closer the match between the PDFs (black and red solid lines in Fig.~\ref{fig:con_pdf_a}) the smaller this distance measure.\\

		Note, that the results presented hereafter are related to a modified number of bins (changed from 93 to 201).
		The number of bins has been adjusted to get a more detailed view of the following figures (conditional PDFs and the Kramers-Moyal coefficients). These detailed illustrations will be described in this readme file in an exemplary manner. A smaller number of bins leads to slightly different results, but the general trend remains the same. 

	\subsection{Estimation of conditional moments}
	In this section, the conditional moments of the velocity increments 
	\begin{eqnarray}
		\label{con_mom}
		M^{(k)}\left(u_r,r,\Delta r\right) =\int_{-\infty}^{\infty} \left(u_{r'}-u_r\right)^k p\left(u_{r'} | u_r \right) du_{r'},
	\end{eqnarray}
	are estimated. 
	Here the most straightforward approach by performing a histogram-based estimation is used. 
	To do so, the user is asked to input the number of scales (steps in the energy cascade) at which Markov analysis needs to be performed.\\

	\texttt{scale\_steps} This pop-up dialog box calculates the possible number of steps between the integral length scale and Taylor length scale which are separated by the Markov length. 
	For these steps, the conditional moments for different $r$ and later on the Kramers-Moyal coefficients (KMCs) the will be estimated.\\

	\texttt{multi\_point} In this pop-up dialog box, it must be selected whether to perform the multipoint analysis or not (see annual review article by Peinke, Tabar \& Wächter \cite{Peinke2018}). 
	To do the Fokker-Planck analysis using the multiscaling approach, the question in this pop-up dialog box must be denied. 
	If multipoint analysis should be performed, an additional condition on the increment must be specified in the next pop-up dialog box.\\
	
\begin{figure}
		\centering
		\includegraphics[width=0.4\textwidth]{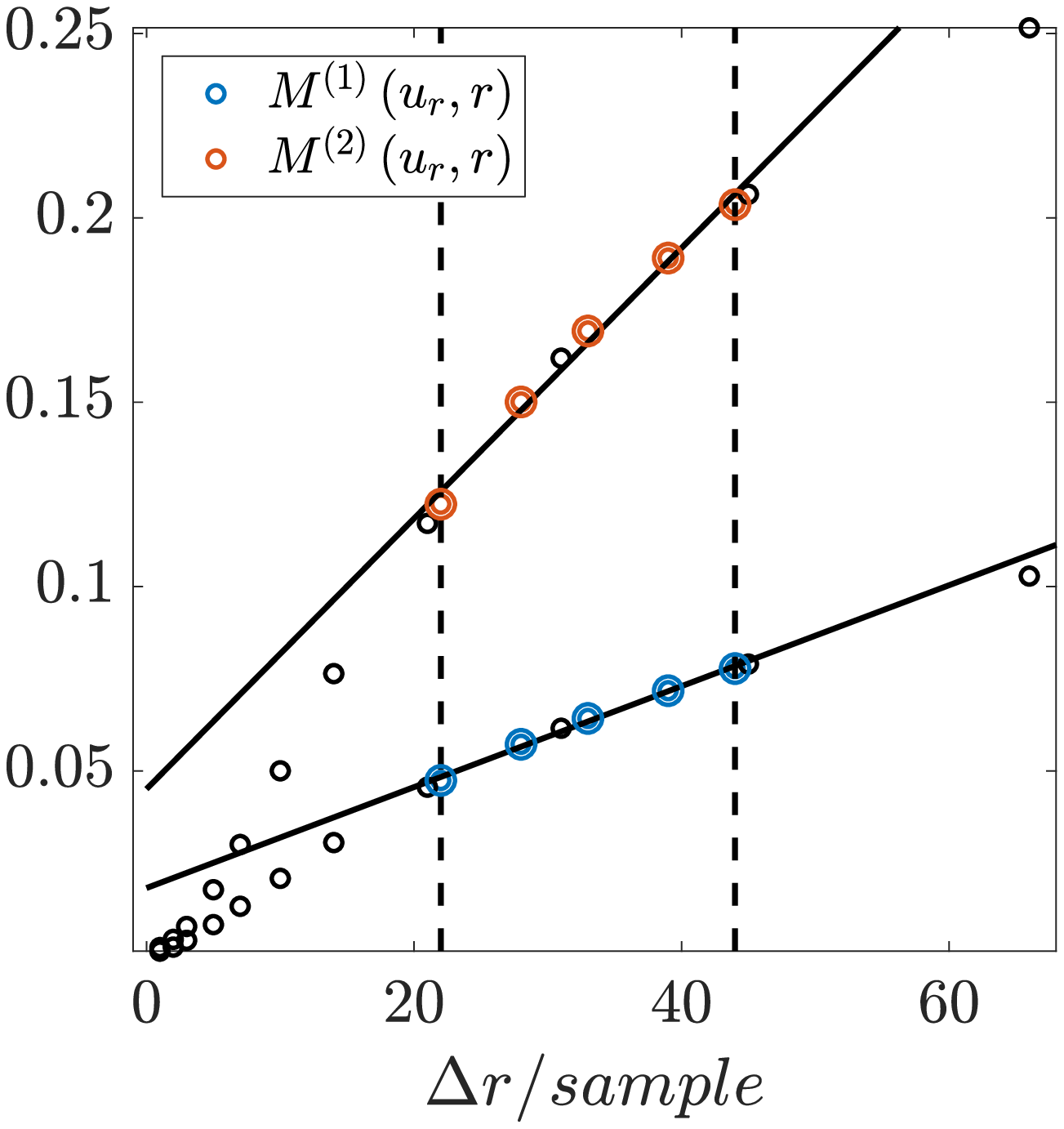}
		\caption {First and second conditional moments $M^{(1,2)}\left(u_r,r,\Delta r\right)$ as a function of the scale separation $\Delta r$. In addition, a linear extrapolation in $\Delta r$ (solid black line) of the first and second order conditional moments is plotted (see Chapter: Estimation of Kramers-Moyal coefficients). The vertical dashed lines and the colored circles limit the range used for the linear fit ($\Delta_{EM}\leq \Delta r\leq 2\Delta_{EM}$).}
		\label{fig:con_mom} 
	\end{figure}

	\texttt{conditional\_moment} This function estimates the $k$-th order conditional moment
	\begin{eqnarray}
		\label{con_mom}
		M^{(k)}\left(u_r,r,\Delta r\right) =\int_{-\infty}^{\infty} \left(u_{r'}-u_r\right)^k p\left(u_{r'} | u_r \right) du_{r'},
	\end{eqnarray}
	$k={1-4}$ for all scales $2\Delta_{EM} <r \leq L$ and for each bin (specified in the function \texttt{scale\_steps} and \texttt{increment\_bin}) for all values of longitudinal velocity increments $u_r$. 
	For a fixed scale $r$ the conditional moments are calculated for 5 different scales separations (colored circles in Fig.~\ref{fig:con_mom}) $\Delta r=r-r'$ within the range of $\Delta_{EM}\leq \Delta r\leq 2\Delta_{EM}$. The condition $r'<r$ is fulfilled.\\
	
	
	\texttt{plot\_conditional\_moment} This function plots in Fig.~\ref{fig:con_mom} the first and second conditional moments $M^{(1,2)}\left(u_r,r,\Delta r\right)$ as a function of the scale separation $\Delta r$. 
	For this purpose, a scale $r$ and the number of a bin (value of the velocity increment $u_r$) condition must be specified. 
	The proposed value in the pop-up dialog box is $r=L$ and $u_r \approx 0$. 
	A possible deviation from a linear law for small values of $\Delta r$ is due to the Einstein-Markov length, as the Markov properties cease to hold for very small scale separations.

	\subsection{Estimation of Kramers-Moyal coefficients (KMCs)}
	\label{KMC}
	In this section, the Kramers-Moyal coefficients $D^{(k)}\left(u_r,r\right)$
	are estimated by means of the conditional moments calculated in the previous section.\\
	
	\texttt{KM\_Calculation} This function calculates the Kramers-Moyal coefficients 
	$D^{(k)}\left(u_r,r\right)$ with $k={1-4}$ for all scales (specified in \texttt{scale\_steps}) and for each bin (specified in the function \texttt{increment\_bin}) for all values of velocity increments by a linear extrapolation in $\Delta r$ of the $k$-th order conditional moments $M^{k}\left(u_r,r\right)$ (see Fig.~\ref{fig:con_mom}) and the function \texttt{KM\_plot\_raw} plots the coefficients (see Fig.~\ref{fig:optimi_a} (a-b)). With $r'<r$:
	\begin{eqnarray}
		D^{(k)}\left(u_r,r\right) = \lim\limits_{r'\rightarrow r}\frac{M^{(k)}\left(u_r,r,\Delta r\right)}{k!\, \left(r'-r\right)}.
		\label{eq.KMcoeff}
	\end{eqnarray}
	Note that some publications use the prefactor $-r$ to implicitly describe a logarithmic scaling of the scale evolution from large to small scales, which is advantageous for complex structures with self-similar properties. 
	This prefactor is omitted here.
	From Eq.~\eqref{con_mom} it can be seen that the smallest scale at which the KMCs can be calculated is slightly larger than $2\Delta_{EM}$.

	\subsection{Pointwise optimization of Kramers-Moyal coefficients}
	\label{KMC_opti}
		As described above, a histogram-based estimation of the Kramers-Moyal coefficients is performed.
		We call this estimation pointwise as for fixed scale $r$ and fixed velocity increment of a bin these coefficients are determined.
		While this approach is one of the most straightforward methods to determine an initial estimate of $D^{(k)}\left(u_r,r\right)$ from data, it is also sometimes characterized by a high degree of uncertainty.
		This is especially true for bins containing a relatively small number of events.
		Furthermore the limit approximation in Eq.~\eqref{eq.KMcoeff} (see also Fig.~\ref{fig:con_mom}) leads to uncertainties in the absolute values of the Kramers–Moyal coefficients, whereas the functional forms of $D^{(k)}\left(u_r,r\right)$ are commonly well estimated. 
	
	In this package, we address this problem by following a two-step optimization. 
	First, a pointwise optimization of the experimentally estimated $D^{(1,2)}\left(u_r,r\right)$ at each scale and value of velocity increment is performed using the short time propagator \cite{Risken} 
	to find the best Fokker-Planck equation to reproduce the conditional PDFs like shown in Fig. \ref{fig:optimi_a} (c, d) (see Section \ref{KMC_opti: conditional PDF}).
	As shown in \cite{Nawroth2007} for experimental data there still remains an uncertainty in the coefficients $D^{(1,2)}$ using this first optimization. 
	This uncertainty we use as freedom to furthermore optimize the KMCs in the following second step. 
	In the second step of the optimization (see details in Section \ref{IFT_opti})
	an optimization of the Kramers-Moyal coefficients towards the integral fluctuation theorem (see details in Section \ref{sec:entropy}) is performed.\\
	
	This procedure to obtain an optimal Fokker-Planck equation can be interpreted as follows: If the condition for a Markov process for the cascade is given, the whole process is uniquely defined by the transition probabilities $p\left(u_{r_{n}}|u_{r_{n-1}}\right)$ for increments from scale $r_{n-1}$ to $r_{n}$. 
	This is the most central feature of the cascade process. 
	In step 1 the best Fokker-Planck equation to model $p\left(u_{r_{n}}|u_{r_{n-1}}\right)$ is found. 
	The remaining uncertainties in the coefficients of the Fokker-Planck equation are used in step 2 to get the best values for the integral fluctuation theorem (IFT). 
	In this way, a Fokker-Planck equation, which is compatible with the Markovian cascade process is obtained.
	As the fulfillment of the IFT is an independent feature but a direct consequence of a Fokker-Planck equation, the obtained Fokker-Planck equation by our procedure is the best solution. 
	The methods presented here allow for given data a critical analysis of these points.

	\subsection{Pointwise optimization of Kramers-Moyal coefficients: conditional PDF}
	\label{KMC_opti: conditional PDF}
		\texttt{KM\_STP\_optimization} This function performs the first step of the pointwise optimization of Kramers-Moyal coefficients $D^{(1,2)}\left(u_r,r\right)$ at each scale and value of velocity increment to minimize possible uncertainties in the absolute values of the Kramers–Moyal coefficients. 
		The purpose of this optimization is to find the best Fokker-Planck equation to reproduce the conditional PDFs as these are the essential part of the Markov process. 
		This optimization procedure is proposed in \cite{kleinhans2005iterative,Nawroth2007,Reinke_2018} and it includes the reconstruction of the conditional probability density functions $p\left(u_{r'} | u_r \right)$ via the short time propagator of Eq.~\eqref{eq.P_STP} with $r'<r$
			\begin{eqnarray}
					\label{eq.P_STP}
					p_{stp}\left(u_{r'} | u_r \right) = & &\frac{1}{\sqrt{4\pi D^{(2)}(u_{r},r) \Delta r}}\\
					& & exp\left( -\frac{\left( u_{r'}-u_r-D^{(1)}(u_{r},r)\Delta r \right)^2}{4 D^{(2)}(u_{r},r) \Delta r}\right). \nonumber
				\end{eqnarray}
		The scale step size $\Delta r=\Delta_{EM}$ leads to consistent results. Smaller steps than $\Delta_{EM}$ do not significantly improve the results. 
		The aim of this optimization is to minimize a weighted mean square error function in logarithmic space \cite{feller1968} (analogous to Kullback–Leibler entropy)
		\begin{eqnarray}
			\xi =\frac{\sum_{-\infty}^{\infty}\sum_{-\infty}^{\infty}\left(p_{exp} + p_{stp}\right)\left(
				ln\left(p_{exp}	\right) - ln\left(p_{stp}\right)\right)^2}
			{\sum_{-\infty}^{\infty}\sum_{-\infty}^{\infty}\left(p_{exp} + p_{stp}\right)\left(
				ln^2\left(p_{exp}	\right) + ln^2\left(p_{stp}\right)\right)}.
		\end{eqnarray}
		This error function is a logarithmic measure of the difference between the experimental $p_{exp}$ and reconstructed $p_{stp}$ conditional probability density function. 
		The optimization procedure systematically changes $D^{(1,2)}\left(u_r,r\right)$ until the error function is minimized. 
		This optimization use the function \texttt{fmincon} implemented in MATLAB. 
		
\begin{figure*}
			\centering
			\includegraphics[width=0.39\textwidth]{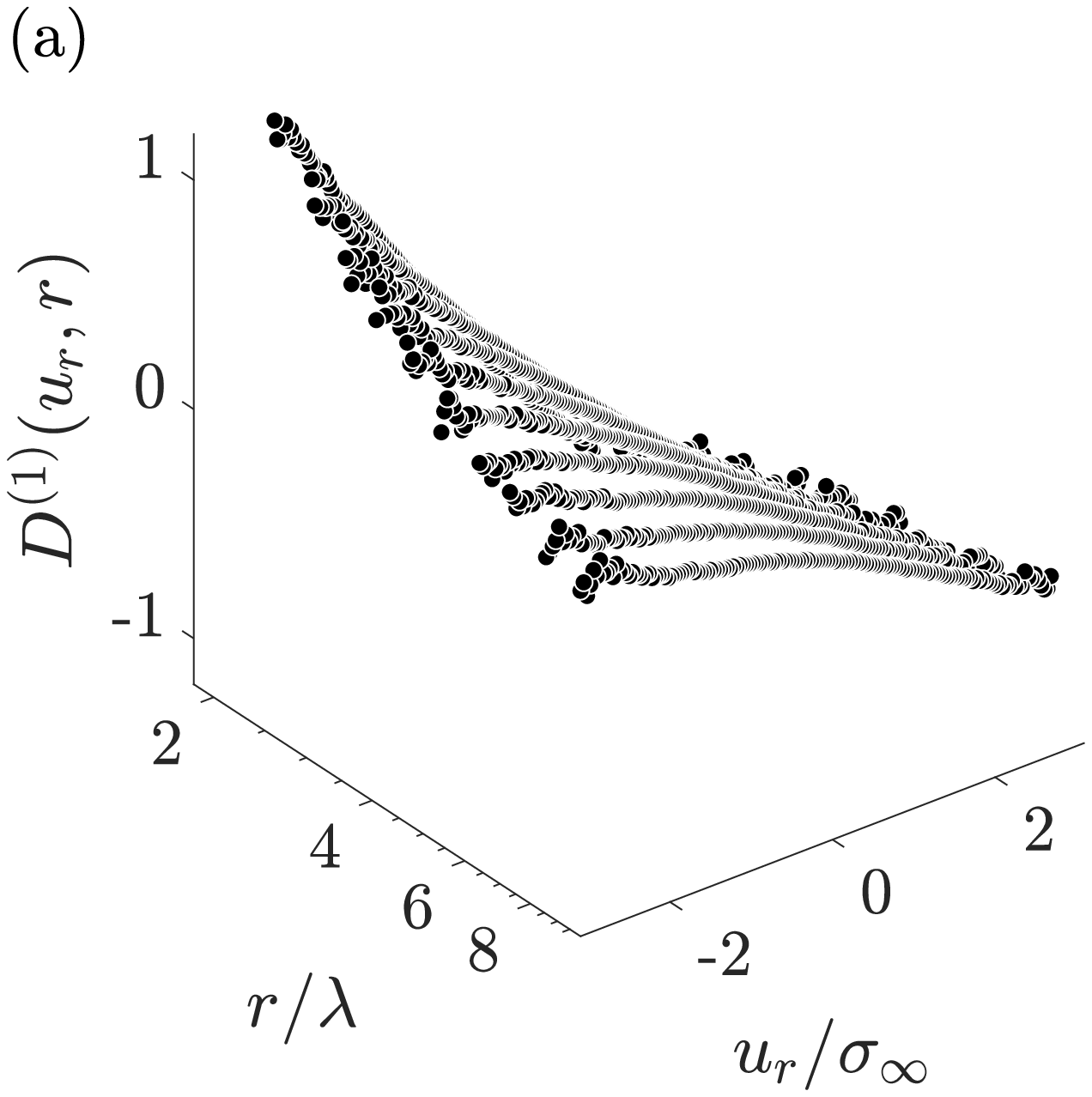}
			\includegraphics[width=0.39\textwidth]{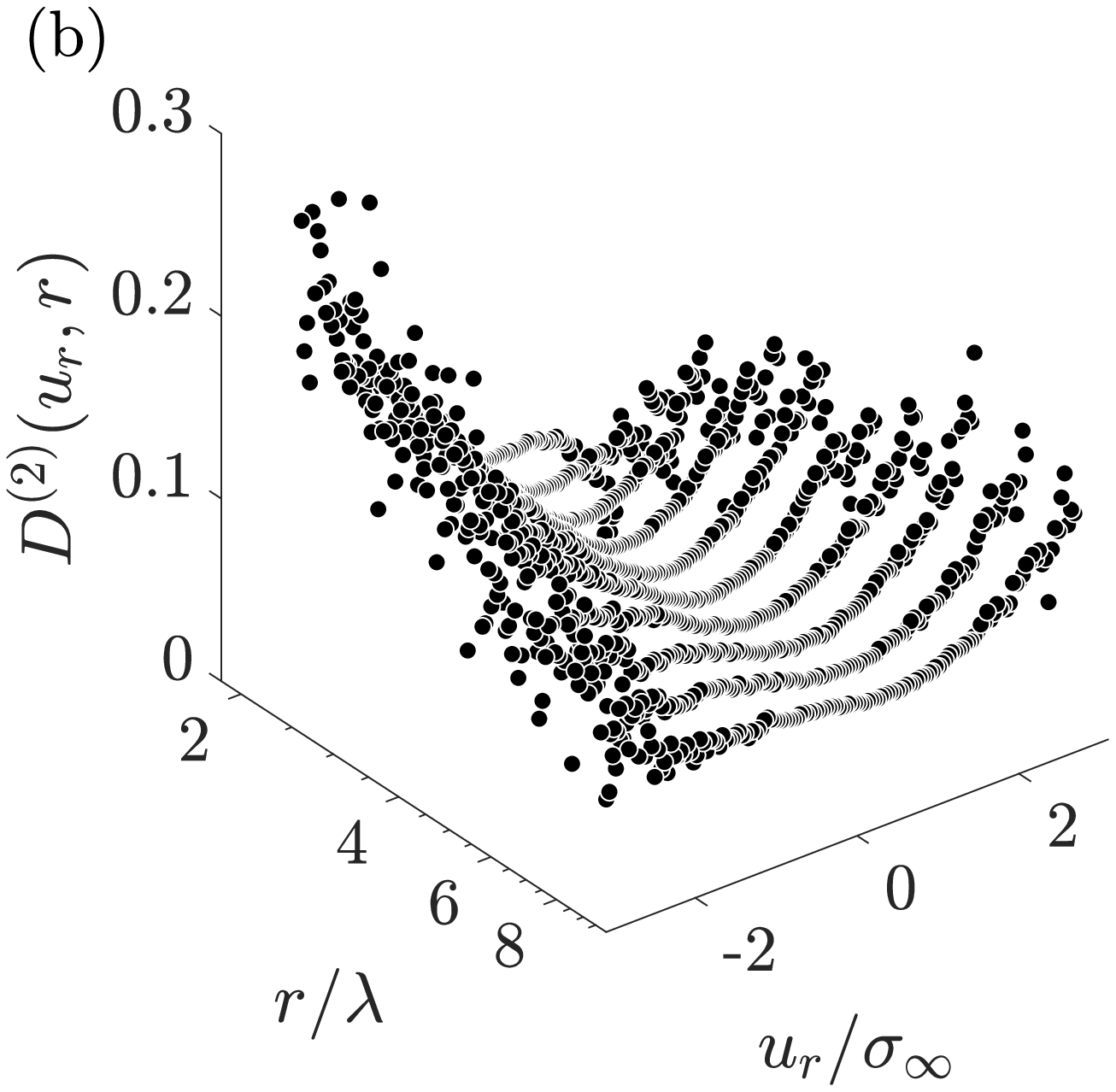}
			\includegraphics[width=0.38\textwidth]{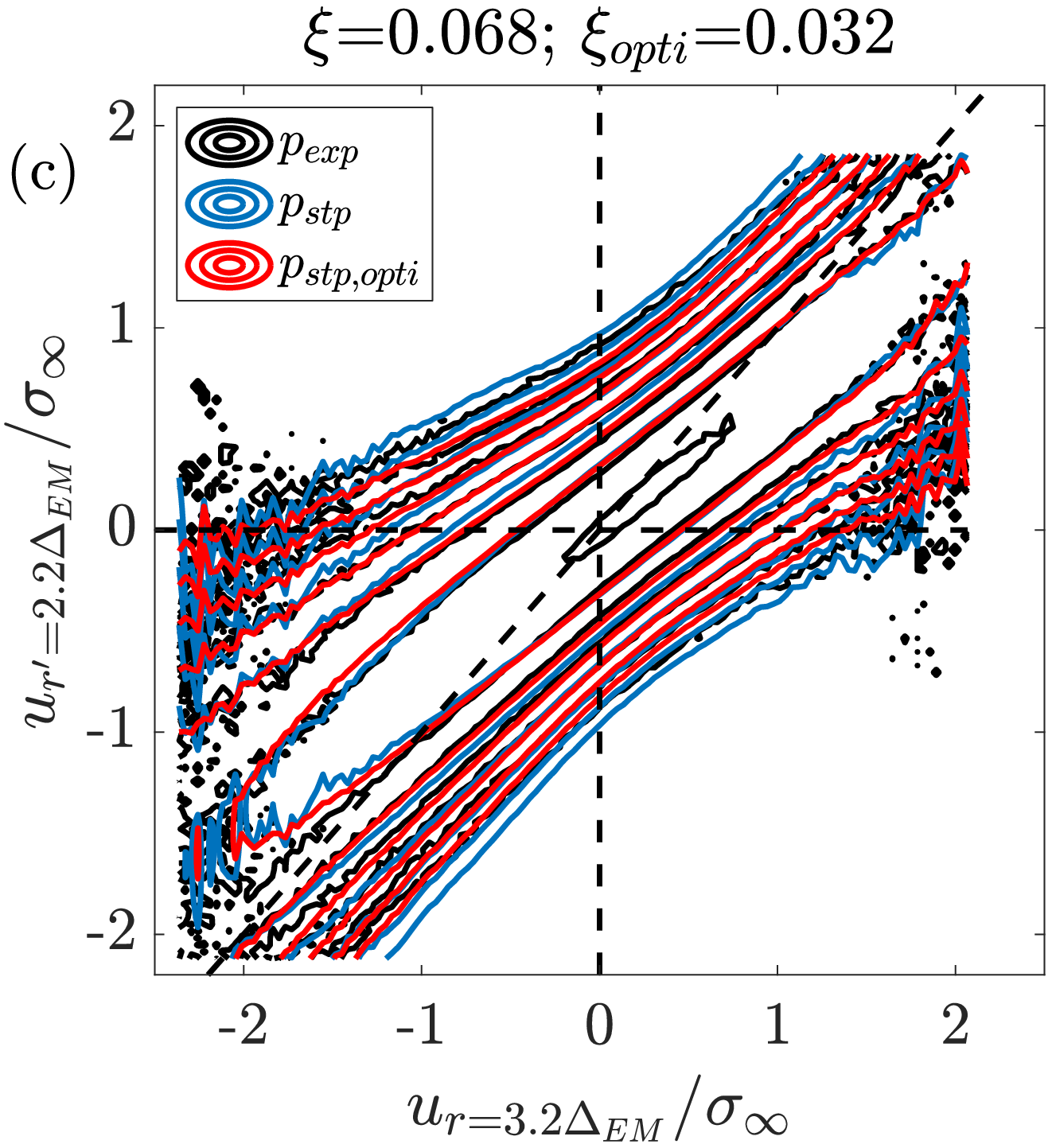}\hspace{0.3cm}
			\includegraphics[width=0.36\textwidth]{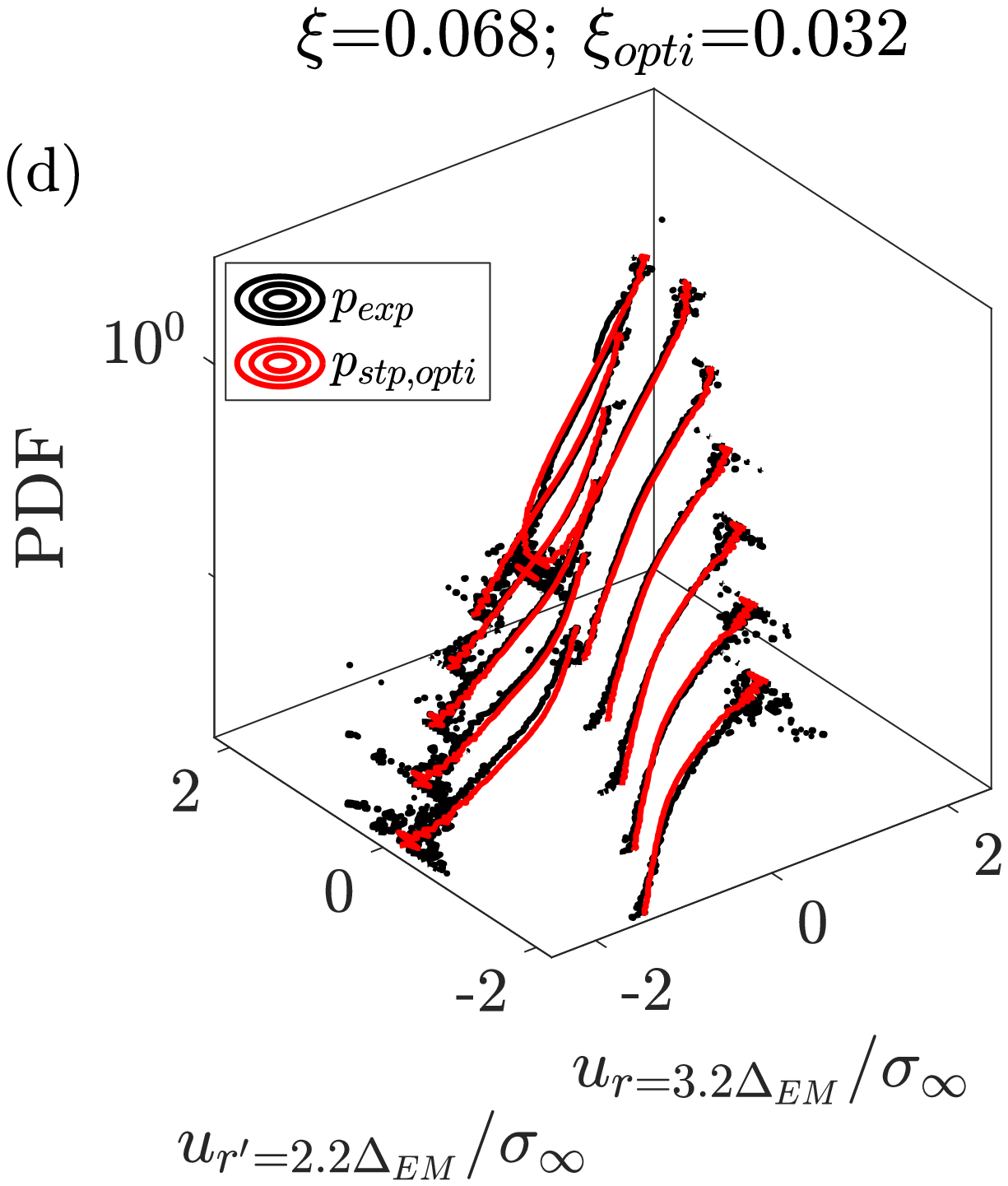}\\
			\includegraphics[width=0.39\textwidth]{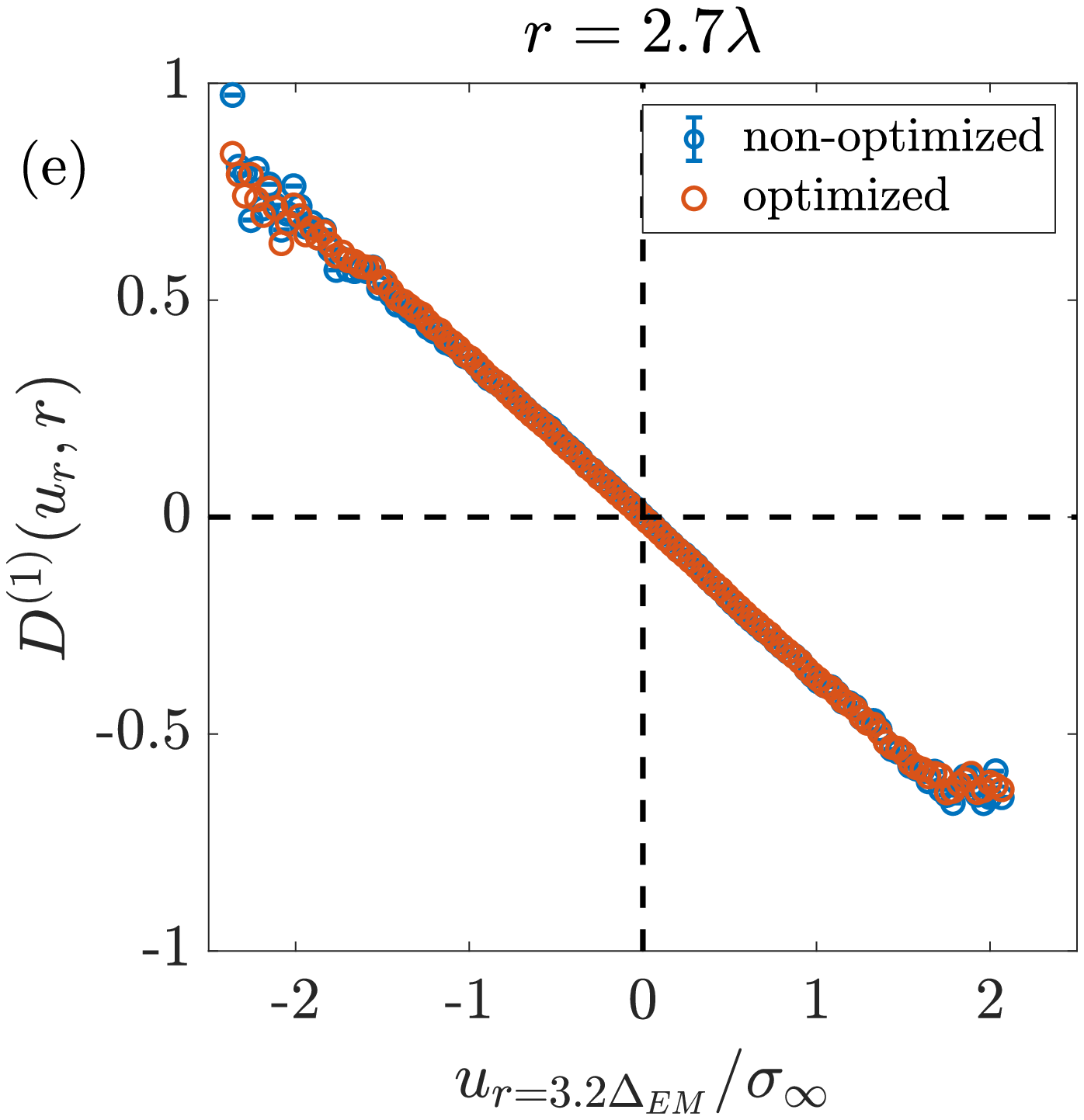}
			\includegraphics[width=0.39\textwidth]{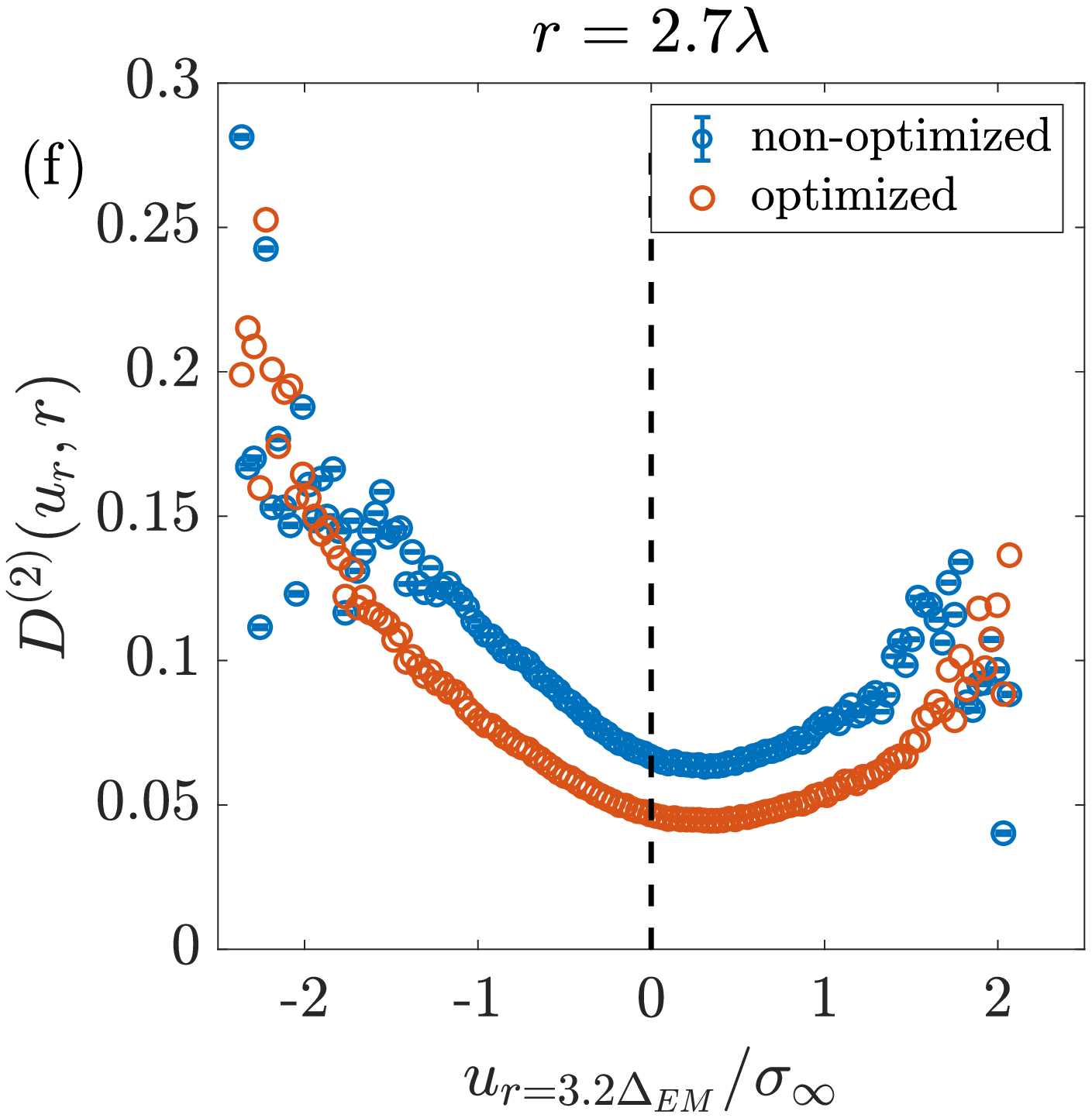}
			\caption {Non-optimized Kramers-Moyal coefficients (a) $D^{(1)}\left(u_r,r\right)$ and (b) $D^{(2)}\left(u_r,r\right)$ with respect to scale $r$ (with $L\approx 10 \, r/\lambda$) and velocity increment obtained by the linear extrapolation method of Fig. \ref{fig:con_mom}. (c) Contour plots showing experimental $p_{exp}$ (black), non-optimized $p_{stp}$ (blue) and optimized conditonal PDFs $p_{stp,opti}$ (red, first step of the pointwise optimization) 
				using the short time propagator (see Eq.~\eqref{eq.P_STP}) of velocity increments for a pair of two scales with $r'<r$ each of which is separated by $\Delta_{EM}$. (d) is a three-dimensional view of (c)
				(only $p_{exp}$ and $p_{stp,opti}$ are shown). (d, e): Non-optimized and optimized (first step of the pointwise optimization) Kramers-Moyal coefficients $D^{(1,2)}\left(u_r,r\right)$ with respect to velocity increment $u_r$ for a fixed scale $r=2.7 \lambda = 3.2 \Delta_{EM}$ obtained by the optimization algorithm.}
			\label{fig:optimi_a} 
		\end{figure*}

		In addition, this function generates a pop-up dialog box whether an example optimization is required. 
		If the pop-up dialog box is denied, then this function straightaway performs the optimization for all scales and all velocity increments without plotting anything. 
		If the pop-up dialog box is confirmed, then the conditional PDFs will be plotted using different representations (see Fig.~\ref{fig:optimi_a} (c-f)) to see the differences between optimized, non-optimized and experimental conditional PDFs. 
		Note, if the variable \texttt{scale\_steps} is equal to 9 then it is possible to enter any scale number from 1 up to scale number 9 (smallest respectively the largest scale).\\
		%
		%
		%
		%

\begin{figure*}
			\centering
			\includegraphics[width=0.402\textwidth]{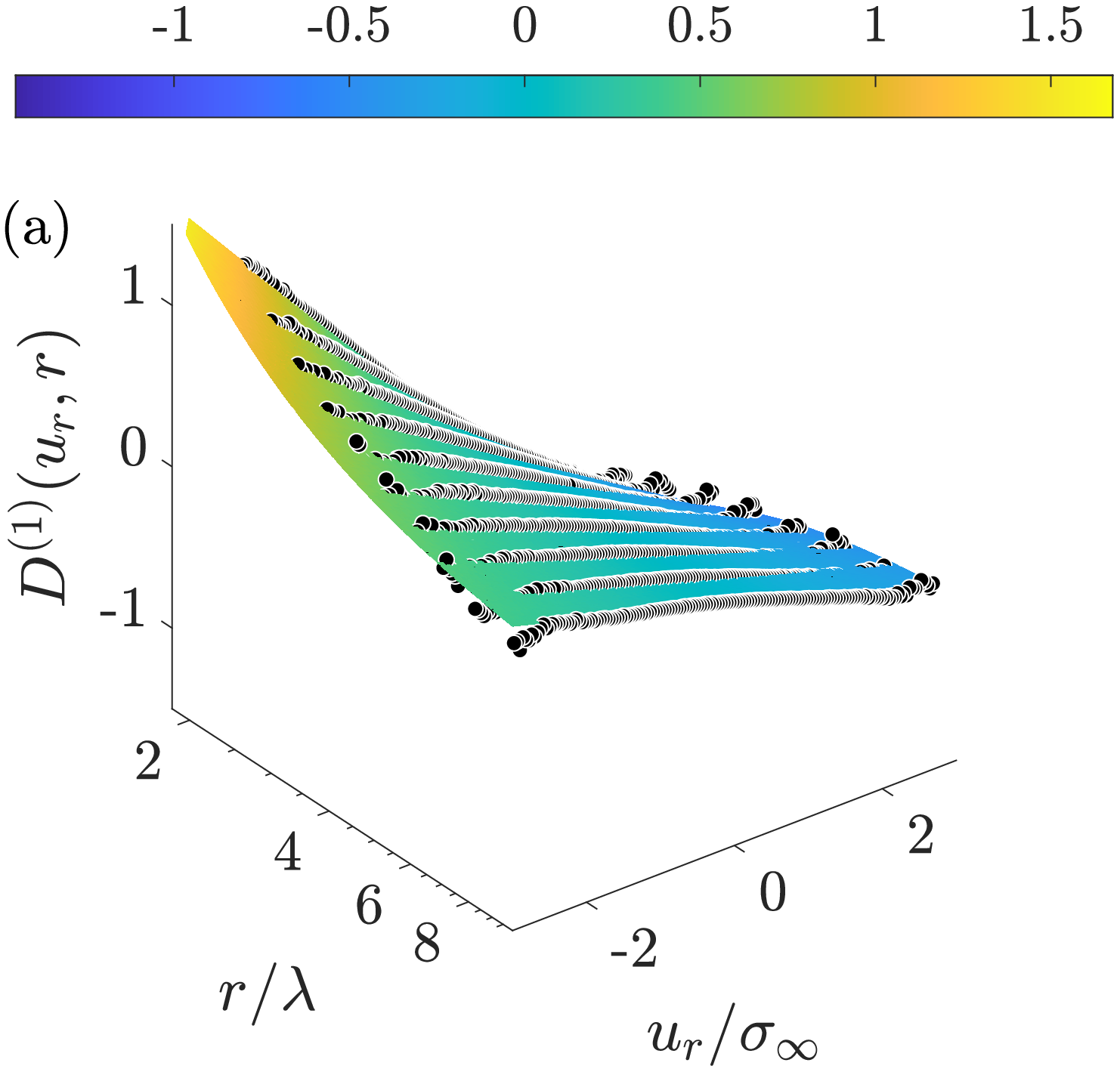}
			\includegraphics[width=0.41\textwidth]{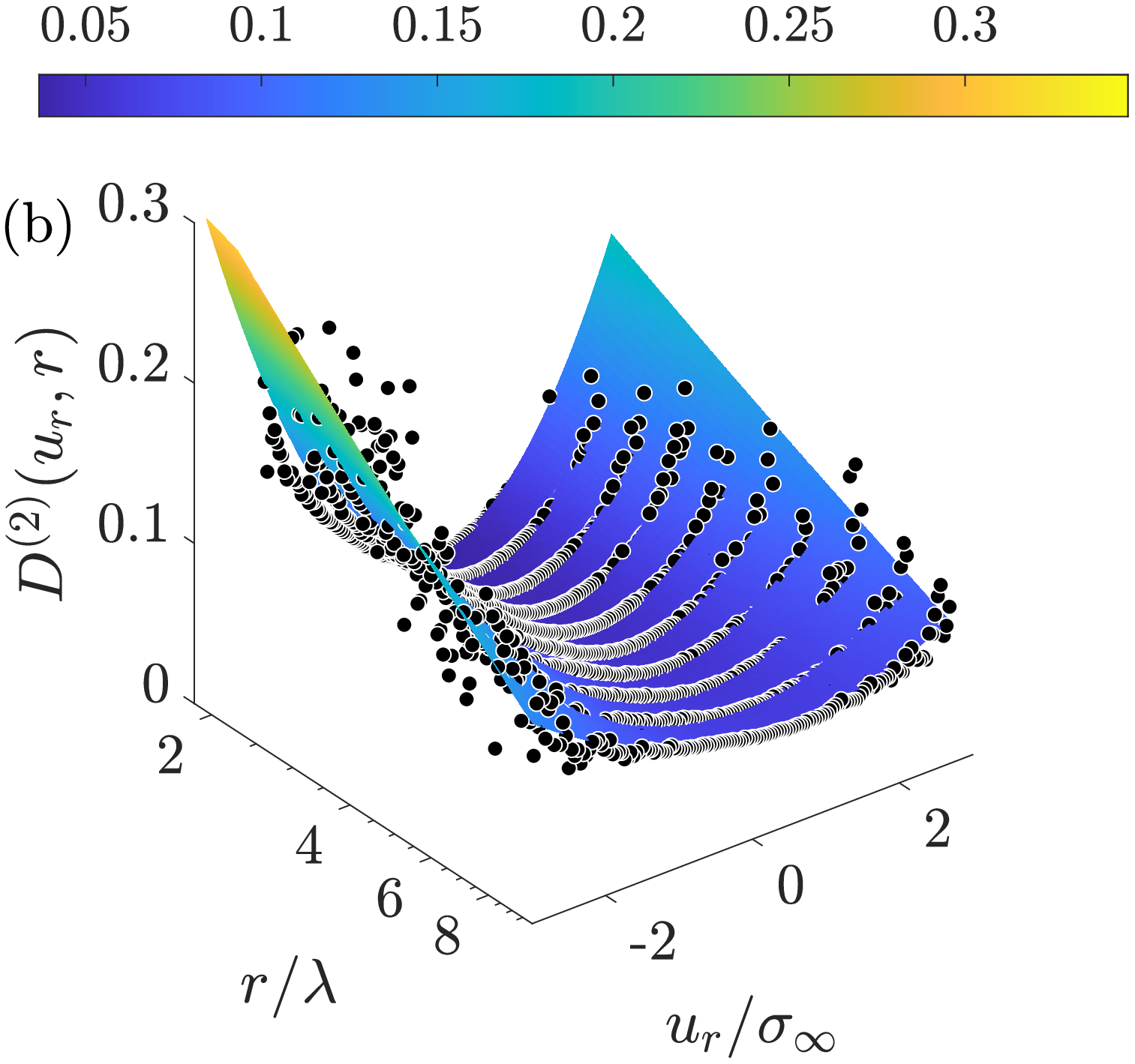}
			\includegraphics[width=0.7\textwidth]{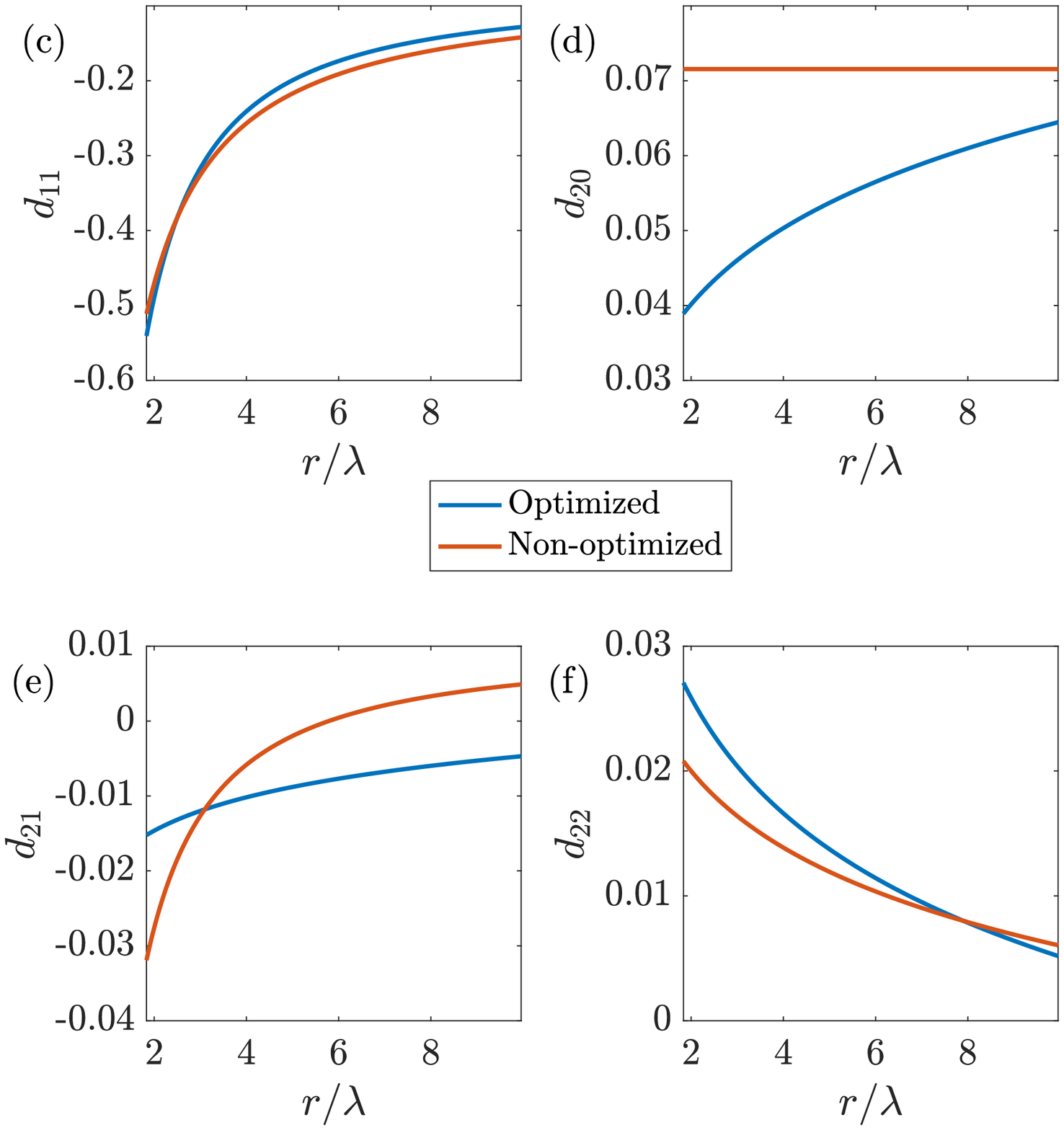}
			\caption {Optimized Kramers-Moyal coefficients (a) $D^{(1)}\left(u_r,r\right)$ and (b) $D^{(2)}\left(u_r,r\right)$ (black filled circles) and the surface fits with a linear function for $D^{(1)}\left(u_r,r\right)$ and a parabolic function for $D^{(2)}\left(u_r,r\right)$ (see Eq.~\eqref{eq:D1andD2_1} and Eq.~\eqref{eq:D1andD2_2}) with respect to scale $r$ and velocity increment $u_r$. The colorbars indicate the mapping of the surface fits data values into the colormap.
			(c - f)Coefficients $d_{ij}(r)$ of the optimized Kramers-Moyal coefficients using the surface fits with a linear function for $D^{(1)}\left(u_r,r\right)$ (see Eq.~\eqref{eq:D1andD2_1} and (c) for $d_{11}(r)$) and a parabolic function for $D^{(2)}\left(u_r,r\right)$ (see Eq.~\eqref{eq:D1andD2_2} and (d) for $d_{20}(r)$, (e) for $d_{21}(r)$ and (f) for $d_{22}(r)$) with respect to scale.}
			\label{Fit_KM} 
		\end{figure*}

		\texttt{FIT\_KM} This function plots in Fig.~\ref{Fit_KM} (a,b) the optimized $D^{(1,2)}\left(u_r,r\right)$.
		In addition this function performs the surface fits with a linear function for $D^{(1)}\left(u_r,r\right)$ and a parabolic function for $D^{(2)}\left(u_r,r\right)$ to the optimized and non-optimized KMCs interpreted in the Itô convention~\cite{GardinerBook}
		\begin{eqnarray} 
			\label{eq:D1andD2_1}
			D^{(1)}(u_{r},r) &=& d_{11}(r)u_{r},\\
			D^{(2)}(u_{r},r) &=& d_{22}(r)u_{r}^2+d_{21}(r)u_{r}+d_{20}(r).
			\label{eq:D1andD2_2}
		\end{eqnarray}
		
		Since the comparison of the $D^{(1,2)}\left(u_r,r\right)$ on the basis of three dimensional figures like in Fig.~\ref{fig:optimi_a} (a-b) is very cumbersome and difficult, these surface fits are used to quantify the dependencies on the increment and the scale.
		The coefficients $d_{ij}(r)$ in the fits are functions of scale~$r$ of the form 
		\begin{eqnarray} 
			d_{ij}(r)=\alpha (r/\lambda)^{\beta}+\gamma.
		\end{eqnarray} 
		Using this type of fit it is possible to quantify intermittency while using different formalism in the literature.
		The constraints of this surface fits were set in a physically and mathematically meaningful way: $d_{11}\leq 0$, $d_{20}\geq 0$ and $d_{22}\geq 0$. 
		After fitting, this function plots in Fig.~\ref{Fit_KM} (c-f) the parameters $d_{11}$, $d_{20}$, $d_{21}$ and $d_{22}$ as a function of scale
		for optimized and non-optimized $D^{(1,2)}\left(u_r,r\right)$.

		%


	\section{Part III: Entropy analysis}
	\label{sec:entropy}
\begin{figure}
		\centering
		\includegraphics[width=0.45\textwidth]{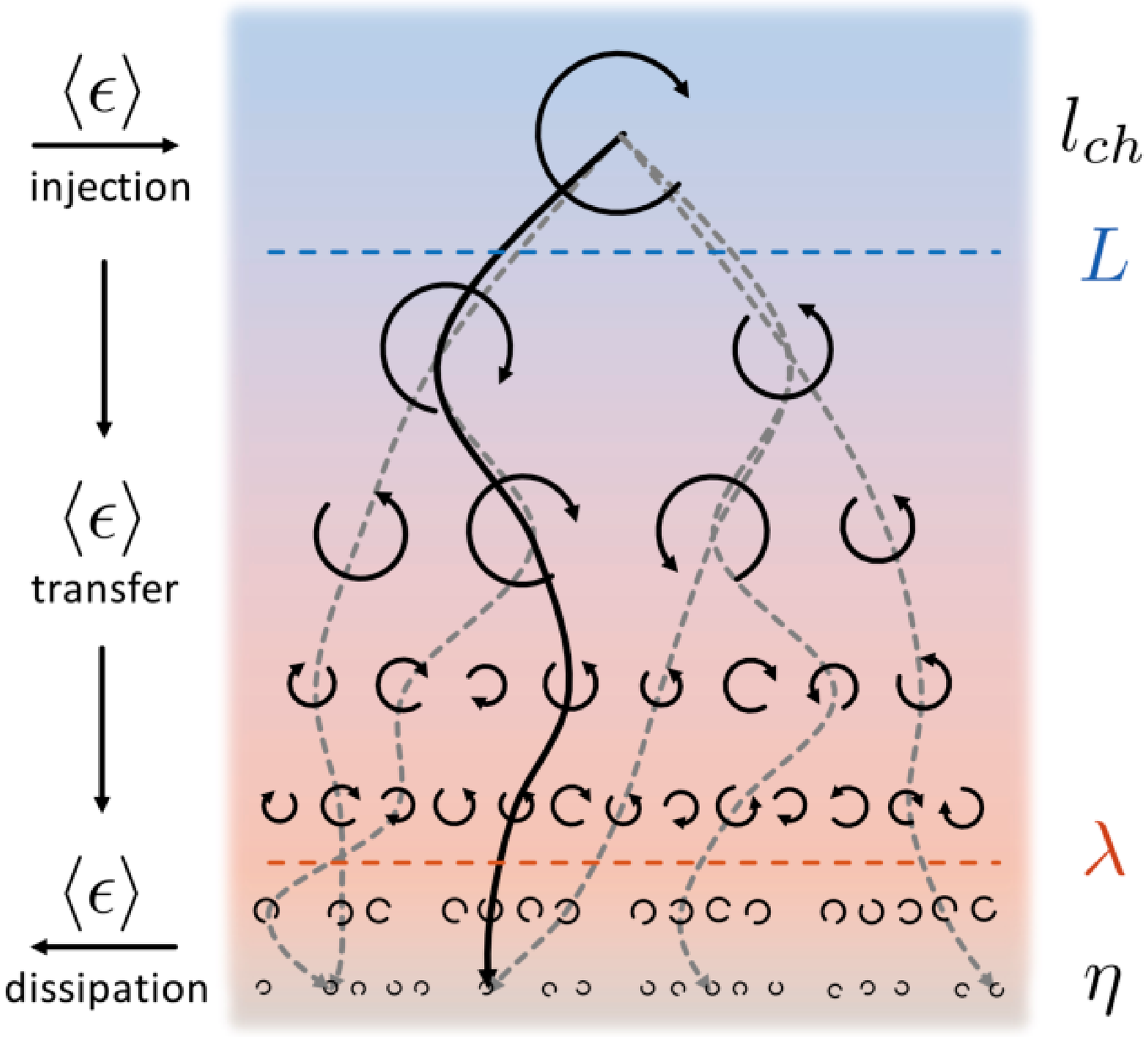}
		\caption{Schematic representation of the phenomenologically inspired turbulent energy cascade process. Shown in grey dashed lines are different cascade trajectories taken by the cascade process between the integral length scale $L$ (initial scale) and the Taylor length scale $\lambda$ (final scale). The black solid trajectory represents the preferential path through the cascade, called instanton.}
	\label{fig:traje}
	\end{figure}

	In the remaining part of the script, the calculation leading towards the integral fluctuation theorem will be done.
	For the complete evolution process through the hierarchy of length scales $r$ from the integral length scale $L$ to the Taylor length scale $\lambda$, so-called cascade trajectories
	\begin{eqnarray}
		\left[u(\cdot) \right]=\{u_L,\dots,u_{\lambda}\}
		\label{eq:cascade trajec}
	\end{eqnarray}
	can be extracted based on the velocity increments $u_r$. 
	The notation $\left[u(\cdot) \right]$ indicates the entire path from the initial scale to the final scale (path through the state space) instead of a distinct value $u_r$.
	The effective dynamics of the cascade trajectories through scales are stochastic in scale due to the stochasticity of the variable $u_r$ itself.
	Such a path can be described by a Langevin equation, corresponding to the above-mentioned Fokker-Planck equation.
	An exemplification of some cascade trajectories or ``cascade paths'' are illustrated in Fig.~\ref{fig:traje}. 

	It should be kept in mind that a single trajectory $\left[u(\cdot) \right]$ is not one specific repeated break-up of turbulent structures due to the non-linear interactions in the flow.  
	Instead,  it is assumed that these individual trajectories are probes of the spatial structures of the flow field being composed of numerous different and simultaneously evolving cascade processes.
	Therefore, in this investigation a single trajectory should not be taken as one isolated large eddy evolving down-scale, rather the main assumption is that a large number of these trajectories reflect the statistics caused by the cascade process.
	
	In the spirit of non-equilibrium stochastic thermodynamics \cite{Seifert_2012} it is possible to associate with every individual cascade trajectory $\left[u(\cdot) \right]$ a total entropy variation $\Delta S_{tot}$ \cite{Seifert_2005,Seifert_2012,Sekimoto_2010,Nickelsen_2013,Reinke_2018}. 
	This entropy defined in the following is a stochastic entropy variation associated with the evolution of the cascade process given in terms of a Langevin or Fokker-Planck equation. 
	Therefore, entropy is defined here as a statistical or rather information-theoretic quantity, that is not equivalent to otherwise defined thermodynamic entropy of the fluid.

	\subsection{Validation of Integral Fluctuation Theorem}
	In this section, the validity of the integral fluctuation theorem based on the estimation of the total entropy production for each of the turbulence cascade trajectories is addressed. 
	The validity of the integral fluctuation theorem follows if a system is described by Fokker-Planck equation \cite{Seifert_2012}.\\
	
	\texttt{trajec} Based on velocity increments $u_r$ the cascade trajectories $\left[u(\cdot) \right]=\{u_L,\dots,u_{\lambda}\}$ from the integral length $L$ to the Taylor length $\lambda$ can be extracted from the data series of velocities $v(t)$. 
	A pop-up dialog box is generated to select if the start and end of the cascade trajectory should be adjusted. 
	If the pop-up dialog box is denied, then $\left[u(\cdot) \right]$ start at the integral length $L$ and end  at the Taylor length $\lambda$. 
	If this is confirmed, at the beginning of function \texttt{checkFT} a pop-up dialog box is generated to specify whether the start and/or the end of the cascade trajectory should be adjusted in multiples of integral respectively Taylor length scale.

\begin{figure}[h]
		\centering
		\includegraphics[width=0.5\textwidth]{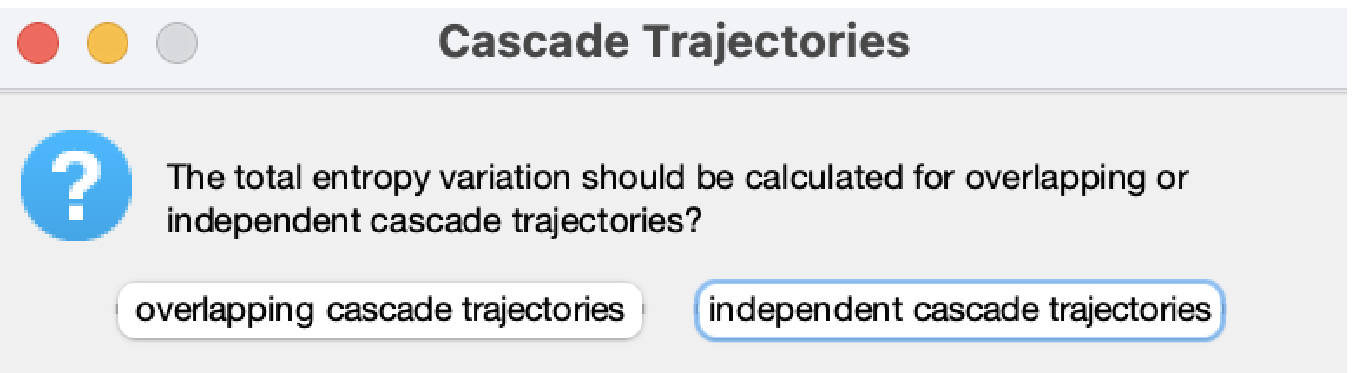}
		\caption{Question dialog box that allows the user to select whether the total entropy should be calculated for overlapping or independent cascade trajectories.}
		\label{fig:trajectory_switch}
	\end{figure}

	In addition a pop-up dialog box (see Fig. \ref{fig:trajectory_switch}) is generated to select whether the total entropy should be calculated for overlapping or independent cascade trajectories.
	Independent cascade trajectories are obtained by splitting the velocity time series $u(t)$ into intervals of integral length scales and calculating the velocity increments with respect to the point at the beginning of these intervals.
	For overlapping cascade trajectories these intervals of integral length scales are not independent but overlap.\\

	\texttt{dr\_ind}
	A pop-up dialog box is generated to define the separation of scales/step increment (in samples) referred to the sequence from large to small scales in the cascade trajectory. The proposed value in the pop-up dialog box is equal 
	to the Einstein-Markov length $\Delta_{EM}$.
	
	\newpage
	\texttt{data\_length} A pop-up dialog box is generated to select the percentage of the data length that should be used to perform the calculation of the total entropy variation (for example, the first 20 \% of the data).\\

	\texttt{checkFT} The set of measured cascade trajectories results in a set of total entropy variation values $\Delta S_{tot}$ (the same number of entropy values as the number of trajectories). 
	This function calculates the system entropy
	\begin{eqnarray}
		\Delta S_{sys}\left[u(\cdot) \right] =  - \ln{\left( \frac{p(u_\lambda, \lambda)}{p(u_L, L)} \right)},
	\end{eqnarray}
	medium entropy 
	\begin{eqnarray}
		\Delta S_{med}\left[u(\cdot) \right] &=&  - \int_{L}^{\lambda} \partial_r u_r \partial_{u_r}\varphi(u_r) dr,\\
		&=& + \int_{L}^{\lambda} \partial_r u_r \frac{D^{(1)}(u_{r},r)-\partial_{u_r}D^{(2)}(u_{r},r)/2}{D^{(2)}(u_{r},r)}dr,\nonumber
	\end{eqnarray}
	and the total entropy variation 
	\begin{eqnarray}
		\Delta S_{tot}\left[u(\cdot)\right] = \Delta S_{sys} + \Delta S_{med},
	\end{eqnarray}
	for all the independent cascade trajectories. 
	The numerical differentiation is approximated by the central difference quotient
	\begin{eqnarray}
		\partial_r u_r = \lim\limits_{r' \rightarrow r} \frac{u_{r'} - u_{r}}{r'-r}.
	\end{eqnarray}
	This numerical differentiation is performed for every individual extracted cascade trajectory $\left[u(\cdot) \right]$ in a sequence from large to small scales. 
	The integration in scale is approximated by using rectangles and a mid-point rule discretization of the scale intervals, therefore the integral takes the average of beginning and end of the discretization interval. 
	The probabilities of starting and ending of the cascade trajectories, $u_L$ and $u_{\lambda}$, can be estimated from the given data. The results depend slightly on the discretization rules and convention. However, the overall statements do not depend on it.\\

	\texttt{plot\_entropy} This function plots in Fig.~\ref{IFT_opti_IFT} (a) the empirical average $\langle e^{\mathrm{-}\Delta S_{tot}} \rangle_N$ of $\Delta S_{tot}$ as a function of the number, $N$ (sample size), of cascade trajectories $\left[u(\cdot) \right]$ with error bars. 
	In addition in Fig.~\ref{IFT_opti_IFT} (b), the probability density functions of the system, medium and total entropy are plotted together with the value of $\langle\Delta S_{tot}\rangle$, which should be larger than 0. 
	The  integral fluctuation theorem (IFT) expresses the integral balance between the entropy-consuming ($\Delta S_{tot}<0$) and the entropy-producing ($\Delta S_{tot}>0$) cascade trajectories and states 
	\begin{eqnarray}
		\langle e^{-\Delta S_{tot}} \rangle_{\left[u(\cdot) \right]} = \int e^{-\Delta S_{tot}}p\left(\Delta S_{tot}\right) d\Delta S_{tot} = 1.
		\label{eq:IFT}
	\end{eqnarray}

	\subsection{Pointwise optimization of Kramers-Moyal coefficients: IFT}
	\label{IFT_opti}
	This function performs a pointwise optimization of $D^{(1,2)}\left(u_r,r\right)$ towards the integral fluctuation theorem, which utilizes the experimental uncertainty in the determination of the Kramers-Moyal coefficients.
	The separation of scales/step increment (in samples) associated with the sequence from large to small scales in the (independent) cascade trajectories is set to a minimum step increment of 1 sample.  
	Note, we use here a separation that is less than or equal to the Einstein-Markov length of $\Delta_{EM}$.\\

	\texttt{iter} this pop-up dialog box is generated to enter the maximum number of iteration which will be performed for the optimization.\\
	
	\texttt{tol\_D1}, \texttt{tol\_D2} this pop-up dialog box is generated to specify the constraints/tolerance in percent of the coefficients $d_{ij}(r)$ which will be used to perform the optimization.\\
	
\begin{figure*}
		\centering
		\includegraphics[width=0.7\textwidth]{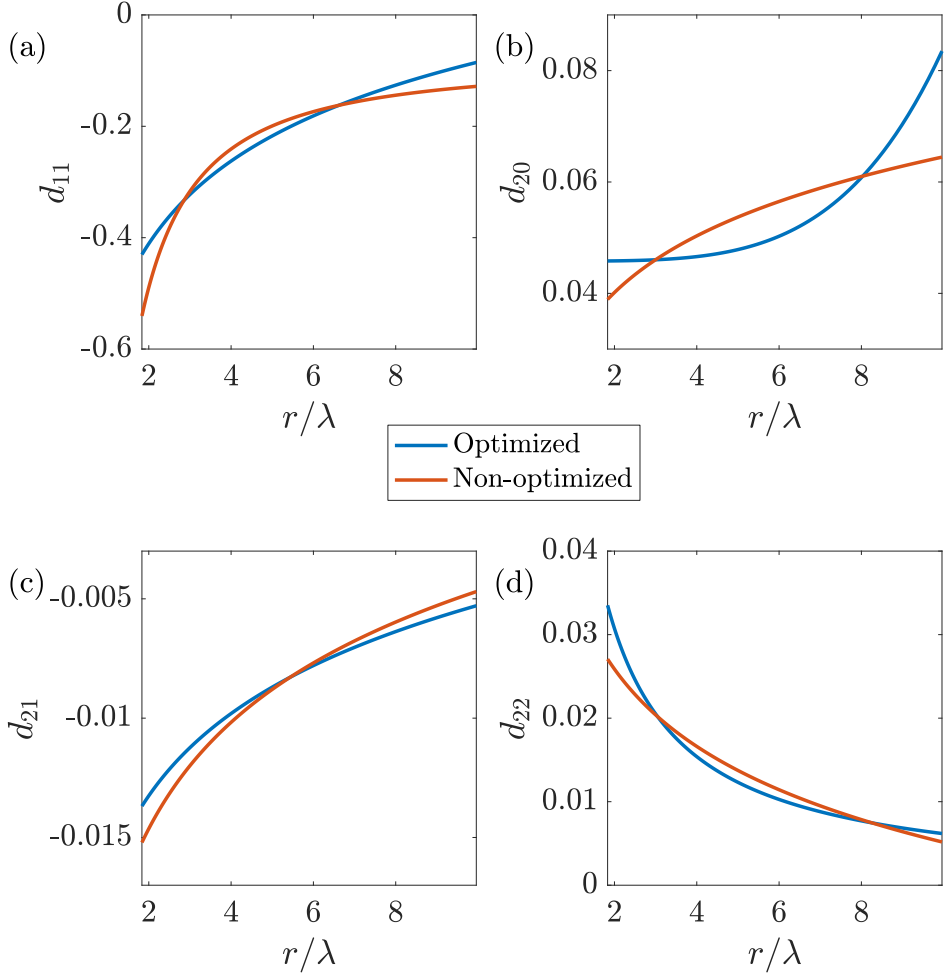}
		\caption {Coefficients $d_{ij}(r)$ of the optimized Kramers-Moyal coefficients using the surface fits with a linear function for $D^{(1)}\left(u_r,r\right)$ (see Eq.~\eqref{eq:D1andD2_1} and (a) for $d_{11}(r)$) and a parabolic function for $D^{(2)}\left(u_r,r\right)$ (see Eq.~\eqref{eq:D1andD2_2} and (b) for $d_{20}(r)$, (c) for $d_{21}(r)$ and (d) for $d_{22}(r)$) with respect to scale.}
		\label{IFT_opti__KM_a} 
	\end{figure*}

	 
	\texttt{OPTI\_IFT\_dij} This function performs the optimization of $D^{(1,2)}\left(u_r,r\right)$ at each scale and at each value of velocity increment in order to satisfy the integral fluctuation theorem with minimum possible error and plots the optimized $d_{ij}$ as a function of scale (see Fig.~\ref{IFT_opti__KM_a}). 
	The optimization procedure systematically changes $D^{(1,2)}\left(u_r,r\right)$ until the error function 
	\begin{eqnarray}
		\xi =|1-\langle e^{\mathrm{-}\Delta S_{tot}} \rangle_{max(N)}|
	\end{eqnarray}
	is minimized. 
	Within the optimization process, the user is asked which $d_{ij}(r)$ should be optimized.
	This optimization use the function \texttt{fmincon} implemented in MATLAB.\\

	%

\begin{figure*}
		\centering
		\includegraphics[width=0.4\textwidth]{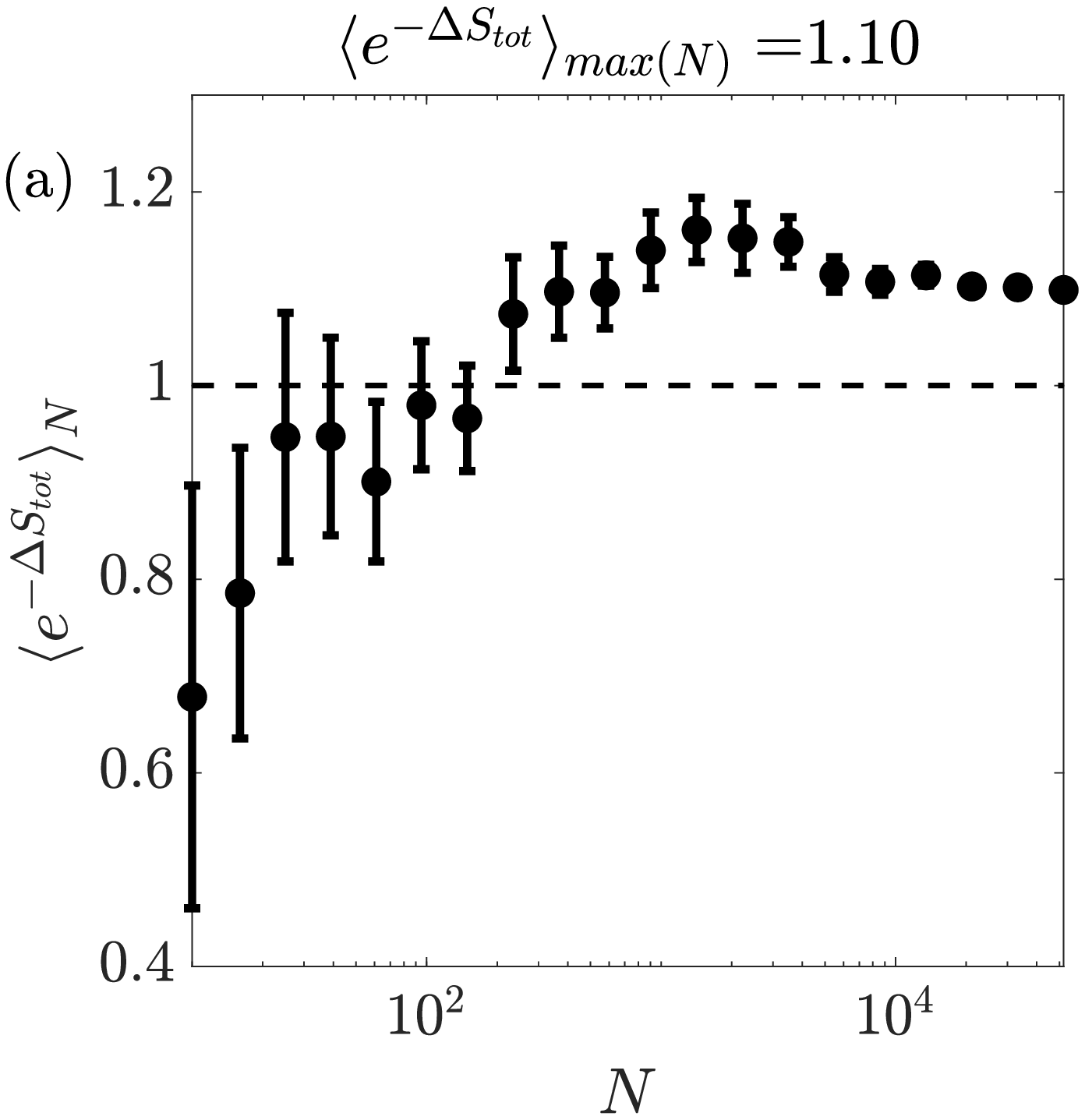}
		\includegraphics[width=0.4\textwidth]{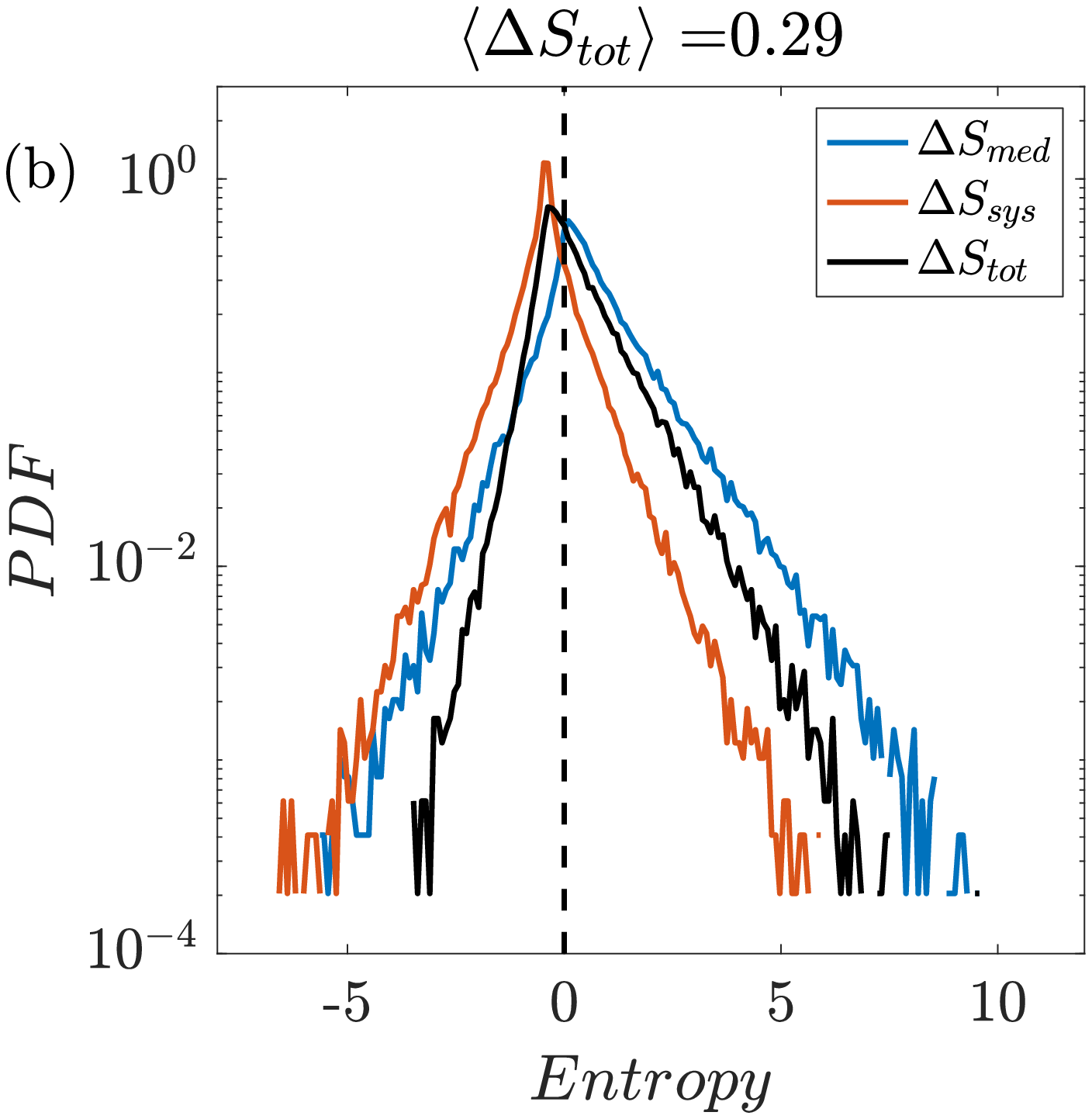}\\
		\includegraphics[width=0.4\textwidth]{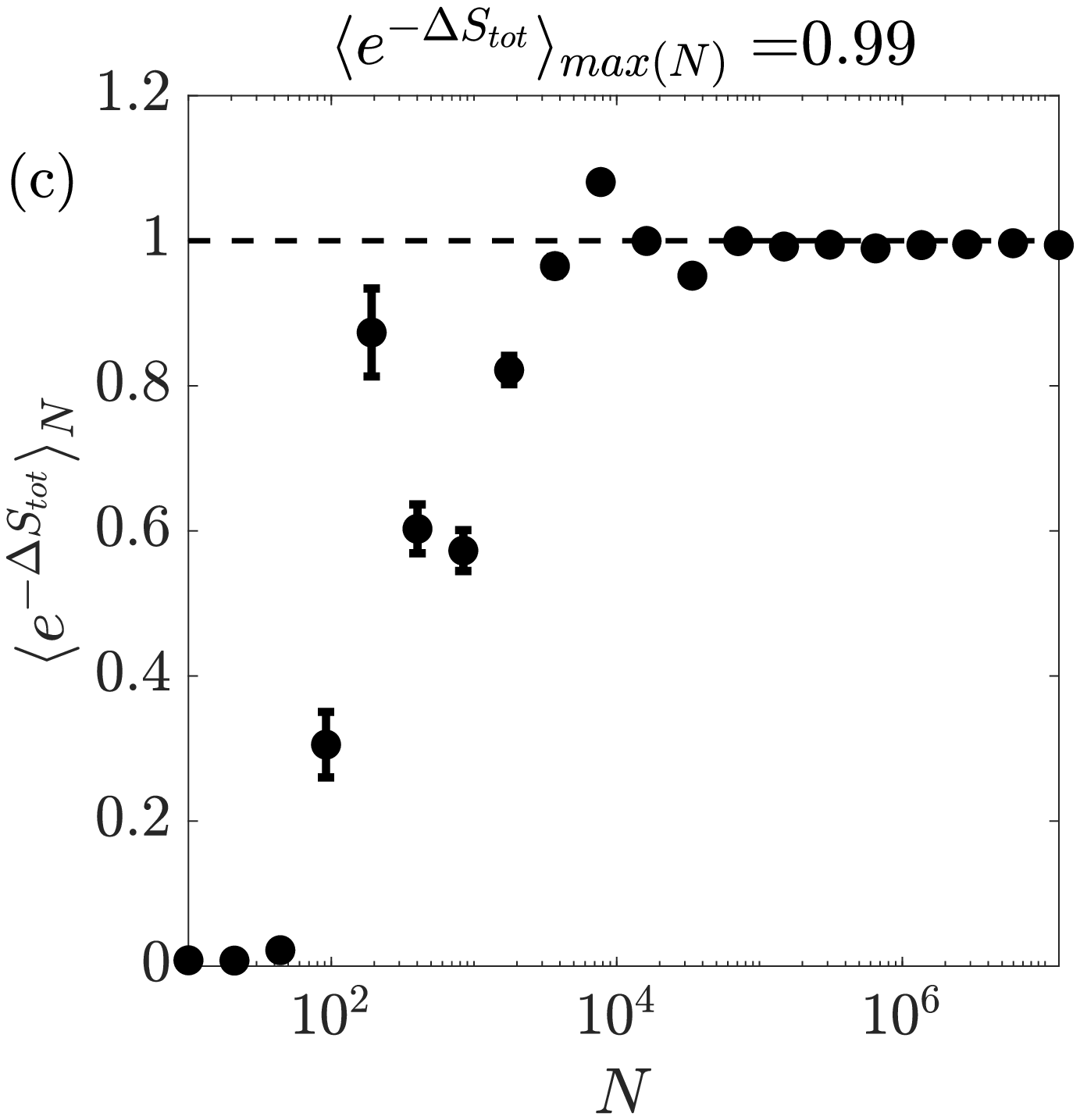}
		\includegraphics[width=0.41\textwidth]{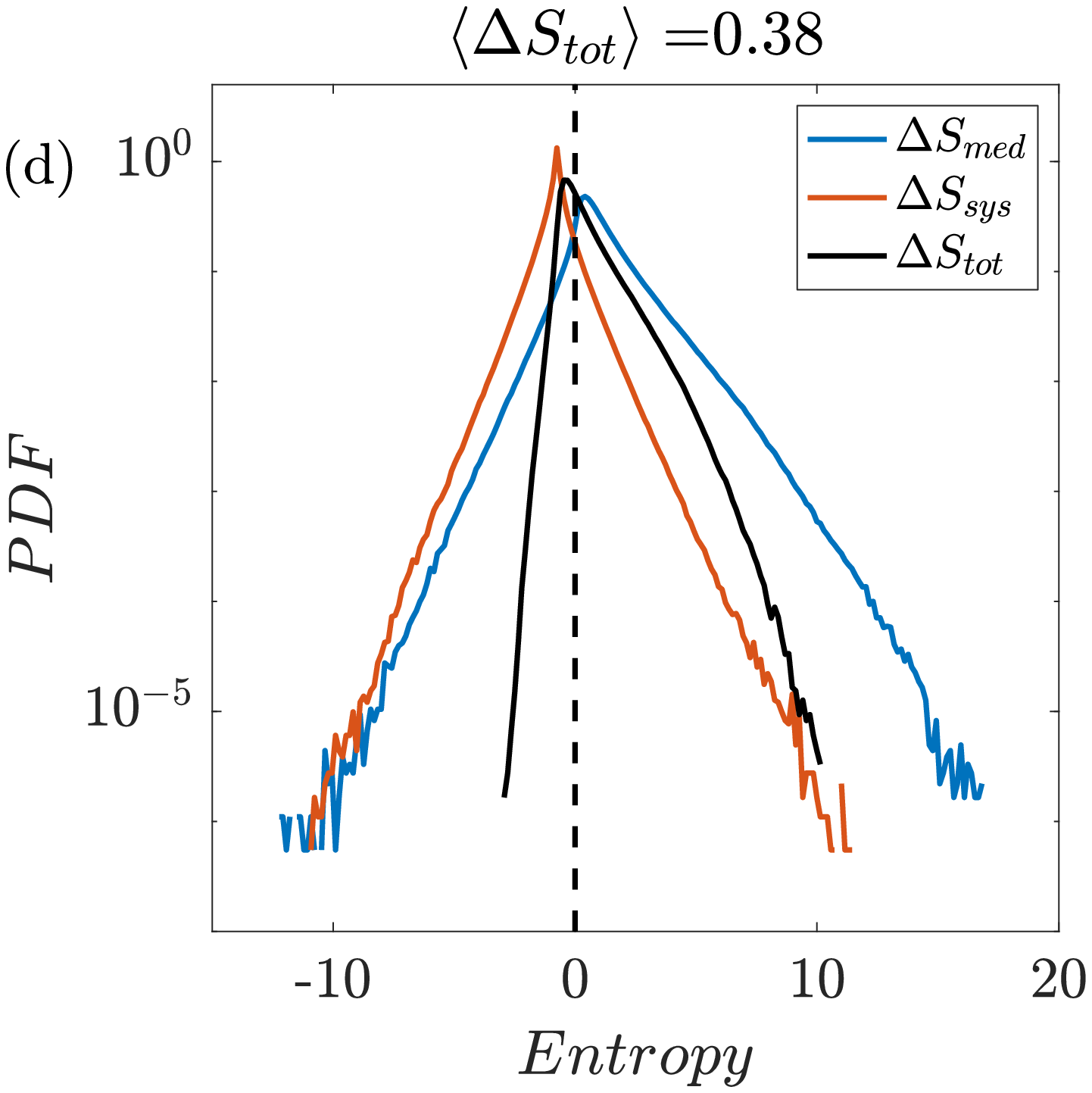}
		\caption {(a, c) empirical average $\langle e^{\mathrm{-}\Delta S_{tot}} \rangle_N$ of $\Delta S_{tot}$ as a function of the number $N$ (sample size) of cascade trajectories $\left[u(\cdot) \right]$ with error bars. According to the integral fluctuation theorem (IFT), the empirical average has to converge to the horizontal dashed line. (b, d) Probability density function of the system $S_{sys}$, medium $S_{med}$ and total entropy variation $S_{tot}$. For (a, b) the optimized coefficients $d_{ij}(r)$ derived from the first step of the pointwise optimization using the short-time propagator \cite{Risken} and (c, d) $d_{ij}(r)$ derived from the second step of the pointwise optimization towards the integral fluctuation theorem are applied.}
		\label{IFT_opti_IFT} 
	\end{figure*}

	Using the function \texttt{checkFT} and \texttt{plot\_entropy} with \texttt{dr\_ind=1} and overlapping cascade trajectories and the optimized Kramers-Moyal coefficients the results presented in Fig.~\ref{IFT_opti_IFT} are obtained for the calculation of $\Delta S_{tot}$.
	
	As it can be seen in Fig.~\ref{IFT_opti__KM_a}, this optimization is a fine-tuning of the coefficients, 
	but by comparing Fig.~\ref{IFT_opti_IFT} (a, b) and \ref{IFT_opti_IFT} (c, d), its impact on the IFT and the PDF of the system, medium and total entropy is clearly evident. 
	In this comparison, it must be taken into account that the separation of scales from large to small scales in the cascade trajectory is different and overlapping cascade trajectories are investigated here.
	%
	%

	\section{PART IV: Consistency check}
	\label{part4}

	In this section, the reconstruction of structure functions i.e. PDFs of the velocity increments at different scales using the estimated KMCs is performed. The comparison of reconstructed and experimental structure functions allows addressing the validity of the estimated KMCs.
	
	The overall scheme is: if the Markov properties are fulfilled and the cascade process can be described by a Fokker-Planck equation (i.e. Pawula theorem is fulfilled), then the knowledge of the  Kramers-Moyal coefficients allow to determine:
	\begin{enumerate}
		\item the conditional PDF given by the short time propagator (consequence of Markov property)
		\item the unconditioned PDF (consequence of the definition of conditioned probabilities) 
		\item the $k$-th order structure function (consequence of the definition).
		\item the entropy production of each cascade path (consequence of the definition)
		\item the fulfillment of the IFT (consequence of the validity of the Fokker-Planck description)
	\end{enumerate}
	
	\texttt{recon\_struc\_pdf} In addition to validating the integral fluctuation theorem, the validity of the estimated drift and diffusion coefficients is subsequently tested via the reconstruction of structure functions and probability density functions of velocity increments at different scales is performed. The comparison of reconstructed and experimental structure functions allows addressing the validity of the estimated KMCs.\\
	
\begin{figure}[h]
			\centering
			\includegraphics[width=0.5\textwidth]{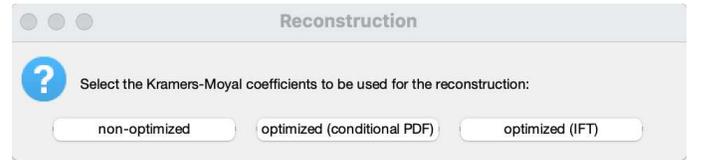}
			\caption{Question dialog box that allows the user to select which the Kramers-Moyal coefficients to be used for the reconstruction.}
			\label{fig:recon_coff}
	\end{figure}

	Initially, a pop-up dialog box (see Fig.~\ref{fig:recon_coff}) is generated to select the Kramers-Moyal coefficients to be used for the reconstruction.

\newpage
	Using the hierarchical ordering of scales \mbox{$L\geq r_0>r_1>...>r_n\geq\lambda$} (each of which is separated by $\Delta_{EM}$) the $k$-th order structure function can be reconstructed at the respective scale, for example $r_0$
	\begin{eqnarray}
		S_{rec}^k(r_0)=\int_{-\infty}^{\infty} u_{r_0}^k p_{rec}(u_{r_0}) du_{r_0}.
	\end{eqnarray}
	The reconstructed unconditional PDF of velocity increments at scale $r_0$ 
	\begin{eqnarray}
		\label{recon_pdf}
		p_{rec}(u_{r_0})=\int_{-\infty}^{\infty} p_{stp}(u_{r_0}|u_{L}) p(u_{L})  du_{L}
	\end{eqnarray}
	is estimated by integrating the reconstructed conditional PDF $p_{stp}(u_{r_0}|u_{L})$, using the short time propagator in Eq.~\eqref{eq.P_STP}, that solely depends on the Kramers-Moyal coefficients $D^{(1,2)}$. 
	
	From Eq.~\eqref{recon_pdf}, it can be seen that the unconditional PDF at the larger scale is needed to perform the reconstruction at the next smaller scale in the hierarchical ordering.
	In a second pop-up dialog box (see Fig.~\ref{fig:recon_coff}) the user must select from the following two scenarios:
		\begin{figure}[h]
			\centering
			\includegraphics[width=0.5\textwidth]{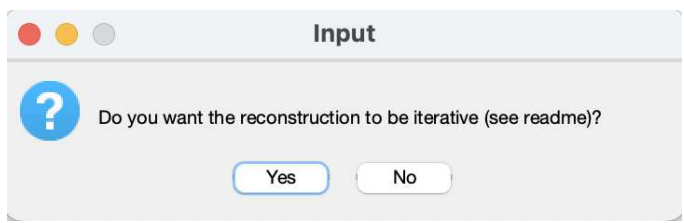}
			\caption{Question dialog box that allows the user to select if the reconstruction should be iterative (2. scenario) or not (1. scenario).}
			\label{fig:recon_iter}
	\end{figure}
	
	\begin{itemize}
		\item In the first scenario, Eq.~\eqref{recon_pdf} is solved using the unconditional PDF at the larger scale estimated directly from the experimental data at each scale independently.
		In the case the optimized $D^{(1,2)}$ (conditional PDF) or optimized $D^{(1,2)}$ (IFT) is chosen, this method often results in very good agreement, because the optimization of the drift and diffusion coefficient is performed at each scale independently.
		The non-optimized $D^{(1,2)}$ leads to a poorer agreement in the experimental and reconstructed structure functions and PDF's.
		\item In the second scenario, Eq.~\eqref{recon_pdf} is solved using the unconditional PDF at the larger scale from the experimental data only at scale $L$ in the first step and an iterative process is used towards smaller scales. 
		In other words, the entire turbulence cascade is reconstructed using only the initial solution at the largest scale from the experimental data.
		Typically, this method leads to very good agreement at large scales, whereas the error between experiment and reconstruction increases towards smaller scales using the iterative approach. 
		As mentioned earlier, this is because the drift and diffusion coefficients are estimated locally at each scale by optimization.
		Accordingly, in this iterative approach, minor deviations at the largest scale influence the reconstruction at the next smaller scales. 
		This error may be reduced by performing a scale dependent optimization from large to small scales. 
		Overall, the convergence of the statistics at large increment values which are rarely encountered should also be taken into account when making this comparison.
	\end{itemize}
	
\begin{figure*}
		\centering
		\includegraphics[width=0.8\textwidth]{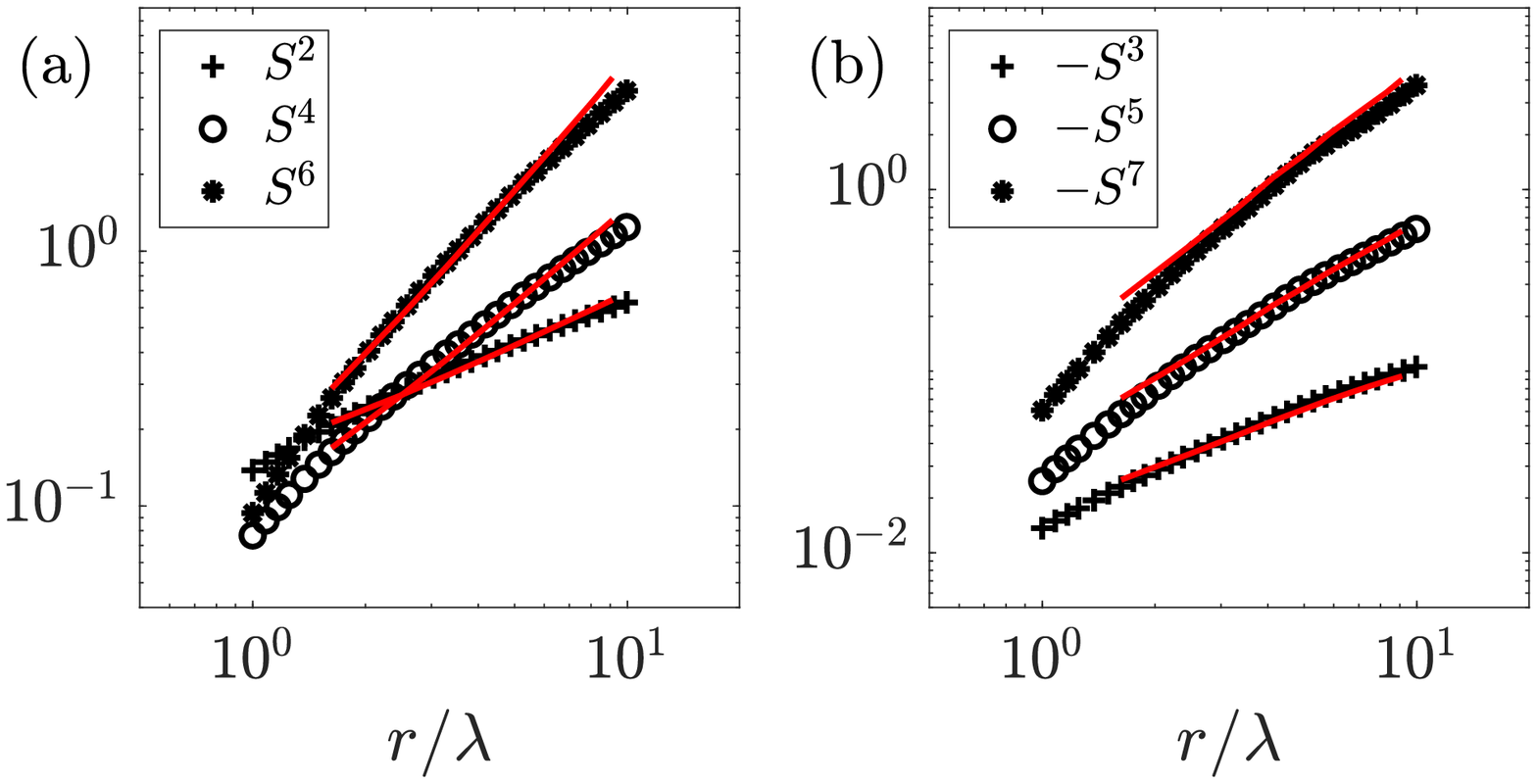}\\
		\includegraphics[width=0.8\textwidth]{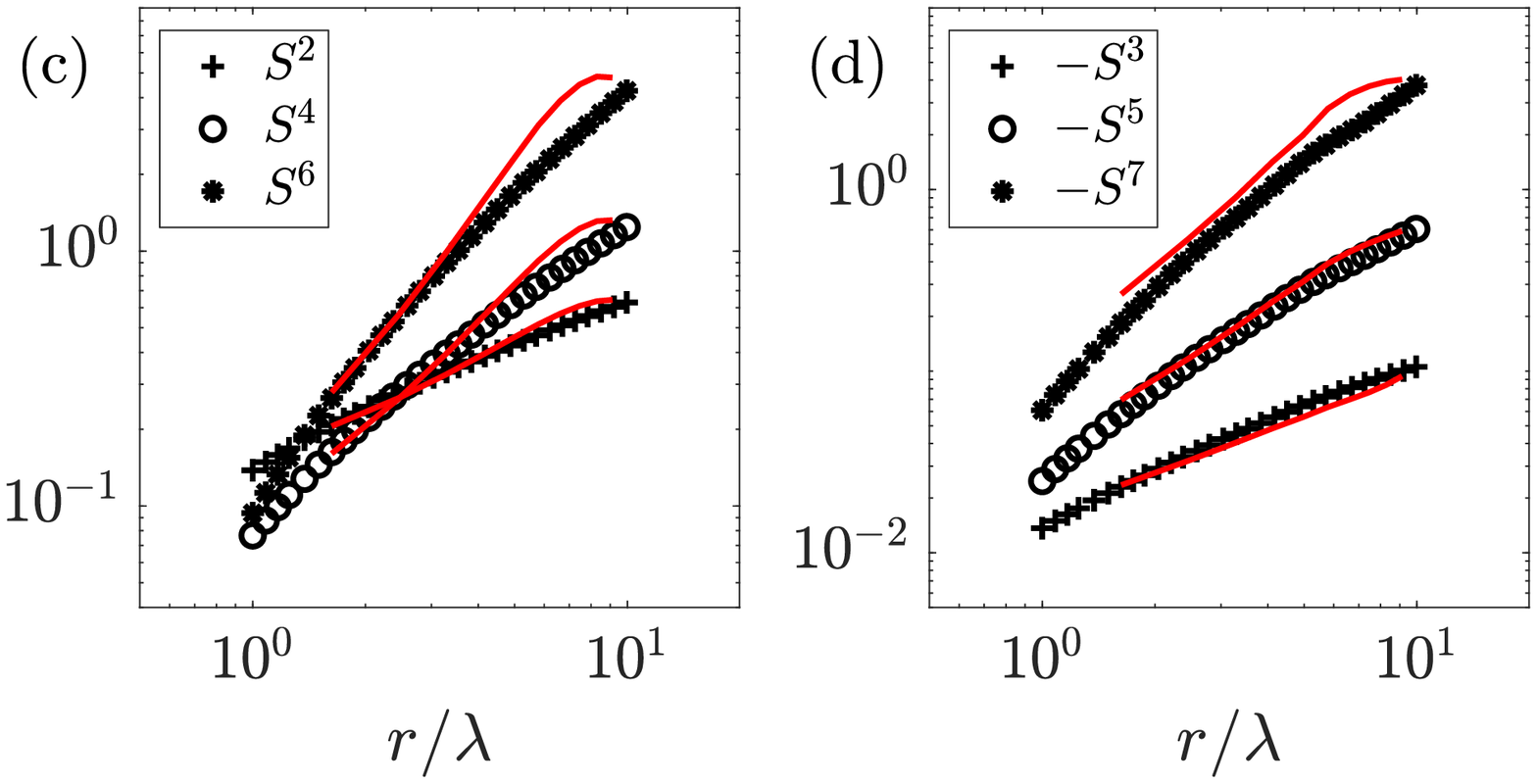}
		\caption {Comparison of experimental (black symbols) and reconstructed (red solid lines) $k$-th order structure function for (a, c) $k={2,4,6}$ and (b, d) $k={3,5,7}$ for scales $\lambda \leq r \leq L$. Reconstruction using the first (top) and second (bottom) scenario.}
		\label{fig:struc_recon_a} 
	\end{figure*}

\begin{figure*}
		\centering
		\includegraphics[width=0.8\textwidth]{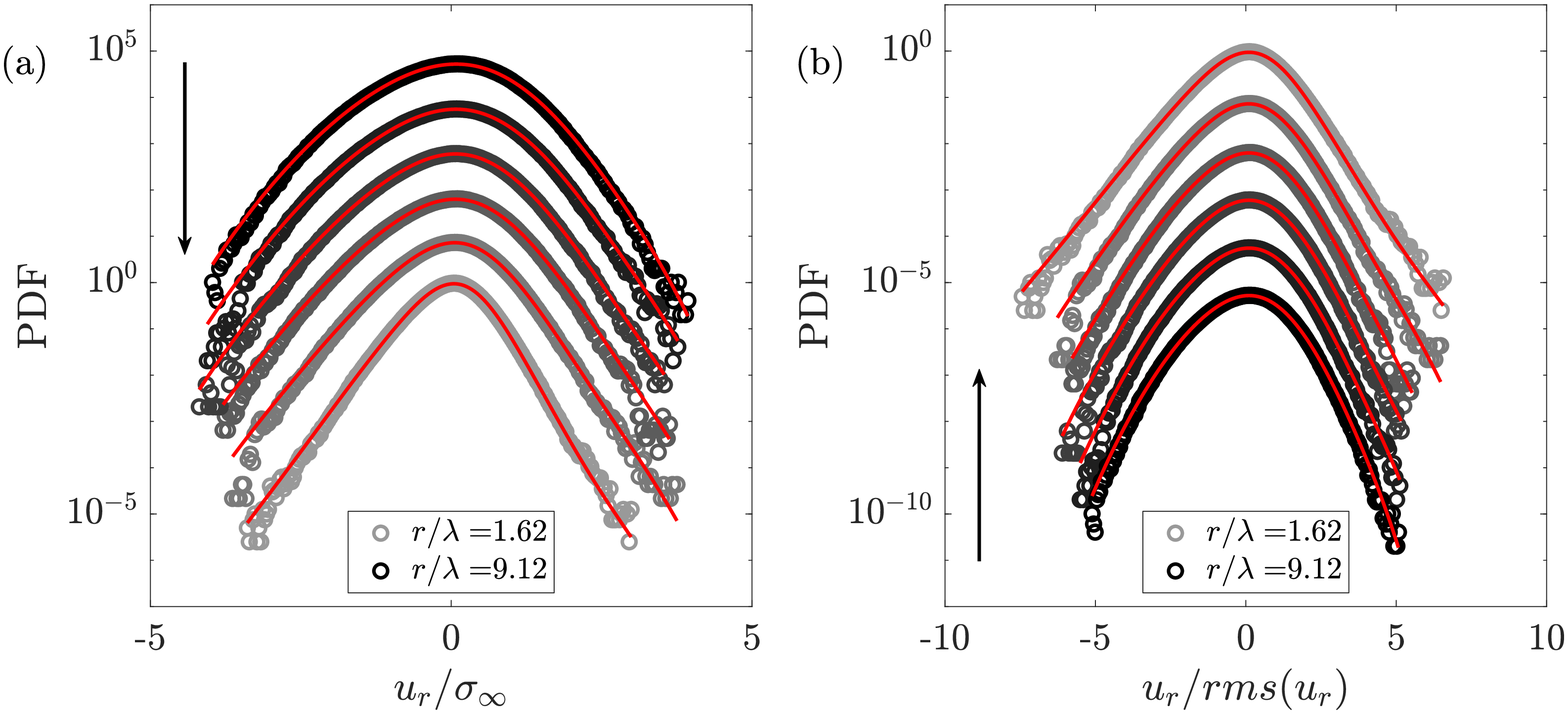}
		\includegraphics[width=0.8\textwidth]{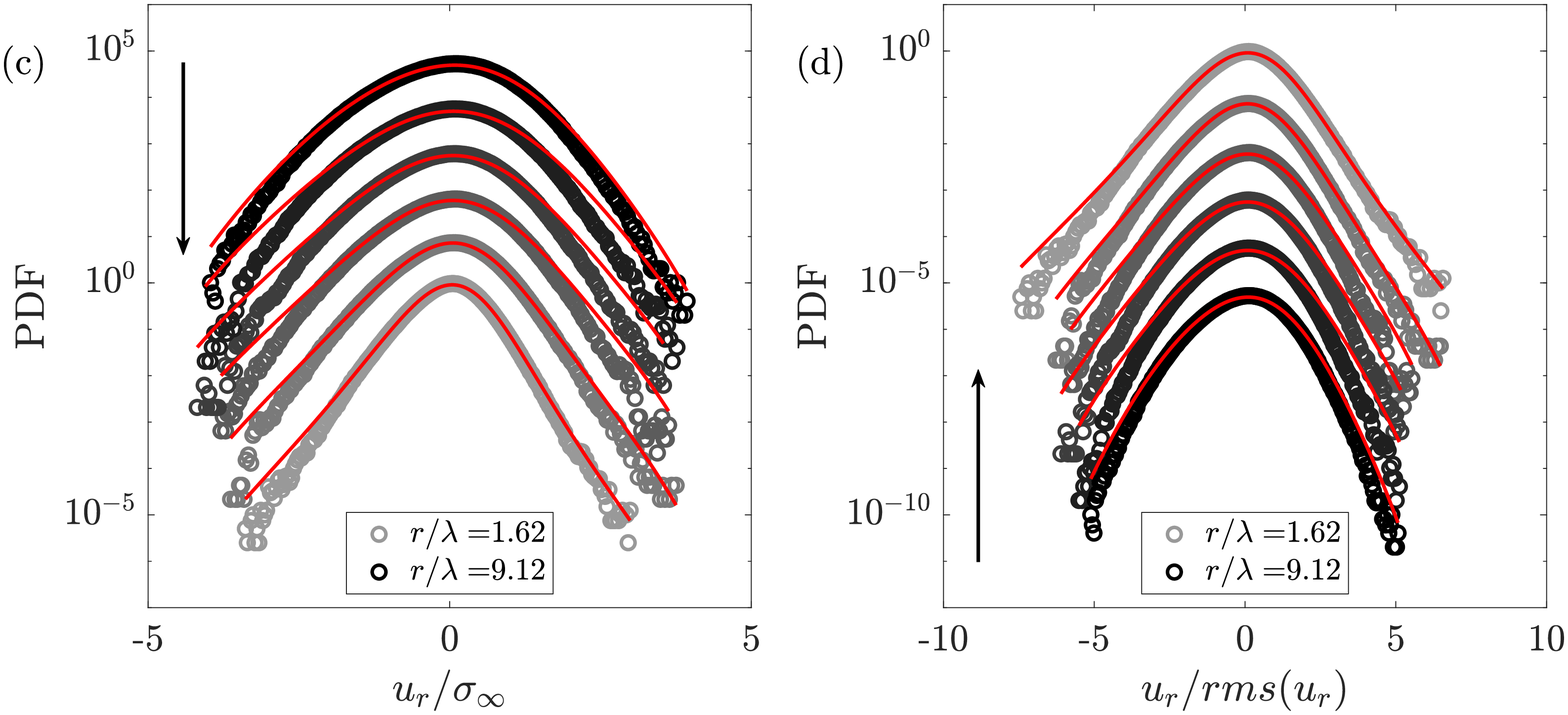}
		\caption {Comparison of experimental (black symbols) and reconstructed (red solid lines) PDF of velocity increments at various scales. The velocity increment $u_r$ is normalized using $\sigma_{\infty}$ (left) or the root mean square of the velocity increment time series at the corresponding scale $rms(u_r)$ (right). For better visualization, the PDFs are shifted in the vertical direction. Reconstruction using the first (a,b) and second (c,d) scenario. The black arrow indicates the direction of decreasing scales.}
		\label{fig:struc_recon_b}
	\end{figure*}

	As just described, the structure function and the probability density functions of velocity increments are obtained from $p_{stp}$. 
	In Fig.~\ref{fig:struc_recon_a} the comparison of experimental and reconstructed $k$-th order structure function with \mbox{$k={2-7}$} for scales $\lambda \leq r \leq L$ are shown.
	In Fig.~\ref{fig:struc_recon_b} the reconstructed probability density functions of velocity increments are plotted.

	\section{Conclusion} \label{sec:summary}
	We present a user-friendly open-source MATLAB\textsuperscript{\textregistered} package which helps the user to perform a standard analysis of given turbulent data and extracts the stochastic equations describing the scale-dependent cascade process in turbulent flows through Fokker-Planck equations along with its application to the Integral fluctuation theorem.
	This user-friendly open source package greatly enhances the practicability and availability of both the standard analyses already established in turbulence research and this new method.
	Moreover, we do not know of a comparably comprehensive collection of standard analyses in turbulence research in a user-friendly package.
	All in all, we believe that this package is of great interest especially to young scientists and newcomers to a stochastic approach to turbulence. 
	The package possesses a high reuse potential for researchers and students in the field of turbulence research.

	This open-source package can also be used by researchers outside of the field of turbulence for the investigation of the statistical properties of complex time series such as financial, surface height or seismic time series to name just a few.
	Using this package may contribute to new insights into the complex characteristics of length/time scale-dependent processes ranging from physics to meteorology, biology systems, finance, economy, surface science and medicine, see also \cite{Friedrich_2011}.
	This paper covers only the discussion of turbulent flows satisfying the assumption of homogeneous, isotropic turbulence and the Taylor hypothesis of frozen turbulence.
	In principle the tools of this package can be applied to turbulence data in other configurations and turbulent flows with non ideal HIT conditions as shown for non fully developed turbulence \cite{luck1999experimental} or turbulence with shear \cite{Reisner_1999,Ali_2019}.
	\begin{acknowledgments}
		The software resulted from funded research. We acknowledge financial support by Volkswagen Foundation (VolkswagenStiftung): 96528.
		We acknowledge the following people for helpful discussions and testing pre-version of the package  \mbox{A. Abdulrazek}, J. Ehrich, A. Engel, J. Friedrich, A. Girard, G. G\"ulker, P. G. Lind, D. Nickelsen, N. Reinke, M. Obligado, \mbox{T. Wester}. 
	\end{acknowledgments}
	
	\section{DATA AVAILABILITY}
	The package is available as free software, under the GNU General Public License (GPL) version 3.
	The package (source code and standalone applications (64-bit) for Windows, macOS and Linux) and an typical dataset can be downloaded from the repository on GitHub or Matlab File Exchange Server to replicate all the results presented in this article. 
	Support is available at  \mbox{\url{github.com/andre-fuchs-uni-oldenburg/OPEN_FPE_IFT}}, where questions can be posted and generally receive quick responses from the authors.\\
	
	\newpage
	\noindent
	{\bf Name:} OPEN\_FPE\_IFT\\
	{\bf Persistent identifier:} GitHub\\ \url{https://github.com/andre-fuchs-uni-oldenburg/OPEN_FPE_IFT}\\
	{\bf Persistent identifier:} Matlab File Exchange Server\\
	\url{https://www.mathworks.com/matlabcentral/fileexchange/80551-open_fpe_ift}\\
	{\bf Publisher:} Andr{\'e} Fuchs\\
	{\bf Version published:} 4.0\\
	{\bf Date published:} 15/06/22\\
	{\bf Operating system:} Windows, macOS and Linux\\
	{\bf Programming language:} MATLAB\\

	\newpage
	\appendix
	\section{List of Nomenclature/Abbreviations} \label{app:technical}
	\paragraph{Latin symbols}$\;$
	\begin{table}[htbp]
		\begin{tabular}{ll}
			$C_\epsilon$ 										& normalized turbulent kinetic energy dissipation rate\\
			$D^{(k)}\left(u_r,r\right)$ 				    & 						$k$-th order Kramers-Moyal coefficients\\
			$E(f)$ 														& energy spectral density in frequency domain\\
			$E(k)$ 														& energy spectral density  in wave number domain\\
			$E_{kin}$ 												& total kinetic energy\\
			$K$ 														& kurtosis\\
			$Fs$ 														& sampling frequency \\
			$L$ 															& integral length scale\\
			$M^{(k)}\left(u_r,r,\Delta r\right)$ 	&	$k$-th order conditional moment\\
			$p(u_r)$ 												 &  probability density function of $u_r$ \\
			$p\left(u_{r'} | u_r \right)$ 					& conditioned PDF of velocity increments  for a pair\\ & of two scales with $r'<r$\\
			$r=-\tau\langle u\rangle$ 					& spatial length scale\\
			$Re$ 														& Reynolds number\\
			$Re_\lambda$ 								& Taylor Reynolds number\\
			$S$ 													& skewness\\
	\end{tabular}
\end{table}

\begin{table}[htbp]
	\begin{tabular}{ll}
			$Ti$ 													& turbulence intensity\\
			$u$ 														& streamwise velocity \\
			$\left<u\right>$ 									& mean streamwise velocity\\
			$\widetilde{u}$ & streamwise velocity fluctuations\\
			$u'$ & standard deviation\\
			$u_{\tau}(t)$ 			& temporal velocity increment at time-scale $\tau$\\
			$u_r(t)=-u_{\tau}(t)$ 							& spatial velocity increment at length scale $r$\\
			$\left[u(\cdot) \right]$						&	cascade trajectory\\
			$k_0$ & shift of the argument of the fit function,\\
			&  i.e. along the wave number axis 
	\end{tabular}
\end{table}

	\newpage
	\paragraph{Greek symbols}$\;$
	\begin{table}[htbp]
		\begin{tabular}{ll}
			$\Delta S_{sys}\left[u(\cdot) \right]$  			& system entropy\\
			$\Delta S_{med}\left[u(\cdot) \right]$ 				&  medium entropy \\
			$\Delta S_{tot}\left[u(\cdot)\right]$ 				& total entropy variation\\	
			$\langle \epsilon\rangle$ 									& mean energy dissipation rate\\
			$\zeta_k$ 																& scaling exponent of  structure functions\\
			$\eta$ 																		& Kolmogorov length scale\\
			$\lambda$ 																& Taylor length scale\\ 
			$\mu$ 																		& intermittency coefficient\\
			$\nu$ 																		& kinematic viscosity of the fluid\\
			$\rho$ 																		& fluid density\\
			$\sigma$ 																& standard deviation of $u$\\
		\end{tabular}
	\end{table}

	\newpage
	\bibliography{biblio}
\end{document}